\documentclass[11pt]{article}

\usepackage[preprint]{acl}

\usepackage{times}
\usepackage{latexsym}

\usepackage[T1]{fontenc}

\usepackage[utf8]{inputenc}

\usepackage{microtype}

\usepackage{inconsolata}

\usepackage{graphicx}
\graphicspath{{figures/}{../figures/}}

\usepackage{booktabs}
\usepackage[table]{xcolor}
\usepackage{multirow}
\usepackage{amsmath}        
\usepackage{amssymb}        
\usepackage{fvextra}   

\title{Beyond the GUI Paradigm: Do Mobile Agents Need the Phone Screen?}

\author{
\parbox{\textwidth}{\centering
\vskip 1.5em
\textbf{Li Gu}$^{1,2*\dag}$ \quad
\textbf{Zihuan Jiang}$^{3*}$ \quad
\textbf{Linqiang Guo}$^{2}$ \quad
\textbf{Zhixiang Chi}$^{3}$ \quad
\textbf{Ziqiang Wang}$^{1,2}$ \\
\textbf{Huan Liu}$^{4}$ \quad
\textbf{Yuanhao Yu}$^{4}$ \quad
\textbf{Tse-Hsun Chen}$^{2}$ \quad
\textbf{Yang Wang}$^{1,2}$ \\[0.8em]
{\small\mdseries
$^1$Mila -- Qu\'ebec AI Institute \quad
$^2$Concordia University \quad
$^3$University of Toronto \quad
$^4$McMaster University \\
$^*$Equal contribution \quad
$^\dag$Project lead
}
}
}

\begin{document}
\maketitle
\begin{abstract}
Recent advances in mobile agents are dominated by the GUI paradigm, in which agents perceive UI information and emit screen interactions. However, mobile platforms also expose a command-line interface (CLI) that provides direct access to device services and data. We argue CLI deserves first-class consideration alongside GUI. We evaluate three coding agents (Claude Code, Terminus-2, mini-swe-agent) across four model APIs on AndroidWorld and MobileWorld without any mobile-specific post-training, comparing against three reproducible GUI baselines (GUI-Owl-1.5-32B, MAI-UI, Qwen3-VL-32B). Claude Code (Opus 4.7) reaches 71.8\% and 51.9\%, outperforming every reproducible GUI baseline (69.3/68.1/57.8\% on AndroidWorld; 43.2/26.3/13.3\% on MobileWorld), while every other CLI configuration remains competitive. To establish the paradigm's ceiling, we provide oracle CLI solutions that reach 88.8\% on AndroidWorld (103/116 tasks CLI-solvable) and 86.3\% on MobileWorld (101/117 tasks CLI-solvable), indicating substantial room for future improvement. To cover everyday user intents beyond the GUI scope, we introduce the \textbf{CLI-Advantage Task Suite}, comprising 45 templates across five categories: bulk operations, multi-condition filtering, aggregation, cross-app workflows, and hidden device state. Every CLI agent outperforms every GUI baseline in all five categories, with substantially fewer steps per task (10.7 vs.\ 18.6). To support future research on mobile CLI agents, we will open-source agent implementations, oracle solutions, the CLI-Advantage suite, and evaluation infrastructure.

\end{abstract}

\section{Introduction}

\begin{figure*}[!t]
\centering
\includegraphics[width=\textwidth]{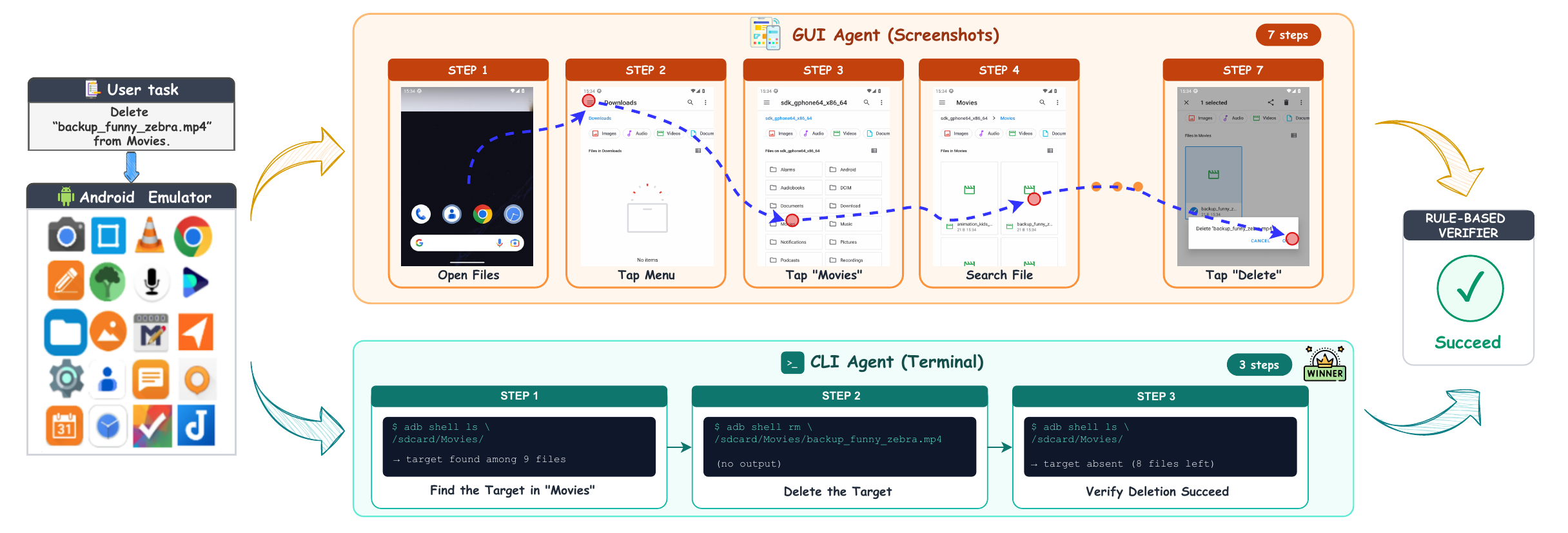}
\caption{\textbf{GUI vs.\ CLI execution traces on a representative mobile task.}
Both agents are asked to delete \texttt{backup\_funny\_zebra.mp4} from the Movies folder
and proceed through three phases---\emph{find the target}, \emph{delete the target},
\emph{verify deletion}. The GUI agent (top) consumes screenshots and emits taps,
unfolding the phases over 7 UI steps (open Files $\rightarrow$ navigate menu
$\rightarrow$ enter \texttt{Movies} $\rightarrow$ search $\rightarrow$ confirm delete).
The CLI agent (bottom) consumes terminal \texttt{stdout} and emits ADB shell commands,
collapsing the same phases into 3 steps (\texttt{ls}, \texttt{rm}, \texttt{ls}).
A rule-based verifier scores both trajectories on their final state,
isolating the interface paradigm as the only variable.}
\label{fig:paradigm_comparison}
\end{figure*}

Mobile agents are autonomous systems that interact with smartphones via natural language to complete user tasks \citep{wu2024foundations, hu2025agents, liu2025llm}. The dominant paradigm treats this as a vision problem: agents observe the phone screen and produce GUI actions, including tapping, swiping, and typing, to navigate applications, mimicking human behavior. Substantial research effort has advanced this paradigm across benchmarks, models, training algorithms, and infrastructure \citep{bai2024digirl, zhou2025mai, wang2025ui, xu2025mobilerl, gu2026generalization, xu2026mobile, team2026ui}.

However, this paradigm is not the only way mobile agents can operate on mobile platforms. Android is a Linux-based platform that exposes a full command-line interface (CLI) through the Android Debug Bridge (ADB), allowing programmatic access to device services and data without ever rendering a pixel \citep{android_adb}. In addition, ADB can expose device states, including battery history, notification logs, and background processes, that never appear on any screen \citep{android_dumpsys}, enabling the CLI paradigm to reach beyond what GUI agents can observe. Meanwhile, frontier coding agents \citep{anthropic2025claudecode, singh2025openai, minimax2026m27} have demonstrated strong capabilities on challenging long-horizon terminal tasks in Linux environments \citep{merrill2026terminalbench}, operating in a text-only modality native to large language models. Motivated by these observations, we argue that this CLI paradigm deserves first-class consideration alongside GUI for mobile agents, and we test the claim by answering two questions. \textbf{Q1 (viability):} Are general coding agents using ADB alone competitive with GUI agents that are post-trained on mobile datasets via Supervised fine-tuning and Reinforcement earning? \textbf{Q2 (superiority):} Are there everyday user intents better served by CLI but ignored by existing GUI-designed benchmarks?
 
To answer Q1, we evaluate three frontier coding agents on two standard mobile-agent benchmarks, AndroidWorld \citep{rawles2025androidworld} and MobileWorld \citep{kong2025mobileworld}, against three reproducible GUI baselines: GUI-Owl-1.5-32B \citep{xu2026mobile}, MAI-UI-8B \citep{zhou2025mai}, and Qwen3-VL-32B \citep{bai2025qwen3}. All experiments are conducted under the same evaluation protocol, with identical tasks, rule-based verifiers, and metrics, and the primary intended variable is the interface paradigm. The coding agents are Claude Code \citep{anthropic2025claudecode}, Terminus-2 \citep{merrill2026terminalbench}, and mini-swe-agent \citep{yang2024swe}, all previously validated on Terminal-Bench, and each receives only terminal stdout as input (no screenshots or accessibility trees) and outputs bash commands, with no mobile-specific training. To bridge the Android platform gap, we manually construct a system prompt containing domain knowledge and strategies derived from public developer documentation, without leakage of task-specific or app-specific information. Motivated by recent advances in harness engineering \citep{tian2026swe, lopopolo2026harness, lee2026meta}, we further design general-purpose tools that wrap raw ADB bash commands to boost agent performance. Finally, to establish an upper bound on the CLI paradigm's capability, we construct oracle bash command solutions through human-LLM collaboration, manually reviewed by three external professionals, showing which tasks are CLI-solvable in AndroidWorld and MobileWorld.

To answer Q2, we evaluate both CLI and GUI agents on tasks that the GUI paradigm was not designed to accommodate. Since existing benchmarks were designed around the GUI paradigm and consequently exclude user intents that no tap sequence can serve, we introduce the \textbf{CLI-Advantage Task Suite}: 45 LLM-generated, human-verified task templates spanning five categories. Three categories have operations the app UI does not expose at scale: \emph{bulk operations} (``delete every \texttt{.apk} file in Downloads''), \emph{multi-condition filtering} (``list expenses above my average spending''), and \emph{aggregation and top-K} (``top-3 expense categories this month''). The remaining two categories address state the app layer does not surface: \emph{cross-app workflows} (``did anyone text me during my meetings yesterday?'') and \emph{hidden device state} (``which apps have been granted location permission''). Each task corresponds to an AndroidWorld-compatible rule-based verifier and a hand-verified oracle CLI solution for reliable assessment.

Our experiments yield strong evidence that the CLI paradigm deserves first-class status alongside GUI for mobile agents. On AndroidWorld (AW) and MobileWorld (MW), Claude Code (Opus 4.7) achieves 71.8\% and 51.9\%, outperforming every reproducible GUI baseline (GUI-Owl-1.5-32B, MAI-UI-8B, and Qwen3-VL-32B at 69.3/68.1/57.8\% on AW and 43.2/26.3/13.3\% on MW). Moreover, mini-swe-agent with Sonnet 4.6 reaches 70.7\% on AndroidWorld, and every other CLI configuration we evaluated (three harnesses and three additional model APIs) is competitive with reproducible GUI baselines, showing a paradigm-level property rather than an artifact of a single agent. Generic ADB tools substantially improve CLI agents with weaker models in both success rate and cost efficiency, but provide no consistent benefit to stronger ones, indicating that tools should be gated on model strength. A trajectory-level error analysis reveals that GUI and CLI agents exhibit distinct failure modes, indicating that the two paradigms require different improvement strategies. On the CLI-Advantage Suite, the gap widens substantially: every CLI agent we evaluated outperforms every GUI baseline (60.7--68.9\% vs.\ 22.2--33.3\%), leading across all five categories, particularly on cross-app workflows where GUI agents cap at $\leq$11\% regardless of model scale. Beyond
success rates, CLI agents also use substantially fewer steps per task than GUI methods (10.7 vs.\ 18.6 mean steps). Finally, oracle CLI agents reach 88.8\% on AndroidWorld (103/116 tasks CLI-solvable) and 86.3\% on MobileWorld (101/117 tasks CLI-solvable), showing that current agents still fall short of the CLI paradigm's 
ceiling. 

\section{Related Work}
\label{sec:related-work}

\paragraph{Mobile GUI agents.}
Mobile GUI agents aim to complete natural-language tasks on smartphones by perceiving screen state and emitting touch, swipe, and type actions, a setting that couples long-horizon planning with fine-grained visual grounding under variable layouts and dynamic app state \citep{wu2024foundations, hu2025agents, liu2025llm}. Prior work falls into two categories. Multi-agent approaches orchestrate proprietary VLMs into pipelines that plan, act, reflect, and remember around a screenshot loop \citep{zhang2023appagent, wang2024mobileagent, wang2024mobileagentv2, xu2025mobileagentv3, wen2024autodroid, guo2025agentsama}. End-to-end approaches instead post-train an open VLM on UI trajectories, first with supervised fine-tuning \citep{qin2025uitars, wang2025ui, wang2025uivenus, team2026ui, xu2026mobile, zhou2025mai, yan2025step, wu2024osatlas, xu2024aguvis, hong2024cogagent, cheng2024seeclick, lin2025showui, you2024ferretui, yang2025magma} and, more recently, with online reinforcement learning in live Android environments \citep{bai2024digirl, shi2025mobilegui, dai2025vdroid, xu2025mobilerl, gu2026generalization}. However, both categories share the same interface: the observation is the rendered UI (screenshot pixels or accessibility tree) and the action space is pixel-level coordinates, which makes performance bounded by grounding accuracy, blind to off-screen system state (battery history, background processes, notification logs), brittle to layout shifts, and unable to express bulk or cross-app operations that the rendered surface does not expose. Different from these works, we drive Android through \texttt{adb} alone --- terminal text as the only observation, shell commands as the action space --- sidestepping these limitations.

\paragraph{Mobile agent benchmarks.}
Mobile agent benchmarks fall into two categories. Offline datasets compile human-annotated Android trajectories and grade agents by per-step action matching against the human references \citep{rawles2023aitw, li2024androidcontrol}, but static demonstrations restrict evaluation to step-level accuracy and cannot capture trajectory-level task success. To address this, online interactive benchmarks boot a live Android emulator and grade trajectory-level success through rule-based or LLM-as-judge verifiers \citep{rawles2025androidworld, xu2025androidlab, wang2024mobileagentbench, chen2025spabench, chai2025a3}; MobileWorld \citep{kong2025mobileworld} further extends this setting with broader app coverage, networked client-server applications, longer-horizon and cross-application tasks, agent-user interaction, and MCP-augmented environments. However, tasks in these benchmarks operationalize only the user intents the screen can surface, introducing a sampling bias that leaves a broad class of real user needs unmeasured. AppWorld \citep{trivedi2024appworld} similarly evaluates daily multi-app tasks through 457 hand-crafted APIs, but operates in a simulated APIs rather than on Android's terminal and is not compared against GUI agents.

\paragraph{Coding and terminal agents.}
Coding agents issue text-based actions through programmatic interfaces such as shell commands and tool calls, evaluated on software-engineering and OS tasks \citep{wang2024codeact, yang2024swe, wang2025openhands}; frontier coding agents \citep{anthropic2025claudecode, singh2025openai, minimax2026m27} achieve strong performance on these benchmarks through text-only interaction. Harness engineering --- the design of tools, system prompts, context management, and feedback loops around an off-the-shelf agent --- has emerged as the principal mechanism for adapting such agents to new domains without retraining \citep{lopopolo2026harness, lee2026meta}. Terminal-Bench \citep{merrill2026terminalbench} demonstrates that coding agents can solve realistic system and developer tasks on Linux through the shell alone, but no prior work evaluates coding agents at operating real Android devices through the OS interface. The closest related work is CoAct-1 \citep{song2025coact}, a hybrid desktop agent on OSWorld \citep{xie2024osworld} whose Orchestrator routes each sub-task to a GUI Operator or a Python/Bash Programmer. However, CoAct-1 neither isolates the two paradigms on identical tasks nor extends to mobile. Different from these works, we compare both paradigms on identical AndroidWorld and MobileWorld tasks --- coding agents through ADB alone, GUI specialists through screenshots --- isolating the interface paradigm as the sole variable.

\section{Experimental Setup}
\label{sec:experimental_setup}

\paragraph{Benchmarks and Baselines.} We evaluate on two interactive mobile-agent benchmarks whose rule-based verifiers check device state directly, ensuring compatibility with terminal-only execution. AndroidWorld~\citep{rawles2025androidworld} provides 116 task templates across 20 Android applications, parameterized by random seeds. MobileWorld~\citep{kong2025mobileworld} extends this setting with broader app coverage spanning networked client-server applications in e-commerce, enterprise communication, and social media, and with longer-horizon, cross-application tasks (27.8 vs.\ 14.3 average steps per task; 62.2\% vs.\ 9.5\% multi-app tasks). We evaluate on its 117 device-state tasks, excluding the user-interaction and MCP \citep{10.1145/3796519} subsets that fall outside the scope of a GUI-vs-CLI comparison. We exclude AndroidLab~\citep{xu2025androidlab} because its XML-based verifier requiring UI state is incompatible with the CLI paradigm. 

Following CoAct-1 \citep{song2025coact}, we report success rate (SR) and average steps per task across three random seeds, with mean and standard deviation reported. For GUI baselines, we evaluate three reproducible GUI agents: two end-to-end specialists post-trained on mobile datasets (GUI-Owl-1.5-32B~\citep{xu2026mobile} and MAI-UI-8B~\citep{zhou2025mai}) and a strong general-purpose vision-language model (Qwen3-VL-32B~\citep{bai2025qwen3}). We distinguish these \emph{reproducible} baselines, re-run end-to-end under our unified evaluation infrastructure, from leaderboard-reported results---UI-Venus-1.5~\citep{team2026ui}, Seed-1.8~\citep{seed2026seed1}, and Step-GUI~\citep{yan2025step}---included in Table~\ref{tab:main_results} for context but not reproducible as their evaluation pipelines are not released. 

\paragraph{CLI Agents.} We use three off-the-shelf coding agents spanning two design categories: a frontier agent (Claude Code~\citep{anthropic2025claudecode}) representing state-of-the-art coding capability, and two minimal-scaffold agents (Terminus-2~\citep{merrill2026terminalbench} and mini-swe-agent~\citep{yang2024swe}) designed for Terminal-Bench command-line tasks with simple harness architectures, chosen to evaluate how effectively such designs transfer to mobile platforms. Specifically, each agent operates in a text-only Android terminal: it receives the task instruction and terminal stdout at each step (e.g., system state, database query responses, or error messages) as input, and emits ADB commands to interact with the device. No visual or structural UI information such as screenshots or accessibility trees is used at any step. Figure~\ref{fig:paradigm_comparison} contrasts CLI and GUI execution on a representative task.

\paragraph{System Prompt and Tools.} To adapt general coding agents to the Android platform, we construct a system prompt derived from public Android developer documentation and engineering practices. The prompt provides four guidance categories: a structured four-phase interaction cycle (discover relevant data, inspect existing state, act through the terminal interface, and verify the result); a prioritized hierarchy of mechanisms for modifying device state; efficiency strategies including batching commands and limiting exploration budgets; and platform-specific patterns for file synchronization and data persistence. Crucially, the prompt contains no task-specific or app-specific information that could leak to CLI agents. Full system prompts for each CLI agent are provided in Appendix~\ref{app:system_prompts}.

Motivated by recent advances in harness engineering~\citep{tian2026swe, lopopolo2026harness, lee2026meta}, we provide general-purpose tools that wrap raw ADB shell commands into higher-level operations. For AndroidWorld, we provide four tools: \texttt{sql} (database queries), \texttt{read-file} and \texttt{write-file} (file I/O), and \texttt{find-files} (file search). For MobileWorld, with more challenging backend services, structured data, and networked APIs, we extend to twelve tools by adding \texttt{content} and \texttt{intent} (Android content providers and intents), \texttt{json-read} and \texttt{json-write} (structured-data parsing), \texttt{http} (HTTP requests), \texttt{pg} (backend PostgreSQL queries), \texttt{backend-exec} (backend container shell), and \texttt{service-status} (backend service listing). Appendix~\ref{app:tools} provides complete tool specifications.

\paragraph{Oracle Construction.} To establish the empirical CLI ceiling on each benchmark, we construct oracle solutions through human--LLM collaboration. Each solution must satisfy two criteria: \emph{solvability}, meaning the command sequence passes the rule-based verifier; and \emph{integrity}, meaning the solution arrives at the answer through principled exploration rather than extracting it from verifier's source code. Each task is attempted up to five times with verifier feedback; every solution is independently reviewed by all three external professionals. This process yields 88.8\% (103/116) CLI-solvable tasks in AndroidWorld and 86.3\% (101/117) in MobileWorld. The remaining 13 tasks on AndroidWorld and 16 on MobileWorld involve interactions outside the CLI paradigm's reach: interacting with visual content (e.g., transcribing receipts from photos, free-hand drawing) or capturing multimodal data (e.g., camera photos or audio recording); on MobileWorld this set additionally contains a handful of in-app panel-only fields (Google Maps detail labels, shopping-cart totals) that no public API exposes. Details can be found in Appendix~\ref{app:oracle}.

\section{CLI-Advantage Task Suite}
\label{sec:cli-advantage}
Real mobile users have everyday needs the touchscreen interface was not designed for: cleaning up leftover files, asking which expense category dominated this month, or finding out which apps have been granted background-location access. Existing mobile-agent benchmarks were designed around the GUI paradigm, so user intents that no tap sequence can serve fall outside their scope. This sampling gap leaves a measurable class of user intent invisible to benchmark scores. We introduce the CLI-Advantage Suite, a set of \textbf{45 task templates across five categories} that target this gap directly, each with an AndroidWorld-compatible rule-based verifier and a hand-verified oracle CLI solution.

\paragraph{Taxonomy.} The 45 task templates fall into five categories of mobile-user intent that current touchscreen UIs serve poorly. Three require operations the app UI does not expose at scale: \textbf{Bulk operations} (10 tasks) apply one change to many items at once (e.g., \textit{``delete every .apk file in Downloads''}); \textbf{Multi-condition filtering} (10) selects items satisfying a conjunction of constraints app filter UIs rarely compose (e.g., \textit{``list expenses above my average spending''}); and \textbf{Aggregation and top-K} (10) computes summaries over many rows no single screen displays (e.g., \textit{``top-3 expense categories by total this month''}). Two require state the app layer does not surface: \textbf{Cross-app workflows} (9) combine state from multiple apps (e.g., \textit{``did anyone text me during my meetings yesterday?''}), and \textbf{Hidden device state} (6) queries system-level facts no app exposes (e.g., \textit{``which apps have been granted location permission''}).

\paragraph{Construction and Verification.} The 45 task templates are LLM-generated and human-verified by three mobile users. Following AndroidWorld's automatic task-parameterization mechanism, each template yields a distinct task instance per seed. With three seeds $\{7, 30, 1234\}$, we construct 135 task instances. Success is checked by a binary rule-based script that reads the final device state, matches the agent's reported answer against ground truth, or both---with no LLM-judge in the loop. We verify each task along three axes. \textit{Solvability}: applying the Section~\ref{sec:experimental_setup} oracle procedure, we construct a hand-verified CLI command sequence per task; all 45 pass the rule-based verifier. \textit{Realism}: three annotators individually rate each task template on a three-level rubric (highly realistic / stretched / contrived); inter-annotator Cohen's $\kappa = 0.91$. \textit{Additivity}: for each task, we compute the cosine similarity to its nearest AndroidWorld neighbour using \texttt{text-embedding-3-large}; the resulting distribution sits to the left of AndroidWorld's within-suite reference, with $0\%$ exceeding similarity $0.7$, as in Figure~\ref{fig:tier4_aw_similarity_density}. Appendix~\ref{app:cli_advantage} lists details and the full task.

\section{Results}
\label{sec:results}

\subsection{Benchmark Performance}
\label{sec:main_results}

\begin{table*}[t]
\centering
\small
\resizebox{\linewidth}{!}{
\renewcommand{\arraystretch}{1.2}
\begin{tabular}{llcccccc}
\toprule
& & \multicolumn{3}{c}{\textbf{AndroidWorld}} & \multicolumn{3}{c}{\textbf{MobileWorld}} \\
\cmidrule(lr){3-5} \cmidrule(lr){6-8}
\textbf{Method} & \textbf{Model API} & \textbf{SR$\uparrow$} & \textbf{SR-CLI$\uparrow$} & \textbf{Avg Step$\downarrow$} & \textbf{SR$\uparrow$} & \textbf{SR-CLI$\uparrow$} & \textbf{Avg Step$\downarrow$} \\
\midrule
\rowcolor{gray!20}\multicolumn{8}{c}{\textit{GUI Agents (Leaderboard-reported)$^\star$}} \\
UI-Venus-1.5-30B-A3B  & -  & 77.6          & -             & -             & 17.1 & - & - \\
Seed-1.8              & -  & 70.7          & -             & -             & 52.1 & - & - \\
Step-GUI-8B           & -  & 67.7          & -             & -             & 16.1 & - & - \\
\midrule
\rowcolor{gray!20}\multicolumn{8}{c}{\textit{GUI Agents (Reproduced)}} \\
GUI-Owl-1.5-32B       & -  & 69.3 (0.5)    & 71.6 (1.5)    & 15.5 (2.5)    & 43.2 (2.6) & 38.1 (2.0) & 34.7 (0.7) \\
MAI-UI-8B             & -  & 68.1 (0.6)    & 69.3 (0.3)    & 18.4 (1.2)    & 26.3 (1.3) & 26.0 (2.0) & 39.7 (0.3) \\
Qwen3-VL-32B          & -  & 57.8 (4.9)    & 62.4 (5.6)    & 17.4 (1.8)    & 13.3 (0.7) & 11.7 (0.3) & 25.6 (0.6) \\
\midrule
\rowcolor{gray!20}\multicolumn{8}{c}{\textit{CLI Agents (Ours)}} \\
Claude Code           & Claude Opus 4.7   & 71.8 (1.8)  & 80.9 (2.0) & 15.2 (0.6) & 51.9 (1.0) & 60.1 (1.1) & 30.2 (0.6) \\
mini-swe-agent        & Claude Sonnet 4.6 & 70.7 (2.3)  & 79.6 (2.6) & 12.1 (0.3) & 43.6 (2.6) & 50.5 (3.0) & 33.5 (0.4) \\
Terminus-2            & GPT-5.3 Codex     & 63.2 (0.5)  & 71.2 (0.6) &  7.1 (0.1) & 36.2 (1.0) & 41.9 (1.1) & 24.5 (1.9) \\
Terminus-2            & MiniMax M2.7      & 55.2 (3.8)  & 62.1 (4.2) & 14.5 (0.5) & 16.8 (1.3) & 19.5 (1.5)) & 32.9 (0.8) \\
\midrule
Oracle (CLI-solvable ceiling) & -     & 88.8          & 100.0         & 3.7  & 86.3 & 100.0 & 5.6 \\
\bottomrule
\end{tabular}}
\caption{\textbf{Main results on AndroidWorld (AW) and MobileWorld (MW).} \textbf{SR} is success rate (\%) over the full task set, mean (std) across 3 seeds. \textbf{SR-CLI} restricts the denominator to the CLI-solvable subset of each benchmark (103/116 on AW; 101/117 on MW); the remaining tasks require GUI interaction. \textbf{Avg Step} is mean agent steps per task. All CLI agents run under \texttt{bash} plus tools that wrap raw \texttt{adb shell} commands. \textbf{Oracle} is the CLI ceiling, $103/116 = 88.8\%$ on AW and $101/117 = 86.3\%$ on MW, derived from ground-truth reference solutions of each task using CLI; Oracle's Avg Step is averaged only over CLI-solvable tasks. $^\star$~Reported in the original paper; evaluation code is unavailable, precluding reproduction. The full leaderboard-extended version is reported in Table~\ref{tab:main_results_full} (Appendix~\ref{app:main_full_extended}).}
\label{tab:main_results}
\end{table*}

\paragraph{CLI agents are competitive with leading GUI agents .}
Claude Code with Opus 4.7 reaches $71.8\%$ on AndroidWorld and $51.9\%$ on MobileWorld (Table~\ref{tab:main_results}), outperforming every reproduced GUI baseline (GUI-Owl-1.5-32B, MAI-UI-8B, and Qwen3-VL-32B) on both benchmarks. On MobileWorld, Opus 4.7 also matches the strongest leaderboard-reported agent, Seed-1.8 ($52.1\%$). On AndroidWorld, the leaderboard-reported UI-Venus-1.5-30B-A3B stands at $77.6\%$, but its evaluation pipeline is not released and we cannot verify this number under our protocol. The CLI lead is sharpest on the CLI-solvable subset (103 tasks on AW, 101 on MW): Opus 4.7 leads GUI-Owl-1.5-32B by roughly $9$ percentage points on AndroidWorld ($80.9\%$ vs.\ $71.6\%$) and roughly $20$ percentage points on MobileWorld ($60.1\%$ vs.\ $38.1\%$).

\paragraph{Competitiveness is a paradigm-level property.}
On AndroidWorld, every CLI configuration in Table~\ref{tab:main_results} lands between $55\%$ and $72\%$ success rate, with mini-swe-agent and Sonnet 4.6 reaching $70.7\%$ and matching GUI-Owl-1.5-32B ($69.3\%$) at more step efficiency. The complete harness $\times$ model-API matrix is reported in Table~\ref{tab:aw_tool_ablation}. On MobileWorld, the picture is more model-dependent: Claude Code with Opus 4.7 and mini-swe-agent with Sonnet 4.6 clear every reproduced GUI baseline, while the two Terminus-2 configurations (with Codex and with MiniMax) trail GUI-Owl, with the gap widening sharply for weaker model APIs on MobileWorld's more challenging, longer-horizon, cross-application tasks. The Oracle CLI ceiling sits at $88.8\%$ on AndroidWorld and $86.3\%$ on MobileWorld, leaving $17$ percentage points of headroom on AndroidWorld and $34$ on MobileWorld.


\begin{table*}[t]
\centering
\footnotesize
\renewcommand{\arraystretch}{1.2}
\begin{tabular}{ll>{\columncolor{gray!20}}cc>{\columncolor{gray!20}}cc>{\columncolor{gray!20}}cc}
\toprule
 & & \multicolumn{2}{c}{\textbf{bash-only}} & \multicolumn{2}{c}{\textbf{+tools}} & \multicolumn{2}{c}{\textbf{$\Delta$}} \\
\cmidrule(lr){3-4} \cmidrule(lr){5-6} \cmidrule(lr){7-8}
\textbf{Harness} & \textbf{Model API} & \textbf{SR$\uparrow$} & \textbf{Cost (\$)$\downarrow$} & \textbf{SR$\uparrow$} & \textbf{Cost (\$)$\downarrow$} & \textbf{SR} & \textbf{Cost} \\
\midrule
\multirow{2}{*}{Claude Code}    & Claude Opus 4.7    & 71.8 (0.5) & 62.15 (1.42) & 71.8 (1.8) & 64.21 (3.47) & $0.0$                                    & {\color{red!60!black}$\uparrow$3.3\%}    \\
                                & Claude Sonnet 4.6  & 67.5 (2.2) & 45.32 (6.24) & 66.1 (4.3) & 44.92 (5.52) & {\color{red!60!black}$\downarrow$1.4}    & {\color{green!50!black}$\downarrow$0.9\%}  \\
\cmidrule(lr){2-8}
\multirow{3}{*}{mini-swe-agent} & Claude Sonnet 4.6  & 69.5 (3.0) & 10.87 (0.40) & 70.7 (2.3) &  9.45 (0.30) & {\color{green!50!black}$\uparrow$1.2}   & {\color{green!50!black}$\downarrow$13.1\%} \\
                                & GPT-5.3 Codex      & 49.7 (0.5) & 11.25 (0.17) & 61.5 (3.0) &  8.65 (0.16) & {\color{green!50!black}$\uparrow$11.8}  & {\color{green!50!black}$\downarrow$23.1\%} \\
                                & MiniMax M2.7       & 57.2 (3.3) &  2.27 (0.18) & 58.3 (2.2) &  2.29 (0.14) & {\color{green!50!black}$\uparrow$1.1}   & {\color{red!60!black}$\uparrow$0.9\%}    \\
\cmidrule(lr){2-8}
\multirow{3}{*}{Terminus-2}     & Claude Sonnet 4.6  & 58.9 (4.1) & 14.19 (2.44) & 59.8 (3.3) & 13.55 (0.76) & {\color{green!50!black}$\uparrow$0.9}   & {\color{green!50!black}$\downarrow$4.5\%}  \\
                                & GPT-5.3 Codex      & 48.9 (3.0) & 11.08 (0.28) & 63.2 (0.5) &  7.97 (0.48) & {\color{green!50!black}$\uparrow$14.3}  & {\color{green!50!black}$\downarrow$28.1\%} \\
                                & MiniMax M2.7       & 55.2 (3.0) &  1.80 (0.05) & 55.2 (3.8) &  1.72 (0.07) & $0.0$                                    & {\color{green!50!black}$\downarrow$4.4\%}  \\
\bottomrule
\end{tabular}
\caption{\textbf{AndroidWorld tool ablation across 3 harnesses and 4 model APIs.} \texttt{bash-only} vs.\ \texttt{+tools} (\texttt{sql}, \texttt{read-file}, \texttt{write-file}, \texttt{find-files}, wrapping raw \texttt{adb shell} commands). All values are mean (std) across 3 seeds; \textbf{SR} is success rate (\%) over all 116 AW tasks; \textbf{Cost} is inference cost in USD. The \textbf{$\Delta$} block reports the change from \texttt{bash-only} to \texttt{+tools}---absolute percentage points for SR, relative percent for Cost; green denotes improvement (SR up, Cost down), red denotes regression (SR down, Cost up). SR columns are shaded grey throughout. Costs are not directly comparable across harnesses for the same model due to differing prompt-cache.}
\label{tab:aw_tool_ablation}
\end{table*}

\paragraph{Tool gains are conditional on model strength.}
Table~\ref{tab:aw_tool_ablation} compares bash-only and \texttt{+tools} modes across three CLI harnesses (Claude Code, mini-swe-agent, Terminus-2) and four model APIs on AndroidWorld. In bash-only mode, GPT-5.3 Codex is the weakest model API ($48.9$--$49.7\%$ SR), well below Claude Sonnet 4.6 ($58.9$--$69.5\%$). Consistent with recent findings that tool gains are conditional on model strength~\cite{lou2026tool}, only GPT-5.3 Codex benefits substantially from the four CLI tools, with success rate rising by $+11.8$~pp on mini-swe-agent and $+14.3$~pp on Terminus-2; other configurations change by at most $1.4$~pp. Inference cost mirrors SR: Codex sees the biggest cost reductions ($\sim 23\%$ on mini-swe-agent and $\sim 28\%$ on Terminus-2), with mini-swe-agent + Sonnet 4.6 next at $\sim 13\%$, while Opus 4.7 costs $\sim 3\%$ more. One mechanism behind these effects is shell quoting: the four CLI tools handle escaping that bash invocations frequently mishandle. Two paired Codex trajectories in Appendix~\ref{app:codex_escaping} illustrate this: \texttt{+tools} converts a bash-only failure into a success (the SR gain) and eliminates retry steps on a single-quote edit (the cost reduction).

\subsection{Trajectory-Level Error Analysis}
\label{sec:failure_modes}

\begin{figure}[!t]
\centering
\includegraphics[width=\linewidth]{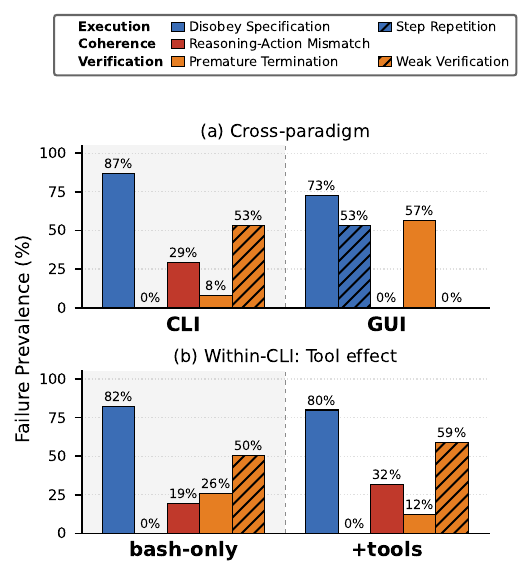}
\caption{\textbf{Trajectory-level failure modes.} Each bar denotes the fraction of failed trajectories exhibiting the corresponding taxonomy category (a trajectory may exhibit multiple). \textbf{(a) Cross-paradigm:} CLI vs.\ GUI on the same 103 CLI-solvable tasks of AndroidWorld. \textbf{(b) Within-CLI tool effect:} GPT-5.3 Codex is evaluated under \texttt{bash-only} vs.\ \texttt{+tools} (four generic, structured tool calls). Color encodes the top-level category (Execution / Coherence / Verification); hatching marks Step Repetition and Weak Verification within their categories.}
\label{fig:tb_failure_modes}
\end{figure}

Since average success rates do not reveal failure patterns, we conduct a trajectory-level error analysis along three comparison axes: (i) GUI vs.\ CLI paradigms; (ii) \texttt{bash-only} vs.\ \texttt{+tools} for GPT-5.3 Codex, selected for its largest tool-boosted SR gain in Table~\ref{tab:aw_tool_ablation}; and (iii) model APIs and harnesses jointly. We adapt the Multi-Agent System Taxonomy (MAST)~\cite{cemri2026why} to our single-agent mobile setting by dropping inter-agent coordination subcategories (following~\cite{merrill2026terminalbench}) and mapping the remaining nodes onto mobile-platform failure modes, yielding a two-level paradigm-agnostic taxonomy: three top-level error classes (\emph{Execution}, \emph{Coherence}, \emph{Verification}), each containing several named failure modes (full taxonomy and per-mode examples in Appendix~\ref{app:failure_modes_taxonomy}). To identify candidate failure-mode clusters, we use Docent~\cite{meng2025docent} to summarize each failed trajectory within each paradigm separately, yielding paradigm-specific clusters. To map clusters onto the named modes in the taxonomy, we apply Claude Opus 4.7 with a rubric calibrated against two human annotators (full protocol and validation in Appendix~\ref{app:failure_modes_pipeline} and~\ref{app:failure_modes_validation}).

Figure~\ref{fig:tb_failure_modes}(a) summarizes failure modes across the two paradigms, reporting percentages over failed trajectories. Both paradigms share \emph{Disobey Specification} as the dominant failure mode ($87\%$ CLI / $73\%$ GUI), but their tails diverge: GUI is dominated by \emph{Step Repetition} and \emph{Premature Termination}, CLI by \emph{Reasoning-Action Mismatch} and \emph{Weak Verification}. These distinct failure signatures suggest that the two paradigms require distinct strategies for improvement. Within CLI, the four tools cut \emph{Premature Termination} from $26\%$ to $12\%$ on GPT-5.3 Codex while raising \emph{Reasoning-Action Mismatch} ($+12$~pp) and \emph{Weak Verification} ($+9$~pp): the tools prevent the agent from quitting early, so failed trajectories now exhibit downstream reasoning and verification errors (Figure~\ref{fig:tb_failure_modes}(b)). Two extensions are detailed in Appendix~\ref{app:failure_modes_extended_results}: the within-CLI failure models across all harness $\times$ model-API combinations (Figure~\ref{fig:tb_model_x_harness}), and a sub-cluster breakdown of \emph{Disobey Specification} that reveals paradigm-specific textures within this otherwise-shared dominant mode (Figure~\ref{fig:tb_variant_clusters}).

\begin{table*}[!t]
\centering
\small
\resizebox{\linewidth}{!}{
\renewcommand{\arraystretch}{1.2}
\begin{tabular}{lccccccc}
\toprule
\textbf{Agent} & \textbf{Bulk} & \textbf{Filter} & \textbf{Aggregation} & \textbf{CrossApp} & \textbf{Hidden} & \textbf{Avg SR} & \textbf{Avg Steps} \\
\midrule
\rowcolor{gray!20}\textit{GUI agents} & & & & & & & \\
Qwen3-VL-32B               & 33.3 $\pm$ 5.8          & 20.0 $\pm$ 10.0         & 36.7 $\pm$ 5.8          &  0.0 $\pm$ 0.0          & 16.7 $\pm$ 0.0          & 22.2 $\pm$ 4.4         & 12.9 $\pm$ 1.1 \\
MAI-UI-8B                  & 56.7 $\pm$ 5.8          & 36.7 $\pm$ 15.3         & 10.0 $\pm$ 10.0         & 11.1 $\pm$ 0.0          & 16.7 $\pm$ 0.0          & 27.4 $\pm$ 6.4         & 23.4 $\pm$ 1.6 \\
GUI-Owl-1.5-32B            & 46.7 $\pm$ 5.8          & 40.0 $\pm$ 10.0         & 43.3 $\pm$ 5.8          & 11.1 $\pm$ 0.0          & 16.7 $\pm$ 16.7         & 33.3 $\pm$ 3.8         & 19.4 $\pm$ 1.0 \\
\midrule
\rowcolor{gray!20}\textit{CLI agents} & & & & & & & \\
Terminus-2 + gpt-5.3-codex & \textbf{76.7 $\pm$ 5.8}  & 56.7 $\pm$ 5.8          & \textbf{83.3 $\pm$ 5.8}  & 25.9 $\pm$ 12.8         & \textbf{55.6 $\pm$ 9.6}  & 60.7 $\pm$ 6.4         & \textbf{5.7 $\pm$ 0.4} \\
mini-swe + minimax-m2.7    & 73.3 $\pm$ 11.5         & 76.7 $\pm$ 5.8          & 80.0 $\pm$ 0.0          & \textbf{48.1 $\pm$ 12.8} & 44.4 $\pm$ 9.6          & 66.7 $\pm$ 4.4         & 15.4 $\pm$ 1.2 \\
Claude Code + Opus 4.7     & \textbf{76.7 $\pm$ 5.8}  & \textbf{86.7 $\pm$ 5.8}  & \textbf{83.3 $\pm$ 5.8}  & 40.7 $\pm$ 6.4          & 44.4 $\pm$ 9.6          & \textbf{68.9 $\pm$ 0.0} & 11.0 $\pm$ 1.4 \\
\bottomrule
\end{tabular}}
\caption{\textbf{Per-category and overall performance on the 45-task CLI-Advantage Suite (3 seeds; mean $\pm$ std; SR in percent)}. Best per column in bold; Bulk and Aggregation ties are bolded jointly.}
\label{tab:cli_advantage_results}
\end{table*}

\subsection{CLI-Advantage Suite}
\label{sec:cli_advantage_results}

We evaluate three CLI agents (bash-only) and three GUI agents on the 45-task CLI-Advantage Suite (Section~\ref{sec:cli-advantage}), with a maximum of 50 steps per task.

\paragraph{All CLI agents lead all GUI agents.} As reported in Table~\ref{tab:cli_advantage_results}, all CLI agents lead all GUI agents in mean success rate by 27 to 47$\%$, and this lead holds across every one of the five categories. CLI agents also use roughly half as many steps per episode on average ($10.7$ vs.\ $18.6$). Among the GUI agents, the two post-trained models (MAI-UI-8B and GUI-Owl-1.5-32B) modestly outperform the general-purpose Qwen3-VL-32B but require more steps to do so. Among CLI agents, Terminus-2 paired with GPT-5.3 Codex is the most step-efficient configuration. Even the strongest CLI agent, Claude Code with Opus 4.7, reaches only $68.9\%$, leaving substantial headroom below the $100\%$ oracle ceiling.

\paragraph{Why do GUI agents fail?} The cross-app workflow category exposes a structural ceiling for the GUI paradigm. Every GUI agent we evaluate caps at no more than $11\%$, suggesting that the bottleneck is maintaining state across multiple apps when each step observes only a single screenshot. The two largest absolute gaps appear in aggregation/top-K and multi-condition filtering, where mean CLI success exceeds mean GUI success by roughly 52 and 41$\%$ respectively. Both categories require set-level operations such as composing predicates and summing across rows, which the CLI expresses naturally with shell pipes and database queries but which touchscreen interfaces were never designed to expose. The gap is narrowest in bulk operations, at roughly 30$\%$, because repeated taps can still complete the task. The CLI advantage therefore grows when tasks demand composition over many items rather than repetition of a single operation.

%
%

\section{Limitations}
\label{sec:limitations}

While the CLI paradigm is competitive with GUI agents, the strongest CLI agents still underperform SoTA GUI systems on the full benchmarks, and all leading CLI agents rely on frontier proprietary APIs---the aggregate API cost of our experiments and ablations is approximately \$8K~USD, well beyond what is acceptable for everyday on-device use. The paradigm is also bounded in scope: a CLI agent can only operate on state reachable from a terminal, so applications that hide internal functionality behind proprietary UI flows and inherently visual workloads (video entertainment, image editing) lie outside its reach. These gaps motivate three complementary directions for future work: harness-level improvements such as richer tool surfaces and better state-discovery prompts; post-training a small coding-agent model specialized for the Android--terminal setting to address efficiency and cost; and a hybrid CLI--GUI agent that partitions workloads across the two paradigms rather than relying on either alone.

\section{Conclusion}
\label{sec:conclusion}

In this work, we investigate whether mobile GUI tasks can be solved purely through a command-line interface (CLI), without on-screen UI information. Evaluating three coding-agent harnesses across four model APIs under a unified protocol on AndroidWorld and MobileWorld, we demonstrate that CLI agents are competitive with leading GUI agents while using substantially fewer steps. To complement existing mobile-agent benchmarks with user intents that fall outside the GUI paradigm, we introduce the CLI-Advantage Suite of 45 task templates across five categories, on which the CLI advantage over GUI agents is even more pronounced in both success rate and step efficiency, indicating that the CLI--GUI gap is paradigm-level rather than a function of model choice. We hope this work provides a strong empirical foundation for future research on CLI agents for mobile platforms.

\bibliography{reference}

@misc{android_adb,
  title = {Android Debug Bridge ({ADB})},
  author = {{Google}},
  howpublished = {\url{https://developer.android.com/tools/adb}},
  year = {2025},
  note = {Accessed: 2026-04-20}
}

@misc{android_dumpsys,
  title = {{dumpsys} system diagnostics},
  author = {{Google}},
  howpublished = {\url{https://developer.android.com/tools/dumpsys}},
  year = {2025},
  note = {Accessed: 2026-04-20}
}

@inproceedings{
merrill2026terminalbench,
title={Terminal-Bench: Benchmarking Agents on Hard, Realistic Tasks in Command Line Interfaces},
author={Mike A Merrill and Alexander Glenn Shaw and Nicholas Carlini and Boxuan Li and Harsh Raj and Ivan Bercovich and Lin Shi and Jeong Yeon Shin and Thomas Walshe and E. Kelly Buchanan and Junhong Shen and Guanghao Ye and Haowei Lin and Jason Poulos and Maoyu Wang and Marianna Nezhurina and Di Lu and Orfeas Menis Mastromichalakis and Zhiwei Xu and Zizhao Chen and Yue Liu and Robert Zhang and Leon Liangyu Chen and Anurag Kashyap and Jan-Lucas Uslu and Jeffrey Li and Jianbo Wu and Minghao Yan and Song Bian and Vedang Sharma and Ke Sun and Steven Dillmann and Akshay Anand and Andrew Lanpouthakoun and Bardia Koopah and Changran Hu and Etash Kumar Guha and Gabriel H. S. Dreiman and Jiacheng Zhu and Karl Krauth and Li Zhong and Niklas Muennighoff and Robert Kwesi Amanfu and Shangyin Tan and Shreyas Pimpalgaonkar and Tushar Aggarwal and Xiangning Lin and Xin Lan and Xuandong Zhao and Yiqing Liang and Yuanli Wang and Zilong Wang and Changzhi Zhou and David Heineman and Hange Liu and Harsh Trivedi and John Yang and Junhong Lin and Manish Shetty and Michael Yang and Nabil Omi and Negin Raoof and Shanda Li and Terry Yue Zhuo and Wuwei Lin and Yiwei Dai and Yuxin Wang and Wenhao Chai and Shang Zhou and Dariush Wahdany and Ziyu She and Jiaming Hu and Zhikang Dong and Yuxuan Zhu and Sasha Cui and Ahson Saiyed and Arinbj{\"o}rn Kolbeinsson and Christopher Michael Rytting and Ryan Marten and Yixin Wang and Jenia Jitsev and Alex Dimakis and Andy Konwinski and Ludwig Schmidt},
booktitle={The Fourteenth International Conference on Learning Representations},
year={2026},
url={https://openreview.net/forum?id=a7Qa4CcHak}
}

@article{wu2024foundations,
  title={Foundations and recent trends in multimodal mobile agents: A survey},
  author={Wu, Biao and Li, Yanda and Zhang, Zhiwei and Wei, Yunchao and Fang, Meng and Chen, Ling},
  journal={arXiv preprint arXiv:2411.02006},
  year={2024}
}

@inproceedings{hu2025agents,
  title={Os agents: A survey on mllm-based agents for computer, phone and browser use},
  author={Hu, Xueyu and Xiong, Tao and Yi, Biao and Wei, Zishu and Xiao, Ruixuan and Chen, Yurun and Ye, Jiasheng and Tao, Meiling and Zhou, Xiangxin and Zhao, Ziyu and others},
  booktitle={Proceedings of the 63rd Annual Meeting of the Association for Computational Linguistics (Volume 1: Long Papers)},
  pages={7436--7465},
  year={2025}
}

@article{liu2025llm,
  title={Llm-powered gui agents in phone automation: Surveying progress and prospects},
  author={Liu, Guangyi and Zhao, Pengxiang and Liang, Yaozhen and Liu, Liang and Guo, Yaxuan and Xiao, Han and Lin, Weifeng and Chai, Yuxiang and Han, Yue and Ren, Shuai and others},
  journal={arXiv preprint arXiv:2504.19838},
  year={2025}
}

@article{team2026ui,
  title={Ui-venus-1.5 technical report},
  author={Team, Venus and Gao, Changlong and Gu, Zhangxuan and Liu, Yulin and Qiu, Xinyu and Shen, Shuheng and Wen, Yue and Xia, Tianyu and Xu, Zhenyu and Zeng, Zhengwen and others},
  journal={arXiv preprint arXiv:2602.09082},
  year={2026}
}

@article{yan2025step,
  title={Step-gui technical report},
  author={Yan, Haolong and Wang, Jia and Huang, Xin and Shen, Yeqing and Meng, Ziyang and Fan, Zhimin and Tan, Kaijun and Gao, Jin and Shi, Lieyu and Yang, Mi and others},
  journal={arXiv preprint arXiv:2512.15431},
  year={2025}
}

@article{xu2026mobile,
  title={Mobile-agent-v3. 5: Multi-platform fundamental gui agents},
  author={Xu, Haiyang and Zhang, Xi and Liu, Haowei and Wang, Junyang and Zhu, Zhaozai and Zhou, Shengjie and Hu, Xuhao and Gao, Feiyu and Cao, Junjie and Wang, Zihua and others},
  journal={arXiv preprint arXiv:2602.16855},
  year={2026}
}

@article{zhou2025mai,
  title={MAI-UI Technical Report: Real-World Centric Foundation GUI Agents},
  author={Zhou, Hanzhang and Zhang, Xu and Tong, Panrong and Zhang, Jianan and Chen, Liangyu and Kong, Quyu and Cai, Chenglin and Liu, Chen and Wang, Yue and Zhou, Jingren and others},
  journal={arXiv preprint arXiv:2512.22047},
  year={2025}
}

@article{wang2025ui,
  title={Ui-tars-2 technical report: Advancing gui agent with multi-turn reinforcement learning},
  author={Wang, Haoming and Zou, Haoyang and Song, Huatong and Feng, Jiazhan and Fang, Junjie and Lu, Junting and Liu, Longxiang and Luo, Qinyu and Liang, Shihao and Huang, Shijue and others},
  journal={arXiv preprint arXiv:2509.02544},
  year={2025}
}

@article{xu2025mobilerl,
  title={Mobilerl: Online agentic reinforcement learning for mobile gui agents},
  author={Xu, Yifan and Liu, Xiao and Liu, Xinghan and Fu, Jiaqi and Zhang, Hanchen and Jing, Bohao and Zhang, Shudan and Wang, Yuting and Zhao, Wenyi and Dong, Yuxiao},
  journal={arXiv preprint arXiv:2509.18119},
  year={2025}
}

@article{shi2025mobilegui,
  title={Mobilegui-rl: Advancing mobile gui agent through reinforcement learning in online environment},
  author={Shi, Yucheng and Yu, Wenhao and Li, Zaitang and Wang, Yonglin and Zhang, Hongming and Liu, Ninghao and Mi, Haitao and Yu, Dong},
  journal={arXiv preprint arXiv:2507.05720},
  year={2025}
}

@article{bai2024digirl,
  title={Digirl: Training in-the-wild device-control agents with autonomous reinforcement learning},
  author={Bai, Hao and Zhou, Yifei and Cemri, Mert and Pan, Jiayi and Suhr, Alane and Levine, Sergey and Kumar, Aviral},
  journal={Advances in Neural Information Processing Systems},
  volume={37},
  pages={12461--12495},
  year={2024}
}

@article{gu2026generalization,
  title={Generalization in Online Reinforcement Learning for Mobile Agents},
  author={Gu, Li and Jiang, Zihuan and Chi, Zhixiang and Liu, Huan and Wang, Ziqiang and Yu, Yuanhao and Berseth, Glen and Wang, Yang},
  journal={arXiv preprint arXiv:2603.07432},
  year={2026}
}

@article{singh2025openai,
  title={Openai gpt-5 system card},
  author={Singh, Aaditya and Fry, Adam and Perelman, Adam and Tart, Adam and Ganesh, Adi and El-Kishky, Ahmed and McLaughlin, Aidan and Low, Aiden and Ostrow, AJ and Ananthram, Akhila and others},
  journal={arXiv preprint arXiv:2601.03267},
  year={2025}
}

@misc{minimax2026m27,
  author       = {{MiniMax}},
  title        = {{MiniMax} {M2.7}: Early Echoes of Self-Evolution},
  year         = {2026},
  month        = mar,
  howpublished = {\url{https://www.minimax.io/news/minimax-m27-en}},
  note         = {Accessed: 2026-05-01}
}

@misc{anthropic2025claudecode,
  author       = {{Anthropic}},
  title        = {Claude {Code}},
  year         = {2025},
  howpublished = {\url{https://www.anthropic.com/claude-code}},
  note         = {Accessed: 2026-05-01}
}

@inproceedings{
rawles2025androidworld,
title={AndroidWorld: A Dynamic Benchmarking Environment for Autonomous Agents},
author={Christopher Rawles and Sarah Clinckemaillie and Yifan Chang and Jonathan Waltz and Gabrielle Lau and Marybeth Fair and Alice Li and William E Bishop and Wei Li and Folawiyo Campbell-Ajala and Daniel Kenji Toyama and Robert James Berry and Divya Tyamagundlu and Timothy P Lillicrap and Oriana Riva},
booktitle={The Thirteenth International Conference on Learning Representations},
year={2025},
url={https://openreview.net/forum?id=il5yUQsrjC}
}

@article{kong2025mobileworld,
  title={MobileWorld: Benchmarking Autonomous Mobile Agents in Agent-User Interactive and MCP-Augmented Environments},
  author={Kong, Quyu and Zhang, Xu and Yang, Zhenyu and Gao, Nolan and Liu, Chen and Tong, Panrong and Cai, Chenglin and Zhou, Hanzhang and Zhang, Jianan and Chen, Liangyu and others},
  journal={arXiv preprint arXiv:2512.19432},
  year={2025}
}

@article{yang2024swe,
  title={Swe-agent: Agent-computer interfaces enable automated software engineering},
  author={Yang, John and Jimenez, Carlos E and Wettig, Alexander and Lieret, Kilian and Yao, Shunyu and Narasimhan, Karthik and Press, Ofir},
  journal={Advances in Neural Information Processing Systems},
  volume={37},
  pages={50528--50652},
  year={2024}
}

@article{tian2026swe,
  title={SWE-Bench Mobile: Can Large Language Model Agents Develop Industry-Level Mobile Applications?},
  author={Tian, Muxin and Wang, Zhe and Yang, Blair and Tang, Zhenwei and Zhu, Kunlun and Dong, Honghua and Li, Hanchen and Xie, Xinni and Wang, Guangjing and You, Jiaxuan},
  journal={arXiv preprint arXiv:2602.09540},
  year={2026}
}

@article{lopopolo2026harness,
  title={Harness engineering: leveraging codex in an agent-first world},
  author={Lopopolo, Ryan},
  journal={OpenAI engineering note},
  year={2026}
}

@article{lee2026meta,
  title={Meta-Harness: End-to-End Optimization of Model Harnesses},
  author={Lee, Yoonho and Nair, Roshen and Zhang, Qizheng and Lee, Kangwook and Khattab, Omar and Finn, Chelsea},
  journal={arXiv preprint arXiv:2603.28052},
  year={2026}
}

@inproceedings{xu2025androidlab,
  title={Androidlab: Training and systematic benchmarking of android autonomous agents},
  author={Xu, Yifan and Liu, Xiao and Sun, Xueqiao and Cheng, Siyi and Yu, Hao and Lai, Hanyu and Zhang, Shudan and Zhang, Dan and Tang, Jie and Dong, Yuxiao},
  booktitle={Proceedings of the 63rd Annual Meeting of the Association for Computational Linguistics (Volume 1: Long Papers)},
  pages={2144--2166},
  year={2025}
}

@article{qin2025uitars,
  title={Ui-tars: Pioneering automated gui interaction with native agents},
  author={Qin, Yujia and Ye, Yining and Fang, Junjie and Wang, Haoming and Liang, Shihao and Tian, Shizuo and Zhang, Junda and Li, Jiahao and Li, Yunxin and Huang, Shijue and others},
  journal={arXiv preprint arXiv:2501.12326},
  year={2025}
}

@article{wang2025uivenus,
  title={Ui-venus technical report: Building high-performance ui agents with rft},
  author={Gu, Zhangxuan and Zeng, Zhengwen and Xu, Zhenyu and Zhou, Xingran and Shen, Shuheng and Liu, Yunfei and Zhou, Beitong and Meng, Changhua and Xia, Tianyu and Chen, Weizhi and others},
  journal={arXiv preprint arXiv:2508.10833},
  year={2025}
}

@article{xu2025mobileagentv3,
  title={Mobile-agent-v3: Fundamental agents for gui automation},
  author={Ye, Jiabo and Zhang, Xi and Xu, Haiyang and Liu, Haowei and Wang, Junyang and Zhu, Zhaoqing and Zheng, Ziwei and Gao, Feiyu and Cao, Junjie and Lu, Zhengxi and others},
  journal={arXiv preprint arXiv:2508.15144},
  year={2025}
}

@inproceedings{wu2024osatlas,
  title={OS-ATLAS: Foundation action model for generalist GUI agents},
  author={Wu, Zhiyong and Wu, Zhenyu and Xu, Fangzhi and Wang, Yian and Sun, Qiushi and Jia, Chengyou and Cheng, Kanzhi and Ding, Zichen and Chen, Liheng and Liang, Paul Pu and others},
  booktitle={International Conference on Learning Representations},
  volume={2025},
  pages={5090--5108},
  year={2025}
}

@article{xu2024aguvis,
  title={Aguvis: Unified pure vision agents for autonomous gui interaction},
  author={Xu, Yiheng and Wang, Zekun and Wang, Junli and Lu, Dunjie and Xie, Tianbao and Saha, Amrita and Sahoo, Doyen and Yu, Tao and Xiong, Caiming},
  journal={arXiv preprint arXiv:2412.04454},
  year={2024}
}

@inproceedings{hong2024cogagent,
  title={Cogagent: A visual language model for gui agents},
  author={Hong, Wenyi and Wang, Weihan and Lv, Qingsong and Xu, Jiazheng and Yu, Wenmeng and Ji, Junhui and Wang, Yan and Wang, Zihan and Dong, Yuxiao and Ding, Ming and others},
  booktitle={Proceedings of the IEEE/CVF conference on computer vision and pattern recognition},
  pages={14281--14290},
  year={2024}
}

@inproceedings{cheng2024seeclick,
  title={Seeclick: Harnessing gui grounding for advanced visual gui agents},
  author={Cheng, Kanzhi and Sun, Qiushi and Chu, Yougang and Xu, Fangzhi and YanTao, Li and Zhang, Jianbing and Wu, Zhiyong},
  booktitle={Proceedings of the 62nd Annual Meeting of the Association for Computational Linguistics (Volume 1: Long Papers)},
  pages={9313--9332},
  year={2024}
}

@inproceedings{lin2025showui,
  title={Showui: One vision-language-action model for gui visual agent},
  author={Lin, Kevin Qinghong and Li, Linjie and Gao, Difei and Yang, Zhengyuan and Wu, Shiwei and Bai, Zechen and Lei, Stan Weixian and Wang, Lijuan and Shou, Mike Zheng},
  booktitle={Proceedings of the Computer Vision and Pattern Recognition Conference},
  pages={19498--19508},
  year={2025}
}

@inproceedings{you2024ferretui,
  title={Ferret-ui: Grounded mobile ui understanding with multimodal llms},
  author={You, Keen and Zhang, Haotian and Schoop, Eldon and Weers, Floris and Swearngin, Amanda and Nichols, Jeffrey and Yang, Yinfei and Gan, Zhe},
  booktitle={European Conference on Computer Vision},
  pages={240--255},
  year={2024},
  organization={Springer}
}

@inproceedings{yang2025magma,
  title={Magma: A foundation model for multimodal ai agents},
  author={Yang, Jianwei and Tan, Reuben and Wu, Qianhui and Zheng, Ruijie and Peng, Baolin and Liang, Yongyuan and Gu, Yu and Cai, Mu and Ye, Seonghyeon and Jang, Joel and others},
  booktitle={Proceedings of the computer vision and pattern recognition conference},
  pages={14203--14214},
  year={2025}
}

@article{wang2024mobileagent,
  title={Mobile-agent: Autonomous multi-modal mobile device agent with visual perception},
  author={Wang, Junyang and Xu, Haiyang and Ye, Jiabo and Yan, Ming and Shen, Weizhou and Zhang, Ji and Huang, Fei and Sang, Jitao},
  journal={arXiv preprint arXiv:2401.16158},
  year={2024}
}

@article{wang2024mobileagentv2,
  title={Mobile-agent-v2: Mobile device operation assistant with effective navigation via multi-agent collaboration},
  author={Wang, Junyang and Xu, Haiyang and Jia, Haitao and Zhang, Xi and Yan, Ming and Shen, Weizhou and Zhang, Ji and Huang, Fei and Sang, Jitao},
  journal={Advances in Neural Information Processing Systems},
  volume={37},
  pages={2686--2710},
  year={2024}
}

@inproceedings{zhang2023appagent,
  title={Appagent: Multimodal agents as smartphone users},
  author={Zhang, Chi and Yang, Zhao and Liu, Jiaxuan and Li, Yanda and Han, Yucheng and Chen, Xin and Huang, Zebiao and Fu, Bin and Yu, Gang},
  booktitle={Proceedings of the 2025 CHI Conference on Human Factors in Computing Systems},
  pages={1--20},
  year={2025}
}

@inproceedings{wen2024autodroid,
  title={Autodroid: Llm-powered task automation in android},
  author={Wen, Hao and Li, Yuanchun and Liu, Guohong and Zhao, Shanhui and Yu, Tao and Li, Toby Jia-Jun and Jiang, Shiqi and Liu, Yunhao and Zhang, Yaqin and Liu, Yunxin},
  booktitle={Proceedings of the 30th annual international conference on Mobile computing and networking},
  pages={543--557},
  year={2024}
}

@article{dai2025vdroid,
  title={V-Droid: Advancing Mobile GUI Agent Through Generative Verifiers},
  author={Dai, Gaole and Jiang, Shiqi and Cao, Ting and Li, Yuanchun and Yang, Yuqing and Tan, Rui and Li, Mo and Qiu, Lili},
  year={2026}
}

@inproceedings{wang2024codeact,
  title={Executable code actions elicit better llm agents},
  author={Wang, Xingyao and Chen, Yangyi and Yuan, Lifan and Zhang, Yizhe and Li, Yunzhu and Peng, Hao and Ji, Heng},
  booktitle={Forty-first International Conference on Machine Learning},
  year={2024}
}

@inproceedings{
wang2025openhands,
title={OpenHands: An Open Platform for {AI} Software Developers as Generalist Agents},
author={Xingyao Wang and Boxuan Li and Yufan Song and Frank F. Xu and Xiangru Tang and Mingchen Zhuge and Jiayi Pan and Yueqi Song and Bowen Li and Jaskirat Singh and Hoang H. Tran and Fuqiang Li and Ren Ma and Mingzhang Zheng and Bill Qian and Yanjun Shao and Niklas Muennighoff and Yizhe Zhang and Binyuan Hui and Junyang Lin and Robert Brennan and Hao Peng and Heng Ji and Graham Neubig},
booktitle={The Thirteenth International Conference on Learning Representations},
year={2025},
url={https://openreview.net/forum?id=OJd3ayDDoF}
}

@article{song2025coact,
  title={Coact-1: Computer-using agents with coding as actions},
  author={Song, Linxin and Dai, Yutong and Prabhu, Viraj and Zhang, Jieyu and Shi, Taiwei and Li, Li and Li, Junnan and Savarese, Silvio and Chen, Zeyuan and Zhao, Jieyu and others},
  journal={arXiv preprint arXiv:2508.03923},
  year={2025}
}

@article{xie2024osworld,
  title={Osworld: Benchmarking multimodal agents for open-ended tasks in real computer environments},
  author={Xie, Tianbao and Zhang, Danyang and Chen, Jixuan and Li, Xiaochuan and Zhao, Siheng and Cao, Ruisheng and Hua, Toh J and Cheng, Zhoujun and Shin, Dongchan and Lei, Fangyu and others},
  journal={Advances in Neural Information Processing Systems},
  volume={37},
  pages={52040--52094},
  year={2024}
}

@inproceedings{trivedi2024appworld,
  title={Appworld: A controllable world of apps and people for benchmarking interactive coding agents},
  author={Trivedi, Harsh and Khot, Tushar and Hartmann, Mareike and Manku, Ruskin and Dong, Vinty and Li, Edward and Gupta, Shashank and Sabharwal, Ashish and Balasubramanian, Niranjan},
  booktitle={Proceedings of the 62nd Annual Meeting of the Association for Computational Linguistics (Volume 1: Long Papers)},
  pages={16022--16076},
  year={2024}
}

@article{rawles2023aitw,
  title={Androidinthewild: A large-scale dataset for android device control},
  author={Rawles, Christopher and Li, Alice and Rodriguez, Daniel and Riva, Oriana and Lillicrap, Timothy},
  journal={Advances in Neural Information Processing Systems},
  volume={36},
  pages={59708--59728},
  year={2023}
}

@article{wang2024mobileagentbench,
  title={Mobileagentbench: An efficient and user-friendly benchmark for mobile llm agents},
  author={Wang, Luyuan and Deng, Yongyu and Zha, Yiwei and Mao, Guodong and Wang, Qinmin and Min, Tianchen and Chen, Wei and Chen, Shoufa},
  journal={arXiv preprint arXiv:2406.08184},
  year={2024}
}

@article{li2024androidcontrol,
  title={On the effects of data scale on ui control agents},
  author={Li, Wei and Bishop, William and Li, Alice and Rawles, Chris and Campbell-Ajala, Folawiyo and Tyamagundlu, Divya and Riva, Oriana},
  journal={Advances in Neural Information Processing Systems},
  volume={37},
  pages={92130--92154},
  year={2024}
}

@inproceedings{chen2025spabench,
  title={Spa-bench: A comprehensive benchmark for smartphone agent evaluation},
  author={Chen, Jingxuan and Yuen, Derek and Xie, Bin and Yang, Yuhao and Chen, Gongwei and Wu, Zhihao and Yixing, Li and Zhou, Xurui and Liu, Weiwen and Wang, Shuai and others},
  booktitle={NeurIPS 2024 Workshop on Open-World Agents},
  year={2024}
}

@article{chai2025a3,
  title={A3: Android Agent Arena for Mobile GUI Agents with Essential-State Procedural Evaluation},
  author={Chai, Yuxiang and Tang, Shunye and Xiao, Han and Lin, Weifeng and Li, Hanhao and Zhang, Jiayu and Liu, Liang and Zhao, Pengxiang and Liu, Guangyi and Wang, Guozhi and others},
  journal={arXiv preprint arXiv:2501.01149},
  year={2025}
}

@article{lou2026tool,
  title={The Tool Illusion: Rethinking Tool Use in Web Agents},
  author={Lou, Renze and Peng, Baolin and Yao, Wenlin and Wu, Qianhui and Cheng, Hao and Nath, Suman and Yin, Wenpeng and Gao, Jianfeng},
  journal={arXiv preprint arXiv:2604.03465},
  year={2026}
}

@inproceedings{
cemri2026why,
title={Why Do Multi-Agent {LLM} Systems Fail?},
author={Mert Cemri and Melissa Z Pan and Shuyi Yang and Lakshya A Agrawal and Bhavya Chopra and Rishabh Tiwari and Kurt Keutzer and Aditya Parameswaran and Dan Klein and Kannan Ramchandran and Matei Zaharia and Joseph E. Gonzalez and Ion Stoica},
booktitle={The Thirty-ninth Annual Conference on Neural Information Processing Systems Datasets and Benchmarks Track},
year={2026},
url={https://openreview.net/forum?id=fAjbYBmonr}
}

@misc{meng2025docent,
  author       = {Meng, Kevin and Huang, Vincent and Steinhardt, Jacob and Schwettmann, Sarah},
  title        = {Introducing Docent},
  year         = {2025},
  month        = {March},
  day          = {24},
  howpublished = {\url{https://transluce.org/introducing-docent}}
}

@article{bai2025qwen3,
  title={Qwen3-vl technical report},
  author={Bai, Shuai and Cai, Yuxuan and Chen, Ruizhe and Chen, Keqin and Chen, Xionghui and Cheng, Zesen and Deng, Lianghao and Ding, Wei and Gao, Chang and Ge, Chunjiang and others},
  journal={arXiv preprint arXiv:2511.21631},
  year={2025}
}

@article{10.1145/3796519,
author = {Hou, Xinyi and Zhao, Yanjie and Wang, Shenao and Wang, Haoyu},
title = {Model Context Protocol (MCP): Landscape, Security Threats, and Future Research Directions},
year = {2026},
publisher = {Association for Computing Machinery},
address = {New York, NY, USA},
issn = {1049-331X},
url = {https://doi.org/10.1145/3796519},
doi = {10.1145/3796519},
note = {Just Accepted},
journal = {ACM Trans. Softw. Eng. Methodol.},
month = feb
}

@inproceedings{guo2025agentsama,
  title={Agent-SAMA: State-Aware Mobile Assistant},
  author={Guo, Linqiang and Liu, Wei and Heng, Yi Wen and Chen, Tse-Hsun Peter and Wang, Yang},
  booktitle={Proceedings of the AAAI Conference on Artificial Intelligence},
  volume={40},
  number={35},
  pages={29459--29467},
  year={2026}
}

@article{seed2026seed1,
  title={Seed1. 8 model card: Towards generalized real-world agency},
  author={Seed, Bytedance},
  journal={arXiv preprint arXiv:2603.20633},
  year={2026}
}



\appendix


\clearpage

\newcommand{\snake}[1]{%
  \begingroup
  \renewcommand{\_}{\char95\penalty0}%
  \texttt{#1}%
  \endgroup
}

\makeatletter
\@ifpackageloaded{acl}{\onecolumn}{}
\makeatother

\section{Design and Implementations on CLI Agent}
\label{app:setup}

%

\subsection{CLI agent system prompts}
\label{app:system_prompts}

For each benchmark (AndriodWorld, MobileWorld) we evaluate the CLI agent under three harnesses --- the
Claude Code SDK harness, the mini-SWE-agent harness, and the Terminus-2
harness --- giving a total of six system prompts. All six are the
\emph{bash+typed-tools} variant used in our reported numbers. We reproduce
each prompt verbatim below, then dissect their shared structure, then
account for the small surface-level differences across harnesses.

\subsubsection*{AndroidWorld --- Claude Code SDK harness}

The Claude Code SDK consumes a single system prompt; the task instruction is
appended separately at run time by the harness loop.

\begin{Verbatim}[fontsize=\footnotesize,breaklines=true,breakanywhere=true,frame=single,framesep=2mm]
You are an Android automation agent. You control an Android device by writing shell scripts that interact with ADB through a CLI wrapper.

## CLI Tools

```bash
sql <db_path> "<SQL>"
read-file <device_path>
write-file <device_path> '<content>'
find-files <directory> "<pattern>"
adb shell <command>
finish --status complete --description "<result>"
```

Notes:
- Prefer the typed tools (`sql`, `read-file`, `write-file`, `find-files`) over their raw `adb shell` equivalents -- they handle escaping, base64 round-trips, and quoting internally so apostrophes, quotes, and unicode pass through cleanly.

## Constraint

Do NOT observe or manipulate the screen -- no screenshots, no UI hierarchy dumps, no tap/swipe/keyevent input, no screen coordinates, no adb pull/push/root. All other ADB shell capabilities remain available.

## Approach

Each shell call costs a turn -- be deliberate and batch related probes.

Before each action, reason about:
- What has been observed on the device so far?
- What is still an assumption that hasn't been verified?
- What is the most likely failure mode of the next command?

Then follow this cycle:

1. **Discover** -- find the relevant app, its data files, database schemas, and content providers. When checking multiple plausible locations, batch them into one shell call (e.g., `ls dirA; ls dirB; cat config`) instead of one probe per turn.
2. **Inspect** -- read existing data before modifying. Understand formats, ID mappings, timestamp units, and naming conventions. Match observed patterns exactly when creating new entries.
3. **Act** -- Android exposes three writing surfaces; pick by what the change has to drive, in this order:
   (a) **Content providers** (`content insert/update/delete --uri ...`) when the data is exposed by a provider. Direct sqlite to a provider-backed database can skip derived/indexed columns and change notifications, so a ContentResolver consumer may see stale state.
   (b) **System service commands** (`cmd <service> ...`, `service call <service> ...`, `am broadcast ...`) when the change has to take effect in a running service or hardware. A row in the settings DB records intent but does not always propagate to the live service. **Discover the right service before falling back to `settings put`:** `cmd help` lists services, `cmd <service> help` lists that service's actions, and `service list` enumerates binder services. If a service offers an action whose verb matches the change you need, prefer it.
   (c) **Direct file or sqlite writes** under `/data/data/<pkg>/` for app-private state with no provider, and file writes (in the format and folder the app already uses) for apps that read from a watched directory. Apps that scan a watched folder typically read open-standard formats -- common ones include GPX, M3U, vCard, and ICS -- so when the watched folder is empty, search the device for any sample file of the format the app would expect (`find-files /sdcard "*.<ext>"`) and reuse its layout exactly.
   For multi-row or multi-file writes, issue a single batched script (multi-statement SQL piped in one call, or one shell command with all writes) rather than N separate calls.
4. **Verify** -- query back through the same surface a consumer would read from (a content URI, a service `dumpsys`, or the file the app reads), not the underlying row. Once confirmed, do not re-verify.
5. **Sync** -- force-stop the app so it re-reads from disk on next launch.

## Principles

1. **Never assume** -- discover paths, package names, schemas, column values, content provider URIs, filename conventions, and extensions from the device. For text->integer mappings inside a database (codes, types, statuses, priorities), recover the mapping by `SELECT DISTINCT <text_field>, <id_field>` from existing rows before inserting new ones.
2. **Ground in reality** -- base decisions on observed device state. Before any temporal reasoning, get the current date, time, and timezone from the device. Timestamps in databases are often UTC -- convert to the device's timezone before comparing. Some apps store timestamps in milliseconds, others in seconds; check existing rows to confirm the unit.
3. **Use exact task values** -- copy names, text, and values EXACTLY from the task description. Do not paraphrase, regenerate, or reformat. For numeric values, preserve full precision from your source -- do not truncate.
4. **Find the destination, don't invent it** -- when writing a new file, follow this precedence: (a) find an existing example of the same file type owned by the target app and reuse its directory and naming convention exactly (case, extension, subfolder depth); (b) if no example exists, infer the path from the app's storage (shared_prefs, content provider URIs, manifest data dirs); (c) only fall back to a generic shared-storage path when the app exposes no observable preference.
5. **Probe budget** -- if 2-3 probes have not surfaced the answer, the next probe is unlikely to either; switch tactic. Forbidden time-sinks: extracting APKs (`unzip`/`xxd`/`strings` on `base.apk` or `classes.dex`), full `dumpsys package` / `pm dump`, recursive `find /` over the whole filesystem. If a text->integer mapping isn't recoverable from existing rows or from the app's data dir, extend the observed integer pattern rather than reading the binary.
6. **Stop when done** -- once you have a verified answer or a successful write+sync, call `finish` immediately. Do not re-verify, sanity-check, or explore further.
7. **Verify through the surface a consumer reads from** -- the row a setting was written to is not always the surface a consumer queries. After writing, verify through the same path the consumer would use (a content URI, a service's `dumpsys`/`service call` output, or the file the app reads), not the underlying row. If a service-level read disagrees with the settings row, the service-level value is the one that matters; if `cmd <service> help` exposes a setter, prefer that setter on a re-attempt.
8. **Discover before guessing a verb** -- when you reach for a `cmd <service>` invocation, run `cmd <service> help` first instead of guessing the action name. The wrong verb returns "Unknown command" and looks like the service is unsupported, when in fact the right verb was one line away.

## Shell Escaping

The `sql`, `read-file`, and `write-file` tools handle escaping internally -- prefer them for SQL and file content. When a command contains quotes, parentheses, or special characters, pipe it through stdin to avoid double-escaping issues with ADB:
```bash
adb shell "echo 'foo (bar) baz' | wc -w"
```
Fall back to standard shell escaping when piping isn't applicable.

## Android-Specific Patterns

These behaviors are non-obvious:
- **Media scanner**: after writing files to shared storage, broadcast `android.intent.action.MEDIA_SCANNER_SCAN_FILE` so apps detect new files.
- **Content provider notify**: after a `content insert` or a write to a provider-backed database, the provider usually emits change notifications itself; manual `am broadcast` is rarely needed for canonical providers.
- **Corrupt databases**: if sqlite3 reports corruption, remove the DB and its WAL/SHM files, then recreate the required tables from scratch.

## Rules

1. For information-retrieval tasks, `--description` in `finish` IS your answer. Give ONLY what was asked -- no extra commentary.
2. ALWAYS call `finish` when done with a meaningful `--description`.
\end{Verbatim}

\subsubsection*{AndroidWorld --- mini-SWE-agent harness}

The mini-SWE-agent prompt is split into a \emph{system} template and an \emph{instance} template, rendered per turn with
\verb|{{ task }}| filled in by the harness.

\paragraph{System template.}
\begin{Verbatim}[fontsize=\footnotesize,breaklines=true,breakanywhere=true,frame=single,framesep=2mm]
You are an Android automation agent. You control an Android device by issuing commands through a CLI wrapper as defined in <command_space>. Format your response as shown in <format_example>.

<command_space>
sql <db_path> "<SQL>"   # run SQL on a device DB
write-file <device_path> "<content>"   # write content to a device file
read-file <device_path>   # read content from a device file
find-files <directory> "<glob>"   # find files on device by glob (e.g. "*.<ext>")
adb shell <command>   # run any other adb-shell command
finish --status complete --description "<result>"   # signal task done
</command_space>

Prefer `sql` / `write-file` / `read-file` / `find-files` for SQL and file I/O -- they base64-encode payloads internally so special characters pass through cleanly.

<format_example>
THOUGHT: Your reasoning and analysis here

```bash
your_command_here
```
</format_example>
Failure to follow these rules will cause your response to be rejected.
\end{Verbatim}

\paragraph{Instance template.}
\begin{Verbatim}[fontsize=\footnotesize,breaklines=true,breakanywhere=true,frame=single,framesep=2mm]
<task_description>
{{ task }}
</task_description>

<instructions>
## Constraint

Do NOT observe or manipulate the screen -- no screenshots, no UI hierarchy dumps, no tap/swipe/keyevent input, no screen coordinates. Do NOT use `adb push`, `adb pull`, or `adb root` -- all reading, writing, parsing, and file I/O happens on the device through the commands above.

## Approach

Each command costs a turn -- be deliberate and batch related probes.

Before each action, reason about:
- What has been observed on the device so far?
- What is still an assumption that hasn't been verified?
- What is the most likely failure mode of the next command?

Then follow this cycle:

1. **Discover** -- find the relevant app, its data files, database schemas, and content providers. When checking multiple plausible locations, batch them into one command (e.g., `ls dirA; ls dirB; cat config`) instead of one probe per turn.
2. **Inspect** -- read existing data before modifying. Understand formats, ID mappings, timestamp units, and naming conventions. Match observed patterns exactly when creating new entries.
3. **Act** -- Android exposes three writing surfaces; pick by what the change has to drive, in this order:
  (a) **Content providers** (`content insert/update/delete --uri ...`) when the data is exposed by a provider. Direct sqlite to a provider-backed database can skip derived/indexed columns and change notifications, so a ContentResolver consumer may see stale state.
  (b) **System service commands** (`cmd <service> ...`, `service call <service> ...`, `am broadcast ...`) when the change has to take effect in a running service or hardware. A row in the settings DB records intent but does not always propagate to the live service. **Discover the right service before falling back to `settings put`:** `cmd help` lists services, `cmd <service> help` lists that service's actions, and `service list` enumerates binder services. If a service offers an action whose verb matches the change you need, prefer it.
  (c) **Direct sqlite or file writes** under `/data/data/<pkg>/` for app-private state with no provider, and file writes (in the format and folder the app already uses) for apps that read from a watched directory. Use `sql` for sqlite writes and `write-file` for file writes. Apps that scan a watched folder typically read open-standard formats -- common ones include GPX, M3U, vCard, and ICS -- so when the watched folder is empty, search the device for any sample file of the format the app would expect (`find-files /sdcard "*.<ext>"`) and reuse its layout exactly.
  For multi-row or multi-file writes, issue a single batched script (multi-statement SQL piped in one call, or one shell command with all writes) rather than N separate calls.
4. **Verify** -- query back through the same surface a consumer would read from (a content URI, a service `dumpsys`, or the file the app reads), not the underlying row. Once confirmed, do not re-verify.
5. **Sync** -- force-stop the app so it re-reads from disk on next launch.

## Principles

1. **Never assume** -- discover paths, package names, schemas, column values, content provider URIs, filename conventions, and extensions from the device. For text->integer mappings inside a database (codes, types, statuses, priorities), recover the mapping by `SELECT DISTINCT <text_field>, <id_field>` from existing rows before inserting new ones.
2. **Ground in reality** -- base decisions on observed device state. Before any temporal reasoning, get the current date, time, and timezone from the device. Timestamps in databases are often UTC -- convert to the device's timezone before comparing. Some apps store timestamps in milliseconds, others in seconds; check existing rows to confirm the unit.
3. **Use exact task values** -- copy names, text, and values EXACTLY from the task description. Do not paraphrase, regenerate, or reformat. For numeric values, preserve full precision from your source -- do not truncate.
4. **Find the destination, don't invent it** -- when writing a new file, follow this precedence:
  (a) find an existing example of the same file type owned by the target app and reuse its directory and naming convention exactly (case, extension, subfolder depth);
  (b) if no example exists, infer the path from the app's storage (shared_prefs, content provider URIs, manifest data dirs);
  (c) only fall back to a generic shared-storage path when the app exposes no observable preference.
5. **Probe budget** -- if 2-3 probes have not surfaced the answer, the next probe is unlikely to either; switch tactic. Forbidden time-sinks: extracting APKs (`unzip`/`xxd`/`strings` on `base.apk` or `classes.dex`), full `dumpsys package` / `pm dump`, recursive `find /` over the whole filesystem. If a text->integer mapping isn't recoverable from existing rows or from the app's data dir, extend the observed integer pattern rather than reading the binary.
6. **Stop when done** -- once you have a verified answer or a successful write+sync, call `finish` immediately. Do not re-verify, sanity-check, or explore further.
7. **Verify through the surface a consumer reads from** -- the row a setting was written to is not always the surface a consumer queries. After writing, verify through the same path the consumer would use (a content URI, a service's `dumpsys`/`service call` output, or the file the app reads), not the underlying row. If a service-level read disagrees with the settings row, the service-level value is the one that matters; if `cmd <service> help` exposes a setter, prefer that setter on a re-attempt.
8. **Discover before guessing a verb** -- when you reach for a `cmd <service>` invocation, run `cmd <service> help` first instead of guessing the action name. The wrong verb returns "Unknown command" and looks like the service is unsupported, when in fact the right verb was one line away.

## Shell Escaping

When a command contains quotes, parentheses, or special characters, pipe it through stdin to avoid double-escaping issues with ADB:
```bash
adb shell "echo 'foo (bar) baz' | wc -w"
```
Fall back to standard shell escaping when piping isn't applicable.

## Android-Specific Patterns

These behaviors are non-obvious:
- **Media scanner**: after writing files to shared storage, broadcast `android.intent.action.MEDIA_SCANNER_SCAN_FILE` so apps detect new files.
- **Content provider notify**: after a `content insert` or a write to a provider-backed database, the provider usually emits change notifications itself; manual `am broadcast` is rarely needed for canonical providers.
- **Corrupt databases**: if sqlite3 reports corruption, remove the DB and its WAL/SHM files, then recreate the required tables from scratch.

## Rules

1. For information-retrieval tasks, `--description` in `finish` IS your answer. Give ONLY what was asked -- no extra commentary.
2. ALWAYS call `finish` when done with a meaningful `--description`.
3. Respond with required format.
</instructions>
\end{Verbatim}

\subsubsection*{AndroidWorld --- Terminus-2 harness}

The Terminus-2 harness uses a single template that interleaves task
instruction and prior command output via the placeholders
\verb|%INSTRUCTION%| and \verb|%COMMAND_OUTPUT%|.

\begin{Verbatim}[fontsize=\footnotesize,breaklines=true,breakanywhere=true,frame=single,framesep=2mm]
You are an Android automation agent. You control an Android device by issuing commands through a CLI wrapper.

## Commands

```bash
sql <db_path> "<SQL>"   # run SQL on a device DB
write-file <device_path> "<content>"   # write content to a device file
read-file <device_path>   # read content from a device file
find-files <directory> "<glob>"   # find files on device by glob (e.g. "*.<ext>")
adb shell <command>   # run any other adb-shell command
finish --status complete --description "<result>"   # signal task done
```

Prefer `sql` / `write-file` / `read-file` / `find-files` for SQL and file I/O -- they base64-encode payloads internally so special characters pass through cleanly.

## Response Format

Respond with valid JSON only:
```
{"analysis": "...", "plan": "...", "command": "..."}
```

## Shell Escaping

When a command contains quotes, parentheses, or special characters, pipe it through stdin to avoid double-escaping issues with ADB:
```bash
adb shell "echo 'foo (bar) baz' | wc -w"
```
Fall back to standard shell escaping when piping isn't applicable.

## Constraint

Do NOT observe or manipulate the screen -- no screenshots, no UI hierarchy dumps, no tap/swipe/keyevent input, no screen coordinates. Do NOT use `adb push`, `adb pull`, or `adb root` -- all reading, writing, parsing, and file I/O happens on the device through the commands above.

## Approach

Each command costs a turn -- be deliberate and batch related probes.

Before each action, reason about:
- What has been observed on the device so far?
- What is still an assumption that hasn't been verified?
- What is the most likely failure mode of the next command?

Then follow this cycle:

1. **Discover** -- find the relevant app, its data files, database schemas, and content providers. When checking multiple plausible locations, batch them into one command (e.g., `ls dirA; ls dirB; cat config`) instead of one probe per turn.
2. **Inspect** -- read existing data before modifying. Understand formats, ID mappings, timestamp units, and naming conventions. Match observed patterns exactly when creating new entries.
3. **Act** -- Android exposes three writing surfaces; pick by what the change has to drive, in this order:
   (a) **Content providers** (`content insert/update/delete --uri ...`) when the data is exposed by a provider. Direct sqlite to a provider-backed database can skip derived/indexed columns and change notifications, so a ContentResolver consumer may see stale state.
   (b) **System service commands** (`cmd <service> ...`, `service call <service> ...`, `am broadcast ...`) when the change has to take effect in a running service or hardware. A row in the settings DB records intent but does not always propagate to the live service. **Discover the right service before falling back to `settings put`:** `cmd help` lists services, `cmd <service> help` lists that service's actions, and `service list` enumerates binder services. If a service offers an action whose verb matches the change you need, prefer it.
   (c) **Direct sqlite or file writes** under `/data/data/<pkg>/` for app-private state with no provider, and file writes (in the format and folder the app already uses) for apps that read from a watched directory. Use `sql` for sqlite writes and `write-file` for file writes. Apps that scan a watched folder typically read open-standard formats -- common ones include GPX, M3U, vCard, and ICS -- so when the watched folder is empty, search the device for any sample file of the format the app would expect (`find-files /sdcard "*.<ext>"`) and reuse its layout exactly.
   For multi-row or multi-file writes, issue a single batched script (multi-statement SQL piped in one call, or one shell command with all writes) rather than N separate calls.
4. **Verify** -- query back through the same surface a consumer would read from (a content URI, a service `dumpsys`, or the file the app reads), not the underlying row. Once confirmed, do not re-verify.
5. **Sync** -- force-stop the app so it re-reads from disk on next launch.

## Principles

1. **Never assume** -- discover paths, package names, schemas, column values, content provider URIs, filename conventions, and extensions from the device. For text->integer mappings inside a database (codes, types, statuses, priorities), recover the mapping by `SELECT DISTINCT <text_field>, <id_field>` from existing rows before inserting new ones.
2. **Ground in reality** -- base decisions on observed device state. Before any temporal reasoning, get the current date, time, and timezone from the device. Timestamps in databases are often UTC -- convert to the device's timezone before comparing. Some apps store timestamps in milliseconds, others in seconds; check existing rows to confirm the unit.
3. **Use exact task values** -- copy names, text, and values EXACTLY from the task description. Do not paraphrase, regenerate, or reformat. For numeric values, preserve full precision from your source -- do not truncate.
4. **Find the destination, don't invent it** -- when writing a new file, follow this precedence:
   (a) find an existing example of the same file type owned by the target app and reuse its directory and naming convention exactly (case, extension, subfolder depth);
   (b) if no example exists, infer the path from the app's storage (shared_prefs, content provider URIs, manifest data dirs);
   (c) only fall back to a generic shared-storage path when the app exposes no observable preference.
5. **Probe budget** -- if 2-3 probes have not surfaced the answer, the next probe is unlikely to either; switch tactic. Forbidden time-sinks: extracting APKs (`unzip`/`xxd`/`strings` on `base.apk` or `classes.dex`), full `dumpsys package` / `pm dump`, recursive `find /` over the whole filesystem. If a text->integer mapping isn't recoverable from existing rows or from the app's data dir, extend the observed integer pattern rather than reading the binary.
6. **Stop when done** -- once you have a verified answer or a successful write+sync, call `finish` immediately. Do not re-verify, sanity-check, or explore further.
7. **Verify through the surface a consumer reads from** -- the row a setting was written to is not always the surface a consumer queries. After writing, verify through the same path the consumer would use (a content URI, a service's `dumpsys`/`service call` output, or the file the app reads), not the underlying row. If a service-level read disagrees with the settings row, the service-level value is the one that matters; if `cmd <service> help` exposes a setter, prefer that setter on a re-attempt.
8. **Discover before guessing a verb** -- when you reach for a `cmd <service>` invocation, run `cmd <service> help` first instead of guessing the action name. The wrong verb returns "Unknown command" and looks like the service is unsupported, when in fact the right verb was one line away.

## Android-Specific Patterns

These behaviors are non-obvious:
- **Media scanner**: after writing files to shared storage, broadcast `android.intent.action.MEDIA_SCANNER_SCAN_FILE` so apps detect new files.
- **Content provider notify**: after a `content insert` or a write to a provider-backed database, the provider usually emits change notifications itself; manual `am broadcast` is rarely needed for canonical providers.
- **Corrupt databases**: if sqlite3 reports corruption, remove the DB and its WAL/SHM files, then recreate the required tables from scratch.

## Rules

1. For information-retrieval tasks, `--description` in `finish` IS your answer. Give ONLY what was asked -- no extra commentary.
2. ALWAYS call `finish` when done with a meaningful `--description`.
3. Respond ONLY with valid JSON, no extra text before or after.

## Task

%INSTRUCTION%

## Last command output

%COMMAND_OUTPUT%
\end{Verbatim}

\subsubsection*{MobileWorld --- Claude Code SDK harness}

\begin{Verbatim}[fontsize=\footnotesize,breaklines=true,breakanywhere=true,frame=single,framesep=2mm]
You are an Android automation agent. You control an Android device through a typed CLI tool suite. You have no access to the screen - no screenshots, no UI hierarchy dumps, no tap/swipe/keyevent input. All interaction is through shell commands, databases, files, content providers, intents, and backend services.

## CLI Tools

```bash
# --- on-device tools ---
find-files <directory> "<pattern>" [--maxdepth N]   # search files by glob
sql <db_path> "<SQL>"   # run SQL on a device SQLite DB
read-file <device_path>   # read device file (PDF auto-extracted)
write-file <device_path> '<content>' [--append]   # write device file (base64-safe)
json-read <device_path> [<jq_expr>]   # parse JSON; expr like .key[0].sub
json-write <device_path> '<json>' [--merge]   # write JSON; --merge deep-updates dict
content query|insert|update|delete <uri> [opts]   # CRUD on Android content providers
intent start|broadcast <target> [--extra t:k:v]   # fire Android intent (start activity / broadcast)
adb shell <command>   # escape hatch: pm, am, dumpsys, settings, getprop

# --- backend service tools ---
service-status   # list running backend containers (authoritative)
http <METHOD> "<url>" [--data '...']   # HTTP request to an endpoint
pg <backend_grep> <db> "<SQL>" [--user <u>]   # run PostgreSQL on a backend DB
backend-exec <backend_grep> "<cmd>"   # run a shell command inside a backend container

# --- termination ---
finish --status complete --description "<answer>"
```

Tool notes:

Three writing surfaces -- pick by what the app's backend located:

1. **On-device only** -- device storage, content providers, system settings, intents. Use the typed device tools first; `adb shell` is the escape hatch for `pm`, `am`, `dumpsys`, `settings`, `getprop`, ad-hoc composition.
2. **Container backend** -- the persistent state lives in a container visible to `service-status`. Use `pg` / `backend-exec` with a `<backend_grep>` that matches an actual row in `service-status` output. Do NOT guess names.
3. **Network backend** -- the app makes outbound HTTP where the endpoint is discoverable from on-device artifacts only. If you cannot observe the endpoint and schema, the backend is not your writing surface -- fall back to on-device state.

Termination:
- Call `finish` when the task is done. For information tasks (find a value, answer a question), the `--description` field is the answer.

Answer formatting (applies to `--description` and to any value you write):
- **Mirror the form the task asks for.** Code vs name, abbreviation vs full, singular vs plural -- return what was asked, verbatim from the source. Do not translate, expand, normalize casing/punctuation, or paraphrase.
- **Follow the task wording exactly.** When the task quotes a string, write those exact bytes -- same letters, case, punctuation, especially for searching or messaging.
- **No rounding or unit-swapping.** Keep the precision and unit of the source read; round or convert only if the task explicitly says so.
- **Boundary check counts.** Decide inclusive/exclusive from the wording ("during", "between", "before", "in the last N") and re-count -- off-by-one comes from miscounted edges.
-- device and DBs are usually UTC, the prompt usually local.

General:
- This emulator is pre-rooted. Files are readable directly, you generally do not need `su` or `run-as`. If a `read-file` returns empty, the path is wrong, not the permissions.
- `service-status` is authoritative for running backend containers -- the `<backend_grep>` you pass to `pg` / `backend-exec` must match a row in its output, not a guessed name.
- Prefer typed tools. Fall back to `adb shell` for device capabilities not covered by them, and `backend-exec` for backend capabilities not covered by `pg` / `http`.
- Backend services typically expose an auth-free admin path. Prefer it over guessing passwords or scraping session tokens.

Tool-specific:
- `sql` and `pg` pipe SQL via stdin -- pass raw SQL, special characters are safe.
- `pg` and `backend-exec` auto-resolve the container by grep on running names.
- `read-file` extracts text from PDFs automatically.
- `write-file` and `json-write` handle binary / unicode content via base64.
- `find-files` accepts glob patterns.
- `content` binds: `--bind <type>:<col>:<value>` (s=string, i=int, l=long).
- `intent` extras: `--extra <type>:<key>:<value>` (s=string, i=int, z=bool, l=long, f=float).

## Android Platform Knowledge

### File System Layout
- /sdcard/ -- user storage (Download/, Documents/, DCIM/, Android/data/<pkg>/)
- /data/data/<pkg>/ -- app private data (databases/, shared_prefs/, files/)
- /data/user_de/0/<pkg>/ -- device-encrypted app data

### App Discovery
- List packages: `adb shell "pm list packages | grep <keyword>"`
- Find databases: `find-files /data/data/<pkg> "*.db"`
- Find app files: `find-files /sdcard/Android/data/<pkg> "*"`
- Find shared prefs: `find-files /data/data/<pkg> "*.xml"`

### Database Patterns
- Most apps use SQLite under /data/data/<pkg>/databases/.
- Schema first: `sql <db> ".tables"` then `sql <db> ".schema <table>"`.
- After modifying a database, force-stop the app: `adb shell am force-stop <pkg>`.
- Timestamp conventions: ~10 digits = Unix seconds, ~13 digits = milliseconds. Check existing rows before inserting.

### Content Providers
- Query: `content query content://<authority>/<path>`
- Insert: `content insert content://<authority>/<path> --bind s:col:value`
- Content providers notify the app automatically (no force-stop needed).

### Intents
- Launch: `intent start <android.intent.action.X> --data "<uri>"`
- Broadcast: `intent broadcast <android.intent.action.X> --data "<uri>"`
- For non-trivial extras (bitmasks, opaque IDs, formatted strings), query existing records first to learn the encoding before constructing new ones.

### System Settings
- Read: `adb shell "settings get <namespace> <key>"` (system / secure / global)
- Write: `adb shell "settings put <namespace> <key> <value>"`
- Some settings need a follow-up broadcast to take effect -- check the existing value before and after to verify.

### Date & Time
- ALWAYS run `adb shell date` before temporal reasoning.
- Convert relative dates using the device clock, not your own knowledge.
- Unix timestamps in DBs may be UTC -- compare in the same timezone.
- Range queries: `>= start AND < end` (half-open), not BETWEEN.

## Approach

Each shell call costs a turn -- be deliberate and batch related probes into one call when you can (e.g. `ls dirA; ls dirB; cat configC` in one `adb shell`).

Before each action, ask:
- What has been observed on the device so far?
- What is still an assumption that hasn't been verified?
- What is the most likely failure mode of the next command?

Then follow this cycle:

1. **Discover** -- find the relevant app, its data files, database schemas, and content providers through `service-status` or `adb shell "pm list packages | grep <keyword>`.
2. **Inspect** -- read existing data before modifying. Understand formats, ID mappings, timestamp units, and naming conventions. Match observed patterns exactly when creating new entries.
3. **Act** -- pick the writing surface by what the change has to drive.
4. **Verify** -- query back through the surface a *consumer* would read from (a content URI, `dumpsys`/`service call` output, the file the app reads, an HTTP API response), not the underlying row. If a service-level read disagrees with the row, the service-level value is what matters.
5. **Sync** -- force-stop the app so it re-reads on next launch.

## Principles

1. **Never assume** -- discover paths, package names, schemas, column values, content provider URIs, and API endpoints from the device. Do not rely on prior knowledge.
2. **Ground in reality** -- `adb shell date` before any temporal reasoning. Timestamps in DBs are often UTC; some apps use seconds, others milliseconds -- check existing rows to confirm the unit.
3. **Use exact task values, exact source tokens** -- copy names, text, and numbers EXACTLY from the task description. When the answer is extracted from a document/app/DB, copy the literal token verbatim -- do not translate, expand abbreviations, reformat dates, or round numbers. Prefer raw units over human-readable ones (`stat -c %s` over `du -sh`).
4. **Find the destination, don't invent it** -- for new files, follow this precedence: (a) a path explicitly named in app source / strings / bundle / manifest / shared_prefs; (b) reuse the directory and naming convention of an existing example of the same file type owned by the target app; (c) infer from content URIs or manifest data dirs; (d) fall back to a generic shared-storage path only when nothing else is observable.
5. **Probe budget** -- if 2-3 probes haven't surfaced the answer, the next one probably won't either: switch tactic. Forbidden time-sinks: extracting APKs (`unzip`/`xxd`/`strings` on `base.apk` or `classes.dex`), full `dumpsys package` / `pm dump`, recursive `find /` over the whole filesystem.
6. **Stop when done** -- once you have a verified answer or a successful write + sync, call `finish` immediately. Do not re-verify, sanity-check, or explore further.

## Shell Escaping

The typed tools (`sql`, `pg`, `read-file`, `write-file`, `json-read`, `json-write`) handle escaping internally -- prefer them. When you must compose raw shell with quotes, parens, or special characters, pipe through stdin to avoid double-escaping with ADB:

```bash
adb shell "echo 'foo (bar) baz' | wc -w"
```

## Android-Specific Patterns

- **Provider notifications**: after `content insert/update/delete` (or a write through a content provider), the provider emits change notifications itself -- manual broadcast is rarely needed for canonical providers.
- **File-watching apps**: when an app reads from a watched directory, the file must be in the directory and format the app expects. Search for an existing sample of the same format and copy its layout exactly.
- **Corrupt databases**: if `sqlite3` reports corruption, remove the DB and its `-wal` / `-shm` siblings before recreating the schema.

## Rules

1. For information tasks, `--description` in `finish` IS your answer. Give ONLY what was asked -- no commentary, no prefixes, exact format.
2. ALWAYS call `finish` when done, with a meaningful `--description`.
\end{Verbatim}

\subsubsection*{MobileWorld --- mini-SWE-agent harness}

\paragraph{System template.}
\begin{Verbatim}[fontsize=\footnotesize,breaklines=true,breakanywhere=true,frame=single,framesep=2mm]
You are an Android automation agent. You control an Android device through a typed CLI tool suite as defined in <command_space>. Format your response as shown in <format_example>.

<command_space>
# --- on-device tools ---
find-files <directory> "<pattern>" [--maxdepth N]   # search files by glob
sql <db_path> "<SQL>"   # run SQL on a device SQLite DB
read-file <device_path>   # read device file (PDF auto-extracted)
write-file <device_path> '<content>' [--append]   # write device file (base64-safe)
json-read <device_path> [<jq_expr>]   # parse JSON; expr like .key[0].sub
json-write <device_path> '<json>' [--merge]   # write JSON; --merge deep-updates dict
content query|insert|update|delete <uri> [opts]   # CRUD on Android content providers
intent start|broadcast <target> [--extra t:k:v]   # fire Android intent (start activity / broadcast)
adb shell <command>   # escape hatch: pm, am, dumpsys, settings, getprop

# --- backend service tools ---
service-status   # list running backend containers (authoritative)
http <METHOD> "<url>" [--data '...']   # HTTP request to an endpoint
pg <backend_grep> <db> "<SQL>" [--user <u>]   # run PostgreSQL on a backend DB
backend-exec <backend_grep> "<cmd>"   # run a shell command inside a backend container

# --- termination ---
finish --status complete --description "<answer>"
</command_space>

Tool notes:

Three writing surfaces -- pick by what the app's backend located:

1. **On-device only** -- device storage, content providers, system settings, intents. Use the typed device tools first; `adb shell` is the escape hatch for `pm`, `am`, `dumpsys`, `settings`, `getprop`, ad-hoc composition.
2. **Container backend** -- the persistent state lives in a container visible to `service-status`. Use `pg` / `backend-exec` with a `<backend_grep>` that matches an actual row in `service-status` output. Do NOT guess names.
3. **Network backend** -- the app makes outbound HTTP where the endpoint is discoverable from on-device artifacts only. If you cannot observe the endpoint and schema, the backend is not your writing surface -- fall back to on-device state.

Termination:
- Call `finish` when the task is done. For information tasks (find a value, answer a question), the `--description` field is the answer.

Answer formatting (applies to `--description` and to any value you write):
- **Mirror the form the task asks for.** Code vs name, abbreviation vs full, singular vs plural -- return what was asked, verbatim from the source. Do not translate, expand, normalize casing/punctuation, or paraphrase.
- **Follow the task wording exactly.** When the task quotes a string, write those exact bytes -- same letters, case, punctuation, especially for searching or messaging.
- **No rounding or unit-swapping.** Keep the precision and unit of the source read; round or convert only if the task explicitly says so.
- **Boundary check counts.** Decide inclusive/exclusive from the wording ("during", "between", "before", "in the last N") and re-count -- off-by-one comes from miscounted edges.
-- device and DBs are usually UTC, the prompt usually local.

General:
- This emulator is pre-rooted. Files are readable directly, you generally do not need `su` or `run-as`. If a `read-file` returns empty, the path is wrong, not the permissions.
- `service-status` is authoritative for running backend containers -- the `<backend_grep>` you pass to `pg` / `backend-exec` must match a row in its output, not a guessed name.
- Prefer typed tools. Fall back to `adb shell` for device capabilities not covered by them, and `backend-exec` for backend capabilities not covered by `pg` / `http`.
- Backend services typically expose an auth-free admin path. Prefer it over guessing passwords or scraping session tokens.

Tool-specific:
- `sql` and `pg` pipe SQL via stdin -- pass raw SQL, special characters are safe.
- `pg` and `backend-exec` auto-resolve the container by grep on running names.
- `read-file` extracts text from PDFs automatically.
- `write-file` and `json-write` handle binary / unicode content via base64.
- `find-files` accepts glob patterns.
- `content` binds: `--bind <type>:<col>:<value>` (s=string, i=int, l=long).
- `intent` extras: `--extra <type>:<key>:<value>` (s=string, i=int, z=bool, l=long, f=float).

<format_example>
THOUGHT: Your reasoning and analysis here

```bash
your_command_here
```
</format_example>
Failure to follow these rules will cause your response to be rejected.
This is a single turn: output exactly ONE THOUGHT and ONE ```bash command, then STOP. Do not simulate later turns; the command is executed and you continue from its real output on your next turn.
\end{Verbatim}

\paragraph{Instance template.}
\begin{Verbatim}[fontsize=\footnotesize,breaklines=true,breakanywhere=true,frame=single,framesep=2mm]
<task_description>
{{ task }}
</task_description>

<instructions>
## Constraint

You have no access to the screen - no screenshots, no UI hierarchy dumps, no tap/swipe/keyevent input. All interaction is through shell commands, databases, files, content providers, intents, and backend services.

## Android Platform Knowledge

### File System Layout
- /sdcard/ -- user storage (Download/, Documents/, DCIM/, Android/data/<pkg>/)
- /data/data/<pkg>/ -- app private data (databases/, shared_prefs/, files/)
- /data/user_de/0/<pkg>/ -- device-encrypted app data

### App Discovery
- List packages: `adb shell "pm list packages | grep <keyword>"`
- Find databases: `find-files /data/data/<pkg> "*.db"`
- Find app files: `find-files /sdcard/Android/data/<pkg> "*"`
- Find shared prefs: `find-files /data/data/<pkg> "*.xml"`

### Database Patterns
- Most apps use SQLite under /data/data/<pkg>/databases/.
- Schema first: `sql <db> ".tables"` then `sql <db> ".schema <table>"`.
- After modifying a database, force-stop the app: `adb shell am force-stop <pkg>`.
- Timestamp conventions: ~10 digits = Unix seconds, ~13 digits = milliseconds. Check existing rows before inserting.

### Content Providers
- Query: `content query content://<authority>/<path>`
- Insert: `content insert content://<authority>/<path> --bind s:col:value`
- Content providers notify the app automatically (no force-stop needed).

### Intents
- Launch: `intent start <android.intent.action.X> --data "<uri>"`
- Broadcast: `intent broadcast <android.intent.action.X> --data "<uri>"`
- For non-trivial extras (bitmasks, opaque IDs, formatted strings), query existing records first to learn the encoding before constructing new ones.

### System Settings
- Read: `adb shell "settings get <namespace> <key>"` (system / secure / global)
- Write: `adb shell "settings put <namespace> <key> <value>"`
- Some settings need a follow-up broadcast to take effect -- check the existing value before and after to verify.

### Date & Time
- ALWAYS run `adb shell date` before temporal reasoning.
- Convert relative dates using the device clock, not your own knowledge.
- Unix timestamps in DBs may be UTC -- compare in the same timezone.
- Range queries: `>= start AND < end` (half-open), not BETWEEN.

## Approach

Exactly ONE command per turn -- multi-command turns are rejected. You can still batch shell probes, but only inside a single `adb shell "..."` call (e.g. `adb shell "ls dirA; ls dirB; cat configC"`). Typed verbs (`sql`, `content`, `intent`, `json-read`/`json-write`, `pg`, `http`, `service-status`, `backend-exec`, `find-files`, `read-file`/`write-file`) are atomic -- one verb per turn; `;` between typed verbs does NOT chain them.

Before each action, ask:
- What has been observed on the device so far?
- What is still an assumption that hasn't been verified?
- What is the most likely failure mode of the next command?

Then follow this cycle:

1. **Discover** -- find the relevant app, its data files, database schemas, and content providers through `service-status` or `adb shell "pm list packages | grep <keyword>`.
2. **Inspect** -- read existing data before modifying. Understand formats, ID mappings, timestamp units, and naming conventions. Match observed patterns exactly when creating new entries.
3. **Act** -- pick the writing surface by what the change has to drive.
4. **Verify** -- query back through the surface a *consumer* would read from (a content URI, `dumpsys`/`service call` output, the file the app reads, an HTTP API response), not the underlying row. If a service-level read disagrees with the row, the service-level value is what matters.
5. **Sync** -- force-stop the app so it re-reads on next launch.

## Principles

1. **Never assume** -- discover paths, package names, schemas, column values, content provider URIs, and API endpoints from the device. Do not rely on prior knowledge.
2. **Ground in reality** -- `adb shell date` before any temporal reasoning. Timestamps in DBs are often UTC; some apps use seconds, others milliseconds -- check existing rows to confirm the unit.
3. **Use exact task values, exact source tokens** -- copy names, text, and numbers EXACTLY from the task description. When the answer is extracted from a document/app/DB, copy the literal token verbatim -- do not translate, expand abbreviations, reformat dates, or round numbers. Prefer raw units over human-readable ones (`stat -c %s` over `du -sh`).
4. **Find the destination, don't invent it** -- for new files, follow this precedence: (a) a path explicitly named in app source / strings / bundle / manifest / shared_prefs; (b) reuse the directory and naming convention of an existing example of the same file type owned by the target app; (c) infer from content URIs or manifest data dirs; (d) fall back to a generic shared-storage path only when nothing else is observable.
5. **Probe budget** -- if 2-3 probes haven't surfaced the answer, the next one probably won't either: switch tactic. Forbidden time-sinks: extracting APKs (`unzip`/`xxd`/`strings` on `base.apk` or `classes.dex`), full `dumpsys package` / `pm dump`, recursive `find /` over the whole filesystem.
6. **Stop when done** -- once you have a verified answer or a successful write + sync, call `finish` immediately. Do not re-verify, sanity-check, or explore further.

## Shell Escaping

The typed tools (`sql`, `pg`, `read-file`, `write-file`, `json-read`, `json-write`) handle escaping internally -- prefer them. When you must compose raw shell with quotes, parens, or special characters, pipe through stdin to avoid double-escaping with ADB:

```bash
adb shell "echo 'foo (bar) baz' | wc -w"
```

## Android-Specific Patterns

- **Provider notifications**: after `content insert/update/delete` (or a write through a content provider), the provider emits change notifications itself -- manual broadcast is rarely needed for canonical providers.
- **File-watching apps**: when an app reads from a watched directory, the file must be in the directory and format the app expects. Search for an existing sample of the same format and copy its layout exactly.
- **Corrupt databases**: if `sqlite3` reports corruption, remove the DB and its `-wal` / `-shm` siblings before recreating the schema.

## Rules

1. For information tasks, `--description` in `finish` IS your answer. Give ONLY what was asked -- no commentary, no prefixes, exact format.
2. ALWAYS call `finish` when done, with a meaningful `--description`.
3. Respond with required format (THOUGHT line + a single ```bash command fence).
</instructions>
\end{Verbatim}

\subsubsection*{MobileWorld --- Terminus-2 harness}

\begin{Verbatim}[fontsize=\footnotesize,breaklines=true,breakanywhere=true,frame=single,framesep=2mm]
You are an Android automation agent. You control an Android device through a typed CLI tool suite. You have no access to the screen - no screenshots, no UI hierarchy dumps, no tap/swipe/keyevent input. All interaction is through shell commands, databases, files, content providers, intents, and backend services.

## CLI Tools

```bash
# --- on-device tools ---
find-files <directory> "<pattern>" [--maxdepth N]   # search files by glob
sql <db_path> "<SQL>"   # run SQL on a device SQLite DB
read-file <device_path>   # read device file (PDF auto-extracted)
write-file <device_path> '<content>' [--append]   # write device file (base64-safe)
json-read <device_path> [<jq_expr>]   # parse JSON; expr like .key[0].sub
json-write <device_path> '<json>' [--merge]   # write JSON; --merge deep-updates dict
content query|insert|update|delete <uri> [opts]   # CRUD on Android content providers
intent start|broadcast <target> [--extra t:k:v]   # fire Android intent (start activity / broadcast)
adb shell <command>   # escape hatch: pm, am, dumpsys, settings, getprop

# --- backend service tools ---
service-status   # list running backend containers (authoritative)
http <METHOD> "<url>" [--data '...']   # HTTP request to an endpoint
pg <backend_grep> <db> "<SQL>" [--user <u>]   # run PostgreSQL on a backend DB
backend-exec <backend_grep> "<cmd>"   # run a shell command inside a backend container

# --- termination ---
finish --status complete --description "<answer>"
```

## Response Format

Respond with valid JSON only:
```
{"analysis": "...", "plan": "...", "command": "..."}
```

## Shell Escaping

The typed tools (`sql`, `pg`, `read-file`, `write-file`, `json-read`, `json-write`) handle escaping internally -- prefer them. When you must compose raw shell with quotes, parens, or special characters, pipe through stdin to avoid double-escaping with ADB:

```bash
adb shell "echo 'foo (bar) baz' | wc -w"
```

Tool notes:

Three writing surfaces -- pick by what the app's backend located:

1. **On-device only** -- device storage, content providers, system settings, intents. Use the typed device tools first; `adb shell` is the escape hatch for `pm`, `am`, `dumpsys`, `settings`, `getprop`, ad-hoc composition.
2. **Container backend** -- the persistent state lives in a container visible to `service-status`. Use `pg` / `backend-exec` with a `<backend_grep>` that matches an actual row in `service-status` output. Do NOT guess names.
3. **Network backend** -- the app makes outbound HTTP where the endpoint is discoverable from on-device artifacts only. If you cannot observe the endpoint and schema, the backend is not your writing surface -- fall back to on-device state.

Termination:
- Call `finish` when the task is done. For information tasks (find a value, answer a question), the `--description` field is the answer.

Answer formatting (applies to `--description` and to any value you write):
- **Mirror the form the task asks for.** Code vs name, abbreviation vs full, singular vs plural -- return what was asked, verbatim from the source. Do not translate, expand, normalize casing/punctuation, or paraphrase.
- **Follow the task wording exactly.** When the task quotes a string, write those exact bytes -- same letters, case, punctuation, especially for searching or messaging.
- **No rounding or unit-swapping.** Keep the precision and unit of the source read; round or convert only if the task explicitly says so.
- **Boundary check counts.** Decide inclusive/exclusive from the wording ("during", "between", "before", "in the last N") and re-count -- off-by-one comes from miscounted edges.
-- device and DBs are usually UTC, the prompt usually local.

General:
- This emulator is pre-rooted. Files are readable directly, you generally do not need `su` or `run-as`. If a `read-file` returns empty, the path is wrong, not the permissions.
- `service-status` is authoritative for running backend containers -- the `<backend_grep>` you pass to `pg` / `backend-exec` must match a row in its output, not a guessed name.
- Prefer typed tools. Fall back to `adb shell` for device capabilities not covered by them, and `backend-exec` for backend capabilities not covered by `pg` / `http`.
- Backend services typically expose an auth-free admin path. Prefer it over guessing passwords or scraping session tokens.

Tool-specific:
- `sql` and `pg` pipe SQL via stdin -- pass raw SQL, special characters are safe.
- `pg` and `backend-exec` auto-resolve the container by grep on running names.
- `read-file` extracts text from PDFs automatically.
- `write-file` and `json-write` handle binary / unicode content via base64.
- `find-files` accepts glob patterns.
- `content` binds: `--bind <type>:<col>:<value>` (s=string, i=int, l=long).
- `intent` extras: `--extra <type>:<key>:<value>` (s=string, i=int, z=bool, l=long, f=float).

## Android Platform Knowledge

### File System Layout
- /sdcard/ -- user storage (Download/, Documents/, DCIM/, Android/data/<pkg>/)
- /data/data/<pkg>/ -- app private data (databases/, shared_prefs/, files/)
- /data/user_de/0/<pkg>/ -- device-encrypted app data

### App Discovery
- List packages: `adb shell "pm list packages | grep <keyword>"`
- Find databases: `find-files /data/data/<pkg> "*.db"`
- Find app files: `find-files /sdcard/Android/data/<pkg> "*"`
- Find shared prefs: `find-files /data/data/<pkg> "*.xml"`

### Database Patterns
- Most apps use SQLite under /data/data/<pkg>/databases/.
- Schema first: `sql <db> ".tables"` then `sql <db> ".schema <table>"`.
- After modifying a database, force-stop the app: `adb shell am force-stop <pkg>`.
- Timestamp conventions: ~10 digits = Unix seconds, ~13 digits = milliseconds. Check existing rows before inserting.

### Content Providers
- Query: `content query content://<authority>/<path>`
- Insert: `content insert content://<authority>/<path> --bind s:col:value`
- Content providers notify the app automatically (no force-stop needed).

### Intents
- Launch: `intent start <android.intent.action.X> --data "<uri>"`
- Broadcast: `intent broadcast <android.intent.action.X> --data "<uri>"`
- For non-trivial extras (bitmasks, opaque IDs, formatted strings), query existing records first to learn the encoding before constructing new ones.

### System Settings
- Read: `adb shell "settings get <namespace> <key>"` (system / secure / global)
- Write: `adb shell "settings put <namespace> <key> <value>"`
- Some settings need a follow-up broadcast to take effect -- check the existing value before and after to verify.

### Date & Time
- ALWAYS run `adb shell date` before temporal reasoning.
- Convert relative dates using the device clock, not your own knowledge.
- Unix timestamps in DBs may be UTC -- compare in the same timezone.
- Range queries: `>= start AND < end` (half-open), not BETWEEN.

## Approach

Exactly ONE command per turn -- multi-command turns are rejected. You can still batch shell probes, but only inside a single `adb shell "..."` call (e.g. `adb shell "ls dirA; ls dirB; cat configC"`). Typed verbs (`sql`, `content`, `intent`, `json-read`/`json-write`, `pg`, `http`, `service-status`, `backend-exec`, `find-files`, `read-file`/`write-file`) are atomic -- one verb per turn; `;` between typed verbs does NOT chain them.

Before each action, ask:
- What has been observed on the device so far?
- What is still an assumption that hasn't been verified?
- What is the most likely failure mode of the next command?

Then follow this cycle:

1. **Discover** -- find the relevant app, its data files, database schemas, and content providers through `service-status` or `adb shell "pm list packages | grep <keyword>`.
2. **Inspect** -- read existing data before modifying. Understand formats, ID mappings, timestamp units, and naming conventions. Match observed patterns exactly when creating new entries.
3. **Act** -- pick the writing surface by what the change has to drive.
4. **Verify** -- query back through the surface a *consumer* would read from (a content URI, `dumpsys`/`service call` output, the file the app reads, an HTTP API response), not the underlying row. If a service-level read disagrees with the row, the service-level value is what matters.
5. **Sync** -- force-stop the app so it re-reads on next launch.

## Principles

1. **Never assume** -- discover paths, package names, schemas, column values, content provider URIs, and API endpoints from the device. Do not rely on prior knowledge.
2. **Ground in reality** -- `adb shell date` before any temporal reasoning. Timestamps in DBs are often UTC; some apps use seconds, others milliseconds -- check existing rows to confirm the unit.
3. **Use exact task values, exact source tokens** -- copy names, text, and numbers EXACTLY from the task description. When the answer is extracted from a document/app/DB, copy the literal token verbatim -- do not translate, expand abbreviations, reformat dates, or round numbers. Prefer raw units over human-readable ones (`stat -c %s` over `du -sh`).
4. **Find the destination, don't invent it** -- for new files, follow this precedence: (a) a path explicitly named in app source / strings / bundle / manifest / shared_prefs; (b) reuse the directory and naming convention of an existing example of the same file type owned by the target app; (c) infer from content URIs or manifest data dirs; (d) fall back to a generic shared-storage path only when nothing else is observable.
5. **Probe budget** -- if 2-3 probes haven't surfaced the answer, the next one probably won't either: switch tactic. Forbidden time-sinks: extracting APKs (`unzip`/`xxd`/`strings` on `base.apk` or `classes.dex`), full `dumpsys package` / `pm dump`, recursive `find /` over the whole filesystem.
6. **Stop when done** -- once you have a verified answer or a successful write + sync, call `finish` immediately. Do not re-verify, sanity-check, or explore further.

## Android-Specific Patterns

- **Provider notifications**: after `content insert/update/delete` (or a write through a content provider), the provider emits change notifications itself -- manual broadcast is rarely needed for canonical providers.
- **File-watching apps**: when an app reads from a watched directory, the file must be in the directory and format the app expects. Search for an existing sample of the same format and copy its layout exactly.
- **Corrupt databases**: if `sqlite3` reports corruption, remove the DB and its `-wal` / `-shm` siblings before recreating the schema.

## Rules

1. For information tasks, `--description` in `finish` IS your answer. Give ONLY what was asked -- no commentary, no prefixes, exact format.
2. ALWAYS call `finish` when done, with a meaningful `--description`.

## Task

%INSTRUCTION%

## Last command output

%COMMAND_OUTPUT%
\end{Verbatim}

\makeatletter
\@ifpackageloaded{acl}{\twocolumn}{}
\makeatother

\subsubsection*{Anatomy of a prompt}

All six prompts share the same skeleton, organised around the four
categories of guidance described in Section~\ref{sec:experimental_setup}:

\begin{enumerate}
  \item \textbf{Four-phase interaction cycle:} discover relevant data,
        inspect existing state, act through the terminal interface, and
        verify the result. The cycle appears verbatim under the
        \emph{Approach} heading in each prompt and is preceded by three
        preflight questions (``what has been observed,'' ``what is still
        an assumption,'' ``what is the most likely failure mode'').
        Where a write touches a cached app process, the Verify phase
        concludes with a force-stop sync step so the app re-reads on
        next launch.
  \item \textbf{Prioritised hierarchy of mechanisms for modifying
        device state.} A ranked list of write paths to try in order:
        content providers first (so derived columns and change
        notifications fire), then system service commands
        (\texttt{cmd <service>}, \texttt{service call},
        \texttt{am broadcast}) for live-service effects, then direct
        SQLite or file writes under \texttt{/data/data/<pkg>/} for
        app-private state with no provider. On MobileWorld the same
        hierarchy is restated as three writing surfaces (on-device,
        container backend, network backend) so the agent picks the
        layer by where the consumer reads.
  \item \textbf{Efficiency strategies.} Two rules: batch related
        probes into a single shell call rather than one probe per turn,
        and respect a small probe budget --- if 2--3 probes have not
        surfaced the answer, switch tactic rather than continuing with
        forbidden time-sinks (extracting APKs, full \texttt{dumpsys
        package}, recursive \texttt{find /}). Together they cap
        exploration cost while keeping the agent deliberate per call.
  \item \textbf{Platform-specific patterns for file synchronisation
        and data persistence.} A short list of non-obvious Android
        behaviours: media-scanner broadcast after writes to shared
        storage, the content-provider self-notify semantics that make
        manual broadcasts unnecessary for canonical providers, and the
        recovery procedure for a corrupt SQLite file (remove the DB
        plus its WAL/SHM siblings, then recreate the schema). On
        MobileWorld this category also covers the device file-system
        layout, database discovery patterns, and the half-open
        range-query convention for UTC timestamps.
\end{enumerate}

Around these four behavioural categories each prompt also carries a
small set of structural blocks: a one-sentence agent role; a fenced
\emph{command space} listing the callable verbs and argument shapes
(the AW listing names the four core wrappers from
Appendix~\ref{app:tools} plus \texttt{adb shell} and \texttt{finish};
the MW listing adds the eight MW-only verbs); an explicit no-screen
constraint forbidding screenshots, UI hierarchy dumps,
taps/swipes/keyevents, and \texttt{adb pull} / \texttt{push} /
\texttt{root}; a shell-escaping note recommending stdin pipes to avoid
double-escaping inside \texttt{adb shell}; and a final response-format
block whose contents depend on the harness (see below).

\paragraph{No task-specific or app-specific information that could leak
to CLI agents.}
Every category above describes either (a) a generic device-control
capability (\texttt{sql}, \texttt{read-file}, \texttt{adb shell},
\texttt{content query}, \dots), (b) a generic Android platform
convention that holds for any app (\texttt{/data/data/<pkg>/},
content-provider notify semantics, half-open range queries on UTC
timestamps), or (c) a generic agent-behaviour principle (never assume,
copy task values verbatim, stop on completion). The task instruction
itself enters through a placeholder (\verb|{{ task }}|,
\verb|%INSTRUCTION%|, or the harness's post-system message) and the
prompt body never names a specific benchmark task, application
package, file path read by a verifier, or expected answer.

\subsubsection*{Across-harness differences}

Within a benchmark, the three harness prompts are \emph{almost identical}
in content: the command list, constraint, approach cycle, principles,
shell-escaping rule, Android patterns, and rules are word-for-word the
same. The differences are confined to two surface dimensions:

\begin{enumerate}
  \item \textbf{Output structure.} Each harness imposes a different
        response shape. The Claude Code SDK harness uses native tool
        calls and therefore needs no response-format block --- the agent
        invokes the shim binaries directly via the SDK's Bash tool. The
        mini-SWE-agent harness asks for a free-form \texttt{THOUGHT:} line
        followed by exactly one fenced \verb|```bash| command. The
        Terminus-2 harness asks for a single JSON object
        \texttt{\{"analysis":\ "\dots",\ "plan":\ "\dots",\ "command":\
        "\dots"\}}.
  \item \textbf{Section ordering and partitioning.} The Claude SDK prompt
        is a single block. The mini-SWE-agent prompt is split into a
        system template (role + command space + format example) and an
        instance template (constraint + approach + principles + rules,
        rendered per turn with the task interpolated) following its original design. The Terminus-2
        prompt is a single template that places \emph{Response Format}
        and \emph{Shell Escaping} immediately after the command list and
        defers the \emph{Constraint} block to after them; the AndroidWorld
        instance also moves \emph{Android Platform Knowledge} before the
        approach cycle. These reorderings are cosmetic --- no behavioural
        content is added or removed.
\end{enumerate}

\subsection{Tool specifications}
\label{app:tools}

Each benchmark exposes a fixed CLI tool suite to the agent; the suite is
identical across the three harnesses (Claude Code SDK, mini-SWE-agent,
Terminus-2). Following the framing in Section~\ref{sec:experimental_setup},
all tools are \emph{general-purpose wrappers around raw ADB shell
commands} that lift a raw shell verb into a higher-level operation:
AndroidWorld receives four tools; MobileWorld extends the suite to
twelve by adding eight more that reach the parts of the environment a
device shell alone cannot. Table~\ref{tab:cli-tools} lists every tool,
marks the benchmark(s) on which it is active, and summarises what it
does.

\begin{table*}[!t]
\centering
\small
\begin{tabular}{l c c p{0.7\linewidth}}
\toprule
\textbf{Tool} & \textbf{AW} & \textbf{MW} & \textbf{What it does} \\
\midrule
\multicolumn{4}{l}{\textit{Core file / SQL wrappers (AW shipped suite)}} \\
\midrule
\texttt{sql}             & $\checkmark$ & $\checkmark$ & Run SQL on a device SQLite DB; pipes the query via base64+stdin into \texttt{adb shell sqlite3} so quotes / unicode survive nested parses. \\
\texttt{read-file}       & $\checkmark$ & $\checkmark$ & Read a device file via base64-round-tripped \texttt{adb shell}. On MW, \texttt{.pdf} paths are additionally extracted host-side through a \texttt{pdftotext}/PyMuPDF/pypdf/strings fallback chain. \\
\texttt{write-file}      & $\checkmark$ & $\checkmark$ & Write content to a device path via base64-encoded \texttt{adb shell}; \texttt{--append} supported. \\
\texttt{find-files}      & $\checkmark$ & $\checkmark$ & \texttt{adb shell find <dir> -name <glob> -maxdepth N}, with the glob re-quoted so the device shell does not pre-expand it. \\
\midrule
\multicolumn{4}{l}{\textit{MW extensions --- Android platform IPC}} \\
\midrule
\texttt{content}         &              & $\checkmark$ & Wrap the \texttt{adb shell content} \texttt{query}/\texttt{insert}/\texttt{update}/\texttt{delete} verbs with typed \texttt{--bind} fields and standard \texttt{--projection}~/ \texttt{--where}~/ \texttt{--sort}. \\
\texttt{intent}          &              & $\checkmark$ & Wrap \texttt{adb shell am} \texttt{start}/\texttt{broadcast} with typed extras (string, int, bool, long, float, uri). \\
\midrule
\multicolumn{4}{l}{\textit{MW extensions --- structured-data parsing}} \\
\midrule
\texttt{json-read}       &              & $\checkmark$ & Parse a JSON file on the device with a jq-style path expression (\texttt{.key[0].sub}). \\
\texttt{json-write}      &              & $\checkmark$ & Validate and write structured JSON to a device path; \texttt{--merge} deep-updates a top-level dict. \\
\midrule
\multicolumn{4}{l}{\textit{MW extensions --- networked APIs}} \\
\midrule
\texttt{http}            &              & $\checkmark$ & HTTP client with namespace-aware routing: rewrites admin loopback ports to the broker URL, runs \texttt{curl} inside the publishing container for other loopback URLs, and uses \texttt{urllib} for external URLs. \\
\midrule
\multicolumn{4}{l}{\textit{MW extensions --- backend services}} \\
\midrule
\texttt{pg}              &              & $\checkmark$ & PostgreSQL client into a nested backend container; auto-discovers the container by grep on \texttt{docker ps}, then pipes SQL via stdin to \texttt{psql}. \\
\texttt{backend-exec}    &              & $\checkmark$ & Run an arbitrary shell command inside a nested backend container (same grep+stdin pattern as \texttt{pg}). \\
\texttt{service-status}  &              & $\checkmark$ & Read-only \texttt{docker ps} against the nested Docker-in-Docker daemon; authoritative list of running backend containers. \\
\bottomrule
\end{tabular}
\caption{CLI tool suite. AW = AndroidWorld, MW = MobileWorld; $\checkmark$
indicates the verb is callable by the agent on that benchmark.}
\label{tab:cli-tools}
\end{table*}

\paragraph{Core four (AW + MW).}
The four wrappers shipped with both benchmarks --- \texttt{sql},
\texttt{read-file}, \texttt{write-file}, \texttt{find-files} --- each
lift a single \texttt{adb shell} verb into a higher-level operation: SQL
queries, file I/O, and file search. The wrapper layer handles the
multi-layer shell parses, the base64 round-tripping for quotes / unicode
/ binary content, and the glob re-quoting for device-side \texttt{find}.
Anything an agent can do through these four it could in principle also
do through a sufficiently careful \texttt{adb shell} invocation; the
wrapper just removes a class of quoting bugs.

\paragraph{MW extensions (eight more tools).}
MobileWorld presents a more challenging setting with backend services,
structured data, and networked APIs, so eight additional wrappers are
exposed there:
\begin{itemize}
  \item \emph{Android platform IPC} --- \texttt{content} and
        \texttt{intent} lift the standard content-provider and
        \texttt{am start}/\texttt{broadcast} surfaces into typed
        \texttt{--bind} / \texttt{--extra} forms.
  \item \emph{Structured-data parsing} --- \texttt{json-read} and
        \texttt{json-write} add a structured-JSON read/write path on
        top of \texttt{adb shell cat} and \texttt{write-file} so the
        agent does not have to compose jq inside a nested shell.
  \item \emph{Networked APIs} --- \texttt{http} reaches MobileWorld's
        in-environment HTTP endpoints by routing requests through the
        correct network namespace, so the agent's
        \texttt{localhost:<port>} resolves to the publishing container
        rather than the runner's loopback.
  \item \emph{Backend services} --- \texttt{pg}, \texttt{backend-exec},
        and \texttt{service-status} reach into the nested
        Docker-in-Docker daemon where MobileWorld's backend services
        (Mastodon, Mattermost, $\ldots$) live; without them the agent
        cannot interact with the backend half of MobileWorld at all.
\end{itemize}
AndroidWorld has no backend services and no JSON-as-app-state contract,
so these eight verbs are not exposed there.

\paragraph{The raw surface and termination signal.}
Two additional verbs sit outside the wrapper count. The raw
\texttt{adb shell <cmd>} pass-through is always available on both
benchmarks as the underlying surface the wrappers sit on top of, used
as an escape hatch when no typed verb fits. The harness termination
signal \texttt{finish --status complete --description "<answer>"} marks
the episode terminated and records the agent's answer for IR tasks; it
is a signalling primitive rather than a device or backend operation.

\paragraph{No task- or app-specific information.}
Consistent with the system-prompt claim in
Appendix~\ref{app:system_prompts}, the tool suite contains no
task-specific or app-specific information that could leak to CLI agents.
Every verb in Table~\ref{tab:cli-tools} maps to a generic systems
primitive --- shell, SQLite, file system, glob, content provider,
intent, HTTP, JSON, \texttt{docker ps}, \texttt{psql} --- and no verb
names a benchmark concept. The wrappers treat all agent-supplied
arguments as opaque strings; they do not parse the task instruction, do
not dispatch on the task name, do not read the verifier source, and do
not return verifier internals. The same binary serves every task in the
benchmark. Hazardous verbs that could exfiltrate state into or out of
the simulated device (\texttt{adb pull}, \texttt{adb push},
\texttt{adb root}) are blocked by a benchmark-wide deny pattern.

\subsection{Oracle construction protocol}
\label{app:oracle}

This appendix documents how we built the CLI-solvable subset of each
benchmark and the ground-truth trajectories used as reference solutions
through human--LLM collaboration. The subset counts cited in
Section~\ref{sec:experimental_setup} are \textbf{88.8\% (103/116)} on
AndroidWorld and \textbf{86.3\% (101/117)} on MobileWorld; the
remaining 13 tasks on AW and 16 tasks on MW involve interactions
outside the CLI paradigm's reach --- interacting with visual content
(e.g., transcribing receipts from photos, free-hand drawing) or
capturing multimodal data (e.g., camera photos or audio recording),
plus a small number of UI-panel-only data sources on MW (Google Maps
detail panels, in-app cart totals) --- and are enumerated at the end of
this section.

\subsubsection*{(1) The two acceptance criteria}

A task is admitted into the CLI-solvable subset only if its candidate
trajectory satisfies \emph{both}:

\paragraph{Solvability.}
The candidate trajectory must pass the benchmark's own rule-based
verifier when replayed against a freshly reset container at the
canonical seed (seed~7 for AW; per-task default for MW). Solvability is
checked by resetting the container to the canonical initial state,
dispatching each step in the candidate trajectory through the
production CLI tool surface (the same wrappers the scoring agent uses;
see Appendix~\ref{app:tools}), issuing \texttt{finish}, and reading the
binary reward from the container's verifier endpoint. We re-run this
check inside a validation pipeline that dispatches every step in the
published ground-truth reference against a live broker and fails on
any non-zero return code or final score~$\neq$~1.

\paragraph{Integrity.}
Passing the verifier is necessary but not sufficient. We additionally
require:
\begin{enumerate}
  \item \emph{No verifier-internals leakage.} The trajectory must be
        producible from device state and the task goal alone, without
        reading the verifier source, its private fixtures (e.g.\
        parameter-bearing attributes on the task class), or any
        gold-answer file.
  \item \emph{No hardcoded answers.} Numeric and string answers for
        information-retrieval tasks must be \emph{computed} from data
        the verifier itself reads, not hardcoded as constants in the
        trajectory.
\end{enumerate}

\subsubsection*{(2) The oracle-agent loop}

We treat oracle construction itself as a human--LLM collaboration: a
separate \emph{oracle-agent} (Claude Code, running on the host) is
given a far richer context than the scoring agent and proposes, for
each task, a condensed trajectory that a human reviewer then validates.
Each task is attempted up to five times with verifier feedback; the
loop runs in five phases per attempt.

\paragraph{Step 1 --- Setup.}
The oracle-agent runs on the host, outside the per-task evaluation
container. It is granted read access to artifacts the scoring agent
never sees: the full body of past agent trajectories on this task
(across models and harnesses), the system prompts those agents were
run under, and the benchmark's emulator-design and tool-surface
documentation. The oracle-agent's own action space is the same CLI
tool suite from Appendix~\ref{app:tools}, exposed through a single
\emph{oracle-entrypoint} that issues actions against a freshly reset
container; everything else in the rich context is read-only research
material.

\paragraph{Step 2 --- Exploration.}
The oracle-agent first \emph{reads} rather than acts. It scans the
past trajectories for the task and labels each one: which final
verifier-reward did it achieve, which steps did the past agent take,
where did short trajectories diverge from long ones, what error
messages recurred. Successful past trajectories surface a candidate
\emph{path skeleton}; failed past trajectories surface a
\emph{failure-mode list} (wrong URI, wrong table column, missing
\texttt{am force-stop}, wrong timestamp unit, etc.).

\paragraph{Step 3 --- Proposal.}
The oracle-agent condenses the path skeleton and the failure-mode
list into the shortest CLI trajectory it believes will satisfy the
verifier --- replacing GUI-shaped detours with direct
\texttt{sql} / \texttt{content} / \texttt{intent} verbs, collapsing
verification rounds the scoring agents repeated unnecessarily, and
naming exactly the data sources the answer should be computed from.
The proposal is emitted as an ordered step list, not as free text.

\paragraph{Step 4 --- Oracle-acting.}
The oracle-agent then dispatches the proposed trajectory through the
oracle-entrypoint against a freshly reset container, end-to-end, and
records the per-step output plus the final verifier reward. This is
the first time the proposed steps actually touch the device; the
exploration and proposal phases never modify state.

\paragraph{Step 5 --- Human-in-the-loop, one reviewer per round.}
A human reviewer reads the full round output (proposal +
per-step actuation + final reward) and renders one of two verdicts:
\begin{itemize}
  \item \textbf{Invalid.} The verifier rejected the trajectory, or the
        trajectory was incomplete, or the reviewer disagrees with the
        approach. The reviewer gives \emph{general advice only} ---
        ``you skipped the force-stop'', ``the Mattermost message has to
        be sent through the API, not the DB'' --- and never reveals
        verifier internals, expected answer values, or fixture file
        names. The advice is appended to the rich-context bundle and
        the loop re-enters Step 2 with the next round.
  \item \textbf{Valid.} The verifier accepted the trajectory. The
        reviewer then runs the integrity audit from 
        verifier-internals leakage and hardcoded answers. If the trajectory clears both
        checks the round closes and the trajectory is committed to the
        ground-truth reference; if any check fails the round is
        re-opened with the violating step flagged for rewrite, or the
        task is escalated to the manual-verification stage below.
\end{itemize}

The five-phase loop iterates up to five times per task until either a
valid+clean trajectory is committed or the task is judged un-solvable
under the integrity rules and demoted to the GUI-only list.

\subsubsection*{(3) Manual verification}

Each task that exits the oracle-agent loop with a committed trajectory
is then manually verified independently by three external professionals. Each professional re-runs
the solvability check end-to-end against a fresh broker pool and,
separately, audits the integrity of every step under the rejection
signals below.

\paragraph{Reconciliation rule.}
A task is admitted into the CLI-solvable subset only if all three
professionals mark it pass on both solvability and integrity. Any single
rejection demotes the task to the GUI-only list; ties are resolved by
re-running the trajectory live with all three professionals present, and
the demotion stands if the live re-run does not resolve the dispute.

\subsubsection*{(4) Not-CLI-solvable task lists}

The 13 tasks on AW and 16 tasks on MW listed below fall outside the
CLI paradigm's reach: either they require interacting with visual
content (transcribing receipts from photos, free-hand drawing,
image-content classification) or they require capturing multimodal
data (camera photos, audio recording); MW additionally contributes a
handful of cases where the required data point is only rendered in an
in-app panel (Google Maps drive-/walk-time, business detail fields,
shopping-cart totals) with no public API path. We list both
benchmarks separately --- AndroidWorld
(Table~\ref{tab:aw_gui_only_tasks}) and MobileWorld
(Table~\ref{tab:mw_gui_only_tasks}) --- giving the task name and the
precise reason CLI access cannot satisfy the verifier.

\paragraph{AndroidWorld (AW).}
The 13 tasks excluded from the AW CLI-solvable subset are listed in
Table~\ref{tab:aw_gui_only_tasks}; all require either vision (OCR or
image content) or a UI-side capture path (camera shutter, audio
recorder, drawing canvas).

\begin{table*}[!t]
\centering
\footnotesize
\begin{tabular}{l p{0.62\linewidth}}
\toprule
\textbf{Task} & \textbf{Why CLI cannot solve it} \\
\midrule
\texttt{AudioRecorderRecordAudio}            & Requires audio capture through the recorder UI; no programmatic shutter. \\
\texttt{CameraTakePhoto}                     & Requires capturing a real photo through the camera UI. \\
\texttt{ClockTimerEntry}                     & Timer-entry UI has no programmatic equivalent for ``set but do not start''. \\
\texttt{AudioRecorderRecordAudioWithFileName}& Same as \texttt{AudioRecorderRecordAudio}, plus naming requires the recorder save dialog. \\
\texttt{BrowserDraw}                         & Drawing in a Chrome canvas with three colours; no CLI surface. \\
\texttt{BrowserMaze}                         & Navigating a maze widget in Chrome via direction buttons. \\
\texttt{SimpleDrawProCreateDrawing}          & Requires producing pixels in a drawing app and naming the export. \\
\texttt{CameraTakeVideo}                     & Camera video capture through the UI shutter. \\
\texttt{MarkorTranscribeReceipt}             & Transcribe \texttt{receipt.png} content as CSV; the bytes are only legible to OCR/vision. \\
\texttt{BrowserMultiply}                     & Click a button five times in Chrome, read displayed numbers, compute product; no CLI surface and needs OCR. \\
\texttt{ExpenseAddMultipleFromGallery}       & Transcribe expenses from \texttt{expenses.jpg}; needs OCR. \\
\texttt{MarkorTranscribeVideo}               & Transcribe per-frame text from a VLC video; needs vision. \\
\texttt{RecipeAddMultipleRecipesFromImage}   & Transcribe recipes from \texttt{recipes.jpg}; needs OCR. \\
\bottomrule
\end{tabular}
\caption{AndroidWorld --- 13 GUI-only tasks (out of 116; CLI-solvable subset $= 103$).}
\label{tab:aw_gui_only_tasks}
\end{table*}

\paragraph{MobileWorld (MW).}
The 16 tasks excluded from the MW CLI-solvable subset are listed in
Table~\ref{tab:mw_gui_only_tasks}; they include the same vision and
multimodal-capture categories as AW, plus two MW-specific patterns
that AW does not surface: (i) Google Maps panel-only labels
(business phone, neighbour name, walk-/drive-time) that have no
public API path, and (ii) e-commerce cart fields (item list, cart
total) that are only rendered in the in-app cart panel. We treat a
task as GUI-only if either CLI execution cannot in principle satisfy
the verifier, or the verifier nominally accepts a CLI path but the
correct answer is determined by image content that the verifier does
not enforce (e.g.\ a filename-suffix check that happens to align with
the visual content).

\begin{table*}[!t]
\centering
\footnotesize
\begin{tabular}{l p{0.58\linewidth}}
\toprule
\textbf{Task} & \textbf{Why CLI cannot solve it} \\
\midrule
\texttt{MattermostVisualInstructionResponseTask}   & Whiteboard photos in a Mattermost channel: contacts and shift times only legible via OCR. \\
\texttt{CheckConferenceLocationTask}               & Walk-time between two Maps pins is only obtainable from the Google Maps UI; no public API path. \\
\texttt{DownloadSendReceiptTask}                   & Receipt total is only readable by inspecting the receipt image; needs OCR. \\
\texttt{TakeSelfieTask}                            & Requires the camera UI to capture an actual photo; no CLI shutter path. \\
\texttt{GoogleMapsAlibabaPhoneContactTask}         & Business phone is rendered only in the Maps detail panel; no public API. \\
\texttt{GoogleMapsAlibabaSouthNeighborTask}        & Neighbour label is only rendered in the Maps detail panel. \\
\texttt{TextArrivalTimeTask}                       & Drive-time between two cities requires the Google Maps UI; no CLI path. \\
\texttt{ChangeWallpaperTask}                       & Needs vision to identify the sunflower photo among four pushed images. \\
\texttt{SharePhotosTask}                           & Needs vision to identify which gallery images are flowers. \\
\texttt{MastodonManageMultiListTask}               & The Mastodon list-management workflow is exposed only through the in-app UI. \\
\texttt{MattermostSendFileTask}                    & Needs vision to identify the birthday-cake image to upload. \\
\texttt{PhotoManagementTask}                       & Needs vision to identify food photos; the verifier's filename-suffix check only happens to align with the visual content. \\
\texttt{MastodonChangeHeaderTask}                  & Needs vision to identify the tiger photo; the file \texttt{tiger.jpg} only happens to be present. \\
\texttt{MastodonShareLocationTask}                 & Sharing a location in a Mastodon post requires the in-app location picker; no public API path. \\
\texttt{CartManagementTask}                        & Modifying items in the in-app shopping cart requires the cart UI; no backend API exposed for the operations the verifier checks. \\
\texttt{CheckCartPriceTask}                        & Cart total is rendered only in the in-app cart panel; no public API path. \\
\bottomrule
\end{tabular}
\caption{MobileWorld --- 16 GUI-only tasks (out of 117; CLI-solvable subset $= 101$).}
\label{tab:mw_gui_only_tasks}
\end{table*}

\section{CLI-Advantage Suite --- Construction and Extended Results}
\label{app:cli_advantage}

This appendix documents the CLI-Advantage Suite introduced in Section~\ref{sec:cli-advantage}: the full per-category task list (\S\ref{app:cli_advantage_tasks}), the seed-driven randomization mechanism and the two rule-based verifier types (\S\ref{app:cli_advantage_rewards}), the oracle-construction protocol (\S\ref{app:cli_advantage_oracle}), the realism rubric and three-annotator agreement (\S\ref{app:cli_advantage_realism}), and the semantic similarity to AndroidWorld (\S\ref{app:cli_advantage_similarity}). 

\subsection{Full task list}
\label{app:cli_advantage_tasks}

We list all 45 tasks by category, showing the natural-language instruction template each task ships with. Per task, the seed varies the on-device fixture (data values, row counts, file sets, granted permissions, etc.) and, where present, the angle-bracket placeholder fillers (e.g., \texttt{<keyword>}). Section~\ref{app:cli_advantage_rewards} describes the randomization contract and what specifically varies seed-to-seed for each category.

\paragraph{Bulk operations (10 tasks).}
\begin{itemize}\setlength\itemsep{1pt}
\item Delete all \texttt{.tmp} files in the Downloads folder.
\item Rename all files starting with \texttt{Screenshot\_} in the Pictures folder to the format \texttt{YYYYMMDD\_HHMMSS.png} based on each file's modification time.
\item Move all files larger than 50\,MB in the Download folder to the Archive folder; create the Archive folder if it does not exist.
\item Append the footer ``\texttt{-{}-{}-\textbackslash nGenerated by AutoBot}'' to every \texttt{.md} file in the \texttt{Notes} folder in Markor.
\item Change all \emph{Food} entries in Pro Expense to \emph{Entertainment}.
\item Change all overdue tasks in the Tasks app to \emph{High} priority.
\item Delete all events in Simple Calendar Pro whose title contains the word \emph{test} (case-insensitive).
\item Delete duplicate calendar events in Simple Calendar Pro (same title and same start time), keeping only one copy of each.
\item Delete all \texttt{.apk} files in the Downloads folder.
\item Delete all expenses in Pro Expense that are less than \$1.00 (less than 100 cents).
\end{itemize}

\paragraph{Multi-condition filtering (10 tasks).}
\begin{itemize}\setlength\itemsep{1pt}
\item List all contacts that have a birthday set but no phone number; output the contact names.
\item List all contacts that have a phone number but no family name (first name only).
\item List all Joplin notes that contain \texttt{<keyword\_a>} but do NOT contain \texttt{<keyword\_b>}; output the note titles.
\item List all songs in Retro Music by artist \texttt{<artist>} that are longer than 4 minutes; output the song titles.
\item Do all my calendar events this month have reminders set? If any are missing reminders, name them.
\item List all SMS messages in the inbox that contain a URL (\texttt{http://} or \texttt{https://}); output the sender phone numbers.
\item Are there any SMS messages in my inbox from numbers not in my contacts? If so, how many?
\item List all zero-byte (empty) files in the Downloads folder; output the filenames.
\item Which of my expenses are above my average spending? List them.
\item What do I have planned on weekends this month? List all calendar events that fall on a Saturday or Sunday.
\end{itemize}

\paragraph{Aggregation and Top-$K$ (10 tasks).}
\begin{itemize}\setlength\itemsep{1pt}
\item Which 5 notes in Markor were modified most recently in the last 7 days? List the filenames.
\item Which 3 phone numbers have the most SMS messages in your inbox? List their phone numbers.
\item List all groups of contacts that share the same phone number; output each group with the shared phone number and the contact names.
\item What are the top 3 expense categories by total amount this month in Pro Expense? Output the category names.
\item How many suspected duplicate expenses are there in Pro Expense (same date, amount, and category)? Output the number of extras beyond the first of each group.
\item What are the 5 highest-amount expenses in Pro Expense? List the expense names and amounts.
\item What is the total distance of all activities this week in OpenTracks, and which activity covered the longest distance?
\item Which activity in OpenTracks had the highest average speed? Output the activity name and its average speed.
\item What are the 5 longest songs in Retro Music by duration? List their titles.
\item What are the 5 largest files in the Downloads folder? List their filenames and sizes.
\end{itemize}

\paragraph{Cross-app workflows (9 tasks).}
\begin{itemize}\setlength\itemsep{1pt}
\item Calculate this month's total expenses in Pro Expense, write the total (in dollars) to a Markor note \texttt{monthly\_summary.md}, and create a Simple Calendar Pro event titled \texttt{Monthly Expense: \$<total>} on the last day of this month.
\item Which of my contacts texted me but I haven't replied to?
\item Extract all phone numbers mentioned in Markor notes, then list the ones that are NOT in your contacts; output the numbers.
\item Find all events in Simple Calendar Pro whose title contains \texttt{<keyword>}; create a Markor note \texttt{<keyword>\_events.md} listing the event titles and dates, one per line.
\item Export all contacts (name and phone number) to a Markor note \texttt{contacts\_export.md}, one contact per line in the format \texttt{Name: phone\_number}.
\item Did anyone text me during my meetings yesterday? List the senders of all SMS received while a calendar event was active yesterday.
\item Find all SMS messages in the inbox that contain the word \emph{urgent} and create a task in the Tasks app for each one, using the SMS body as the task title.
\item Summarize this week's OpenTracks activities (total distance in km and number of activities) and save it as a Markor note \texttt{weekly\_stats}.
\item Export all Joplin notes whose title contains \emph{meeting} to Markor as separate notes.
\end{itemize}

\paragraph{Hidden device state (6 tasks).}
\begin{itemize}\setlength\itemsep{1pt}
\item List all apps (by package or app name) that have been granted location permission on this device. If none, output \emph{None}.
\item List all apps that have been granted Camera permission on this device. If none, output \emph{None}.
\item What is my phone's current temperature? Is it overheating?
\item What are the 3 most recently installed apps on the device? Output their package names.
\item How long has the device been running since the last reboot? Output the uptime in hours and minutes.
\item Which apps have recently accessed my location in the background?
\end{itemize}

\subsection{Randomization mechanism and rule-based reward}
\label{app:cli_advantage_rewards}

Every task is implemented as a Python class against the AndroidWorld \texttt{TaskEval} interface (the same contract used by the 116 stock AndroidWorld tasks). Each task class provides two methods:
\begin{itemize}\setlength\itemsep{1pt}
\item \texttt{initialize\_task(env)} seeds all on-device fixtures from \texttt{random.Random(self.seed)} and binds any placeholder fillers in the goal text. The same \texttt{(task\_id, seed)} reproduces the same fixture and the same instantiated goal sentence.
\item \texttt{is\_successful(env)} returns a binary reward ($0.0$ or $1.0$) by reading device state and/or the agent's reported answer; the grader is deterministic, never invokes an LLM-judge, and never makes a network call.
\end{itemize}
We publish three evaluation seeds $\{7, 30, 1234\}$. What is stable across all three seeds: task ID, class name, category, goal-sentence template, and the verifier code itself; agents must solve the task class, not memorize values. What varies seed-to-seed differs by category, and is summarized below.

\paragraph{What the seed varies, by category.} For \emph{Bulk} tasks, the fixture is the set of files in a target directory or the set of rows in a target table; the seed picks which items exist, their attributes (sizes, timestamps, category and priority labels, expense amounts), and the size of the subset that must be acted on (e.g., how many files exceed the size threshold, how many expenses fall below \$1.00). For \emph{Filter} and \emph{Aggregation} tasks, the seed populates the same kinds of fixtures and additionally computes the expected answer set on the host (the names that satisfy the predicate; the top-$K$; the duplicate groups), storing them as \texttt{self.\_expected}. For \emph{Cross-app} tasks, the seed sets up both the source-app fixture and the post-action target-app expectation. For \emph{Hidden device state} tasks, the seed configures device-level facts: which packages appear installed, which permissions are granted, the apparent uptime and temperature, and recent-install timestamps. Where the goal text contains angle-bracket tokens (\texttt{<keyword>}, \texttt{<artist>}, etc.), the seed additionally picks the substitution value before the goal is shown to the agent. Two end-to-end worked examples follow.

\paragraph{Two verifier function types.}
Every task bottoms out into one of two verifier shapes, picked by category. CrossApp tasks that both mutate device state and report a summary chain the two checks; we describe only the primitives below.

\begin{enumerate}\setlength\itemsep{6pt}\setlength\parskip{2pt}
\item \textbf{State-check.} The grader re-reads device state after the agent finishes and compares it to the post-action expectation. The agent's natural-language output is ignored.
\begin{small}
\begin{verbatim}
def is_successful(self, env) -> float:
    rows = adb_utils.issue_generic_request(
        "shell", "sqlite3 ... 
        'SELECT ... FROM ... WHERE ...'",
        env)
    return 1.0 if matches_expected(rows) 
    else 0.0
\end{verbatim}
\end{small}

\item \textbf{Cache-match.} The grader inspects the agent's \texttt{FINISH(content=...)} payload. Every expected substring (computed in \texttt{initialize\_task} from the seed) must appear; substring containment, case-sensitive, no whitespace normalization.
\begin{small}
\begin{verbatim}
def is_successful(self, env) -> float:
    return 1.0 if all(
        s in env.interaction_cache
        for s in self._expected
    ) else 0.0
\end{verbatim}
\end{small}
\end{enumerate}

\paragraph{Example 1}
\emph{Goal:} ``Delete all expenses in Pro Expense that are less than \$1.00 (less than 100 cents).''

In \texttt{initialize\_task}, \texttt{random.Random(seed)} populates the Pro Expense SQLite database \texttt{accounting.db} with a base set of $\sim$20 expenses uniformly sampled from $[100,10{,}000]$ cents, plus a seed-derived count $k_{\mathrm{small}} \in \{3,4,5\}$ of expenses with amounts in $[1,99]$ cents. The goal text contains no placeholders; the fixture itself is what the seed varies. After the agent finishes, \texttt{is\_successful} runs \texttt{SELECT COUNT(*) FROM expense WHERE amount < 100} on \texttt{accounting.db} and returns $1.0$ iff the count is zero.

\paragraph{Example 2}
\emph{Goal:} ``Which of my expenses are above my average spending? List them.''

\texttt{initialize\_task} seeds the same expense-style fixture, then computes the expected answer set $E = \{\mathrm{name} : \mathrm{amount}(\mathrm{name}) > \overline{\mathrm{amount}}\}$ on the host and stores it in \texttt{self.\_expected}. The agent must produce a \texttt{FINISH} payload whose text contains every name in $E$ as a substring; partial answers fail. As with the state-check example, $E$ is recomputed on each seed.

\subsection{Oracle construction}
\label{app:cli_advantage_oracle}

Each of the 45 tasks ships with a hand-verified golden-path oracle CLI command sequence that the rule-based verifier accepts. Oracle construction follows the same human--LLM collaborative protocol used for AndroidWorld and MobileWorld and therefore we do not repeat it here. See Appendix~\ref{app:oracle} for the full protocol.

\subsection{Realism rubric and annotator protocol}
\label{app:cli_advantage_realism}

We retain a task only if its goal text reads like a request a real-world mobile user would actually make. Each of the three external mobile-product users  independently rates every candidate task on a three-level rubric.

\paragraph{Rubric anchors.}
\begin{itemize}\setlength\itemsep{2pt}
\item \textbf{Highly realistic} --- phrasing matches how a non-technical user would describe the request in casual conversation. Examples: ``\textit{Delete every \texttt{.apk} file in Downloads}''; ``\textit{Top 3 expense categories this month}''; ``\textit{Which apps have been granted location permission?}''.
\item \textbf{Stretched} --- the intent is plausible, but the surface form leaks engineering vocabulary or assumes more precision than a typical user would volunteer. Examples (paraphrased from excluded candidates): ``\textit{Which note has the highest character count?}''; ``\textit{Total download size in bytes}''.
\item \textbf{Contrived} --- the task uses power-user, developer, or admin framing, or assumes the user already knows the implementation surface. Examples (excluded): ``\textit{Signal strength in dBm}''; ``\textit{Verify all expenses are categorized}''; ``\textit{SMS database storage size}''.
\end{itemize}

\paragraph{Annotator instructions.}
Each annotator received the rubric anchors, a sheet with one representative quote per level, and the 56 candidate goal texts in random order. For every task, the annotator recorded a single label and a one-sentence justification. We did not show annotators the category, task ID, or implementation---only the surface goal text. The final inclusion rule is majority \emph{highly realistic} across the three annotators; ties and minority-realistic tasks were discussed in a single reconciliation pass and dropped if consensus could not be reached.

\paragraph{Agreement.}
We report Cohen's $\kappa$ averaged over the three annotator pairs (Table~\ref{tab:cli_advantage_kappa}), consistent with the headline value reported in Section~\ref{sec:cli-advantage}. Overall pairwise $\kappa = 0.91$; per-category figures are tightest on Bulk and Hidden State (the rubric levels are most distinguishable there) and loosest on Aggregation, where the line between ``highly realistic'' and ``stretched'' turns on whether the user would name the aggregation explicitly. The final yield is \textbf{45 of 56 candidates} retained.

\begin{table}[!t]
\centering
\small
\begin{tabular}{lcc}
\toprule
\textbf{Category} & \textbf{Pairwise $\kappa$} & \textbf{Retained / Candidates} \\
\midrule
Bulk              & 0.95 & 10 / 11 \\
Filter            & 0.91 & 10 / 12 \\
Aggregation       & 0.83 & 10 / 13 \\
Cross-app         & 0.90 & 9 / 10 \\
Hidden state      & 0.95 & 6 / 10 \\
\midrule
\textbf{Overall}  & \textbf{0.91} & \textbf{45 / 56} \\
\bottomrule
\end{tabular}
\caption{Three-annotator agreement on the realism rubric, reported as the mean Cohen's $\kappa$ over the three annotator pairs. The overall value is the headline figure cited in Section~\ref{sec:cli-advantage}.}
\label{tab:cli_advantage_kappa}
\end{table}

\subsection{Similarity to AndroidWorld}
\label{app:cli_advantage_similarity}

\begin{figure}[t]
\centering
\includegraphics[width=\linewidth]{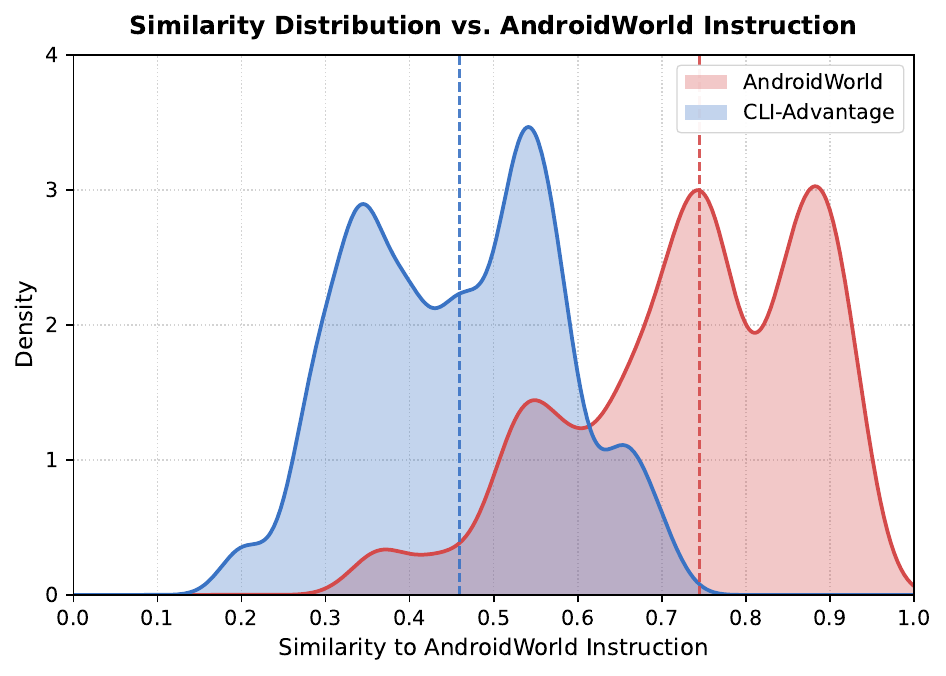}
\caption{Semantic similarity between CLI-Advantage and AndroidWorld instructions. AndroidWorld is shown as a within-suite reference (each task's similarity to its nearest sibling), illustrating the benchmark's task-family regime. CLI-Advantage instructions lie distinctly to its left, with $0\%$ exceeding similarity $0.7$.}
\label{fig:tier4_aw_similarity_density}
\end{figure}

This subsection documents the methodology and density plot (Figure~\ref{fig:tier4_aw_similarity_density}) for the additivity check referenced in Section~\ref{sec:cli-advantage}. Each of the $45$ CLI-Advantage instructions and each of the $116$ AndroidWorld test instructions is embedded with \texttt{text-embedding-3-large}; goal-text fillers are instantiated using seed $7$ before embedding so the comparison runs over concrete strings, not templates. We then form two similarity distributions, both reported as kernel density estimates with bandwidth chosen by Scott's rule:

\begin{itemize}
\item \textbf{CLI-Advantage to AndroidWorld} (blue). For each CLI-Advantage instruction, the cosine similarity to its nearest neighbour in AndroidWorld. $45$ samples, median $\approx 0.45$.
\item \textbf{AndroidWorld within-suite reference} (red). For each AndroidWorld instruction, the cosine similarity to its nearest sibling AndroidWorld instruction (i.e., $k$-NN with $k\!=\!1$, excluding self). $116$ samples, median $\approx 0.75$. This anchors what ``in-family'' looks like under the chosen embedding and metric.
\end{itemize}

The blue distribution is contained in $[0, 0.7]$ --- $0\%$ of CLI-Advantage instructions are as close to an AndroidWorld instruction as AndroidWorld instructions typically are to each other. The suite is additive in the embedding sense, not a redistribution of existing AndroidWorld coverage.

\section{Extended Main Results and Tool Ablation}
\label{app:main_full}

This appendix expands the main-paper results along two axes: per-seed and leaderboard-extended versions of Table~\ref{tab:main_results} on both AndroidWorld and MobileWorld (\S\ref{app:main_full_extended}), and paired qualitative examples illustrating where the four CLI tools change agent behaviour (\S\ref{app:codex_escaping}).

\subsection{Extended results on AndroidWorld and MobileWorld}
\label{app:main_full_extended}

Table~\ref{tab:main_results_full} extends main-paper Table~\ref{tab:main_results} with the full set of leaderboard-reported methods we are aware of on AndroidWorld (AW) and MobileWorld (MW). AW leaderboard numbers are sourced from the public GUI-Owl/MAI-UI benchmark comparison page; MW leaderboard numbers are sourced from the official MobileWorld \texttt{leaderboard.json} snapshot at \url{https://tongyi-mai.github.io/MobileWorld/}, accessed 2026-05-24 (31 entries). Several methods are reported under slightly different display names across the two leaderboards (e.g.\ \texttt{Qwen3-VL-32B-Instruct} on AW vs.\ \texttt{Qwen3-VL-32B} on MW, \texttt{MAI-UI-8b} vs.\ \texttt{MAI-UI-8B}); we collapse these to a single row when the upstream identifier and release date match. \textbf{SR-CLI} and \textbf{Avg Step} columns are only populated for our reproduced GUI baselines and our CLI agents (mean (std) across three seeds, $\{7, 30, 1234\}$); leaderboard-only rows release neither per-task verifier outputs nor step counts, so those cells are marked ``--''. ``$-$'' in the SR columns means the method does not currently appear on that benchmark's leaderboard.

\textbf{Reproduction sanity.} The three reproduced GUI baselines fall within $\sim$$1.5$~pp of the corresponding leaderboard rows: GUI-Owl-1.5-32B (AW $69.3$ reproduced vs.\ $69.8$ leaderboard), MAI-UI-8B (AW $68.1$ vs.\ $70.7$), and Qwen3-VL-32B (AW $57.8$ vs.\ $57.3$). The MW gaps are similarly small for GUI-Owl-1.5-32B (reproduced $43.2$ vs.\ leaderboard $43.9$) and MAI-UI-8B ($26.3$ vs.\ $27.5$); Qwen3-VL-32B reproduces above its leaderboard MW number ($13.3$ vs.\ $11.9$), which we attribute to seed variance ($4.9$~pp std on AW, $0.7$~pp std on MW).

\textbf{Cross-benchmark difficulty drop.} Every leaderboard method that reports both AW and MW drops substantially on MW --- e.g.\ MAI-UI-235B-A22B from $76.7$ to $39.7$, GUI-Owl-1.5-32B-Instruct from $69.8$ to $43.9$, Qwen3-VL-235B-A22B from $63.7$ to $12.8$ --- consistent with MW's longer-horizon, cross-application, backend-driven task design (Section~\ref{sec:experimental_setup}; $27.8$ vs.\ $14.3$ average steps; $62.2$\% vs.\ $9.5$\% multi-app). The single highest leaderboard SR on MW ($63.2$, Seed-2.0-Pro) and the strongest leaderboard end-to-end LLM (Claude-Opus-4.7, $56.4$) both exceed our Claude Code + Opus-4.7 CLI configuration ($51.9$); these leaderboard rows are reported under MW's native screenshot-based protocol and a different harness, so the comparison is across rather than within paradigm.

\begin{table*}[!t]
\centering
\scriptsize
\renewcommand{\arraystretch}{1.05}
\setlength{\tabcolsep}{4pt}
\begin{tabular}{llcccccc}
\toprule
& & \multicolumn{3}{c}{\textbf{AndroidWorld}} & \multicolumn{3}{c}{\textbf{MobileWorld}} \\
\cmidrule(lr){3-5} \cmidrule(lr){6-8}
\textbf{Method} & \textbf{Model API / Notes} & \textbf{SR$\uparrow$} & \textbf{SR-CLI$\uparrow$} & \textbf{Avg Step$\downarrow$} & \textbf{SR$\uparrow$} & \textbf{SR-CLI$\uparrow$} & \textbf{Avg Step$\downarrow$} \\
\midrule
\rowcolor{gray!20}\multicolumn{8}{c}{\textit{Leaderboard-reported --- General-purpose LLMs (no specialized GUI training)}} \\
Seed-1.8                      & ByteDance, end-to-end & 70.7 & -- & -- & 52.1 & -- & -- \\
Seed-2.0-Pro                  & ByteDance, end-to-end & $-$  & -- & -- & 63.2 & -- & -- \\
SeedVL-1.5                    & ByteDance, end-to-end & 62.1 & -- & -- & $-$  & -- & -- \\
Claude-Sonnet-4.5             & Anthropic, end-to-end & 56.0 & -- & -- & 47.8 & -- & -- \\
Claude-Opus-4.6               & Anthropic, end-to-end & $-$  & -- & -- & 44.5 & -- & -- \\
Claude-Opus-4.7               & Anthropic, end-to-end & $-$  & -- & -- & 56.4 & -- & -- \\
Gemini-2.5-Pro                & Google, end-to-end    & 69.7 & -- & -- & $-$  & -- & -- \\
Gemini-3-Pro                  & Google, end-to-end    & $-$  & -- & -- & 51.3 & -- & -- \\
Gemini-3.1-Pro-Preview        & Google, end-to-end    & $-$  & -- & -- & 58.1 & -- & -- \\
GPT-5.5                       & OpenAI, end-to-end    & $-$  & -- & -- & 62.4 & -- & -- \\
OpenAI CUA o3                 & OpenAI, end-to-end    & 31.3 & -- & -- & $-$  & -- & -- \\
Kimi-K2.5                     & Moonshot, end-to-end  & $-$  & -- & -- & 49.6 & -- & -- \\
Kimi-K2.6                     & Moonshot, end-to-end  & $-$  & -- & -- & 55.6 & -- & -- \\
Qwen3.5-397B-A17B             & Alibaba, end-to-end   & $-$  & -- & -- & 42.7 & -- & -- \\
Qwen3-VL-8B-Instruct          & Alibaba, end-to-end   & 47.6 & -- & -- &  9.4 & -- & -- \\
Qwen3-VL-32B-Instruct         & Alibaba, end-to-end   & 57.3 & -- & -- & 11.9 & -- & -- \\
Qwen3-VL-235B-A22B-Instruct   & Alibaba, end-to-end   & 63.7 & -- & -- & 12.8 & -- & -- \\
\midrule
\rowcolor{gray!20}\multicolumn{8}{c}{\textit{Leaderboard-reported --- GUI specialists (mobile post-trained)}} \\
UI-Venus-1.5-30B-A3B          & Ant Group        & 77.6 & -- & -- & 17.1 & -- & -- \\
UI-Venus-72B                  & Ant Group        & $-$  & -- & -- & 16.4 & -- & -- \\
UI-Venus-7B                   & Ant Group        & $-$  & -- & -- &  8.5 & -- & -- \\
MAI-UI-8B                     & Alibaba          & 70.7 & -- & -- & 27.5 & -- & -- \\
MAI-UI-32B                    & Alibaba          & 73.3 & -- & -- & 36.2 & -- & -- \\
MAI-UI-235B-A22B              & Alibaba          & 76.7 & -- & -- & 39.7 & -- & -- \\
UI-TARS-72B-DPO               & ByteDance        & 46.6 & -- & -- & $-$  & -- & -- \\
UI-TARS-1.5                   & ByteDance        & 64.2 & -- & -- & $-$  & -- & -- \\
UI-TARS-2                     & ByteDance        & 73.3 & -- & -- & $-$  & -- & -- \\
Doubao-1.5-UI-TARS            & ByteDance        & $-$  & -- & -- & 26.3 & -- & -- \\
Step-GUI-8B                   & StepFun          & 67.7 & -- & -- & 16.1 & -- & -- \\
GELab-Zero-4B                 & StepFun          & 63.9 & -- & -- & 16.1 & -- & -- \\
GUI-Owl-7B                    & Alibaba          & 66.4 & -- & -- &  7.7 & -- & -- \\
GUI-Owl-32B                   & Alibaba          & $-$  & -- & -- &  8.5 & -- & -- \\
GUI-Owl-1.5-2B-Instruct       & Alibaba          & 67.9 & -- & -- & 32.2 & -- & -- \\
GUI-Owl-1.5-4B-Instruct       & Alibaba          & 69.8 & -- & -- & $-$  & -- & -- \\
GUI-Owl-1.5-8B-Instruct       & Alibaba          & 69.0 & -- & -- & 38.2 & -- & -- \\
GUI-Owl-1.5-8B-Thinking       & Alibaba          & 71.6 & -- & -- & $-$  & -- & -- \\
GUI-Owl-1.5-32B-Instruct      & Alibaba          & 69.8 & -- & -- & 43.9 & -- & -- \\
GUI-Owl-1.5-32B-Thinking      & Alibaba          & 69.8 & -- & -- & $-$  & -- & -- \\
\midrule
\rowcolor{gray!20}\multicolumn{8}{c}{\textit{Leaderboard-reported --- Agentic frameworks (planner + grounder composition)}} \\
GPT-5 $+$ UI-Ins-7B           & OpenAI $+$ Ant Group  & $-$  & -- & -- & 54.0 & -- & -- \\
Gemini-3-Pro $+$ UI-Ins-7B    & Google $+$ Ant Group  & $-$  & -- & -- & 55.6 & -- & -- \\
Claude-4.5-Sonnet $+$ UI-Ins-7B & Anthropic $+$ Ant Group & $-$ & -- & -- & 47.8 & -- & -- \\
\midrule
\rowcolor{gray!20}\multicolumn{8}{c}{\textit{Reproduced GUI baselines (our infrastructure)}} \\
GUI-Owl-1.5-32B   & --                & 69.3 (0.5) & 71.6 (1.5) & 15.5 (2.5) & 43.2 (2.6) & 38.1 (2.0) & 34.7 (0.7) \\
MAI-UI-8B         & --                & 68.1 (0.6) & 69.3 (0.3) & 18.4 (1.2) & 26.3 (1.3) & 26.0 (2.0) & 39.7 (0.3) \\
Qwen3-VL-32B      & --                & 57.8 (4.9) & 62.4 (5.6) & 17.4 (1.8) & 13.3 (0.7) & 11.7 (0.3) & 25.6 (0.6) \\
\midrule
\rowcolor{gray!20}\multicolumn{8}{c}{\textit{CLI agents (ours)}} \\
Claude Code        & Claude Opus 4.7   & 71.8 (1.8)  & 80.9 (2.0) & 15.2 (0.6) & 51.9 (1.0) & 60.1 (1.1) & 30.2 (0.6) \\
mini-swe-agent        & Claude Sonnet 4.6 & 70.7 (2.3)  & 79.6 (2.6) & 12.1 (0.3) & 43.6 (2.6) & 50.5 (3.0) & 33.5 (0.4) \\
Terminus-2            & GPT-5.3 Codex     & 63.2 (0.5)  & 71.2 (0.6) &  7.1 (0.1) & 36.2 (1.0) & 41.9 (1.1) & 24.5 (1.9) \\
Terminus-2            & MiniMax M2.7      & 55.2 (3.8)  & 62.1 (4.2) & 14.5 (0.5) & 16.8 (1.3) & 19.5 (1.5)) & 32.9 (0.8) \\
\midrule
Oracle (CLI ceiling) & --             & 88.8       & 100.0      &  3.7       & 86.3       & 100.0      &  5.6 \\
\bottomrule
\end{tabular}
\caption{\textbf{Full leaderboard-extended version of Table~\ref{tab:main_results}.} SR is success rate (\%); SR-CLI restricts the denominator to the CLI-solvable subset of each benchmark (103/116 on AW; 101/117 on MW; see Appendix~\ref{app:oracle}); Avg Step is mean agent steps per task. Reproduced and CLI rows report mean (std) across three seeds, $\{7, 30, 1234\}$. Leaderboard rows are taken from the AndroidWorld GUI-Owl/MAI-UI comparison page and the MobileWorld official \texttt{leaderboard.json} snapshot (accessed 2026-05-24) and report a single number per benchmark per the conventions of those leaderboards; SR-CLI and Avg Step are unavailable for these rows because per-task and per-step traces are not released. ``--'' marks unavailable metrics; ``$-$'' marks methods absent from a given leaderboard. Methods reported under slightly different display names across the two leaderboards (e.g.\ \texttt{MAI-UI-8b}/\texttt{MAI-UI-8B}, \texttt{Qwen3-VL-32B-Instruct}/\texttt{Qwen3-VL-32B}) are collapsed to a single row.}
\label{tab:main_results_full}
\end{table*}

\subsection{Tool ablation examples}
\label{app:codex_escaping}

We provide two paired Codex trajectories that pin all variables except the toolset (same harness, same model, same seed) and explain the two metric shifts in Table~\ref{tab:aw_tool_ablation}. Table~\ref{tab:case_codex_082} (Task~082) is the SR mechanism: \texttt{bash-only} loses the trajectory to nested-quote and heredoc errors that prevent it from reading the schema, and ultimately writes \texttt{NULL} into fields the verifier expects to be \texttt{''}; \texttt{+tools} sidesteps the same quoting failures with two \texttt{sql} calls, observes the existing-row format, and inserts correctly. Table~\ref{tab:case_codex_052} (Task~052) is the cost mechanism: both runs reach reward~$1$, but \texttt{bash-only} burns ten steps recovering from a single apostrophe in the note body, while \texttt{+tools} writes the file in one \texttt{write-file} call (the wrapper base64-encodes the body internally). Both cases reproduce on a second harness; full traces are in the released artifact.

\begin{table*}[!t]
\centering
\begin{minipage}[t]{0.48\linewidth}
\centering
\textbf{\texttt{bash-only}} \\[2pt]
\textit{17 steps + finish, 221{,}689 input tokens, \textcolor{red!70!black}{reward 0}}
\smallskip

\footnotesize
\setlength{\tabcolsep}{3pt}
\renewcommand{\arraystretch}{1.15}
\begin{tabular}{@{}r@{\hspace{0.4em}}p{\dimexpr\linewidth-2em\relax}@{}}
\toprule
\# & Action $\to$ Result \\
\midrule
1  & \texttt{pm list packages | grep -Ei 'markor|broccoli'; find ... -iname recipes.txt} $\to$ both packages + \texttt{/sdcard/Documents/Markor/recipes.txt} \\
2  & \texttt{cat recipes.txt; run-as broccoli ...} $\to$ recipes read; \texttt{run-as} blocked (not debuggable) \\
3  & probe external dir + \texttt{cmd package resolve-content-provider} (4 guesses) $\to$ ext dir empty; \texttt{Unknown command} \\
4  & list ext files + \texttt{awk -F'|' 'NR==1 || \$4=="4 hrs"'} $\to$ empty; 3 four-hr rows found \\
5  & probe 7 candidate \texttt{content://com.flauschcode.\allowbreak broccoli*} URIs $\to$ 6 \texttt{Could not find provider}, 1 \texttt{No result found} \\
6  & \texttt{find} Broccoli-named files $\to$ only \texttt{recipes.txt} \\
7  & \texttt{logcat -c; force-stop; monkey; logcat -d} $\to$ discovers DB at \texttt{/data/.../broccoli/databases/broccoli} from \texttt{FileUtils: chmod} log \\
8  & \texttt{cmd package query-activities -a SEND -t text/plain} $\to$ matches \texttt{CreateAndEditRecipeActivity} \\
9  & \texttt{am start ... SEND text/plain ... --es TEXT 'Title|desc|...'} $\to$ ``supplied URL is malformed'' (SEND handler scrapes URLs) \\
10 & enumerate $7{\times}7$ action/MIME combos via nested \texttt{"\$out"} $\to$ \textcolor{red!70!black}{\texttt{/system/bin/sh: no closing quote}} \\
11 & retry intent enumeration with cleaner quoting $\to$ only SEND text/plain matches \\
12 & direct \texttt{sqlite3 broccoli '.tables'} $\to$ tables listed \\
13 & \texttt{adb shell "sqlite3 broccoli \textbackslash".schema recipes; .schema categories...\textbackslash""} $\to$ \textcolor{red!70!black}{\texttt{Usage: .schema ?-{}-indent? ?LIKE-PATTERN?}} (nested-quoted dot-cmd) \\
14 & heredoc \texttt{cat <{}<'SQL' | sqlite3 broccoli} $\to$ schema returned; sample queries fail with \texttt{no such column: id} \\
15 & heredoc with \texttt{json\_each(readfile(...))} + nested \texttt{substr}/\texttt{instr} parser $\to$ \textcolor{red!70!black}{\texttt{Error: near "(": syntax error}}; \texttt{json\_each} not built \\
16 & \texttt{awk -F'|' '... printf "INSERT...VALUES(...NULL...NULL...);" ...' | sqlite3} $\to$ \textcolor{red!70!black}{\texttt{/system/bin/sh: no closing quote}} \\
17 & heredoc \texttt{.import recipes.txt \_import\_recipes; INSERT INTO recipes SELECT title, NULL, ..., NULL, ..., 0 FROM \_import\_recipes WHERE preparationTime='4 hrs'} $\to$ 3 rows inserted with \textcolor{red!70!black}{\texttt{imageName=NULL, source=NULL}} \\
18 & \texttt{finish "Inserted 3 recipes ..."} $\to$ \textcolor{red!70!black}{\textbf{Reward 0}} --- \texttt{imageName}/\texttt{source} should be \texttt{''}, not \texttt{NULL} \\
\bottomrule
\end{tabular}
\end{minipage}
\hfill
\begin{minipage}[t]{0.48\linewidth}
\centering
\textbf{\texttt{+tools}} \\[2pt]
\textit{11 steps + finish, \phantom{0}92{,}826 input tokens, \textcolor{green!50!black}{reward 1} ($-58\%$)}
\smallskip

\footnotesize
\setlength{\tabcolsep}{3pt}
\renewcommand{\arraystretch}{1.15}
\begin{tabular}{@{}r@{\hspace{0.4em}}p{\dimexpr\linewidth-2em\relax}@{}}
\toprule
\# & Action $\to$ Result \\
\midrule
1  & \texttt{pm list packages | grep -Ei 'markor|broccoli'; find ... -name recipes.txt} $\to$ both packages + recipes.txt path \\
2  & \texttt{cat recipes.txt; run-as broccoli ...; dumpsys package | grep provider} $\to$ recipes read; \texttt{run-as} blocked \\
3  & probe ext dir + \texttt{find /sdcard *broccoli*} + recipe-file candidates $\to$ only \texttt{recipes.txt} matches \\
4  & \texttt{ls /data/.../broccoli/databases} + \texttt{find ... -name '*.db'} $\to$ \textbf{\texttt{broccoli} DB world-readable} \\
5  & \texttt{sql broccoli "SELECT type,name,sql FROM sqlite\_master ..."} $\to$ full schema in one call; FTS triggers identified \\
6  & \texttt{sql broccoli "SELECT recipeId, title, imageName, ..., source, ... FROM recipes LIMIT 10"} $\to$ \textbf{observes \texttt{imageName=''}, \texttt{source=''}} in every existing row \\
7  & \texttt{sql broccoli "SELECT preparationTime, COUNT(*) FROM recipes GROUP BY preparationTime"} $\to$ 0 existing 4-hr rows; safe to insert \\
8  & \texttt{python3 -{}-version 2>/dev/null || echo NO\_PYTHON} $\to$ \texttt{NO\_PYTHON} \\
9  & \texttt{awk -F'|' 'NR==1 || \$4=="4 hrs" \{print NR"|"\$0\}' recipes.txt} $\to$ exactly 3 source rows (Margherita, Grilled Cheese, Avocado Toast) \\
10 & \texttt{sql broccoli "BEGIN; INSERT INTO recipes (..., imageName, ..., source, ...) VALUES ('...', '', ..., '', ...); ... COMMIT;"} $\to$ \textbf{3 rows inserted with \texttt{imageName=''}, \texttt{source=''}} \\
11 & \texttt{am force-stop com.flauschcode.broccoli} $\to$ --- \\
12 & \texttt{finish "Added the 3 recipes ..."} $\to$ \textcolor{green!50!black}{\textbf{Reward 1}} \\
\bottomrule
\end{tabular}
\end{minipage}
\caption{\textbf{Case 1 --- Task~082 (Broccoli recipes), Codex.} Full \texttt{bash-only} (17-step) and \texttt{+tools} (11-step) trajectories; same harness, same seed. The SR shift ($0 \to 1$) traces to repeated nested-quote and heredoc errors in steps~10--16 of \texttt{bash-only} that prevent it from reading the existing-row format, leading it to insert \texttt{NULL} where the verifier expects \texttt{''} (step~17). \texttt{+tools} reads the schema and existing-row format in two \texttt{sql} calls (steps~5--6) and inserts correctly (step~10).}
\label{tab:case_codex_082}
\end{table*}

\begin{table*}[!t]
\centering
\begin{minipage}[t]{0.48\linewidth}
\centering
\textbf{\texttt{bash-only}} \\[2pt]
\textit{17 steps + finish, 445{,}700 input tokens, \textcolor{green!50!black}{reward 1}}
\smallskip

\footnotesize
\setlength{\tabcolsep}{3pt}
\renewcommand{\arraystretch}{1.15}
\begin{tabular}{@{}r@{\hspace{0.4em}}p{\dimexpr\linewidth-2em\relax}@{}}
\toprule
\# & Action $\to$ Result \\
\midrule
1  & \texttt{pm list packages | grep -i markor} $\to$ \texttt{net.gsantner.markor} \\
2  & \texttt{ls /sdcard/Documents/; ls /sdcard/Markor/; ls /sdcard/ | head -30} $\to$ finds \texttt{Markor} under \texttt{/sdcard/Documents/} \\
3  & \texttt{ls -la /sdcard/Markor/} $\to$ no such directory \\
4  & \texttt{ls -la /sdcard/Documents/} $\to$ confirms \texttt{Markor/} subfolder \\
5  & \texttt{ls -la /sdcard/Documents/Markor/} $\to$ empty \\
6  & \texttt{run-as net.gsantner.markor cat shared\_prefs/*.xml | grep dir|folder|path} $\to$ (no output) \\
7  & \texttt{adb shell 'echo "Lunch ... Cafe L'"'"'amour." > /sdcard/.../eHwd\_helpful\_jacket.md \&\& ls -la ... \&\& cat ...'} $\to$ \textcolor{red!70!black}{\texttt{no closing quote}; \texttt{adb: ls requires an argument}; \texttt{adb: unknown command cat}} (\texttt{\&\&} parsed as separate \texttt{adb} subcommands) \\
8  & \texttt{adb shell 'cat > /sdcard/.../eHwd\_helpful\_jacket.md'} $\to$ (no output, stdin not connected) \\
9  & \texttt{ls -la \&\& cat} chain $\to$ same quoting error \\
10 & \texttt{adb shell ls -la /sdcard/Documents/Markor/} $\to$ file exists but \textbf{0 bytes} \\
11 & \texttt{adb shell 'cat > /sdcard/.../...'} $\to$ still empty \\
12 & \texttt{adb shell ls -la ...} $\to$ 0 bytes \\
13 & \texttt{adb shell 'echo "Lunch meeting ... Cafe L'"'"'amour." > /sdcard/.../...'} (no \texttt{\&\&} chain) $\to$ written \\
14 & \texttt{adb shell ls -la ...} $\to$ \textbf{47 bytes} \\
15 & \texttt{adb shell cat ...} $\to$ \texttt{Lunch meeting with Sarah at 1 PM Cafe L'amour.} \\
16 & \texttt{am broadcast MEDIA\_SCANNER\_SCAN\_FILE} $\to$ \texttt{Broadcast completed: result=0} \\
17 & \texttt{am force-stop net.gsantner.markor} $\to$ --- \\
18 & \texttt{finish "Created note ..."} $\to$ \textcolor{green!50!black}{\textbf{Reward 1}} \\
\bottomrule
\end{tabular}
\end{minipage}
\hfill
\begin{minipage}[t]{0.48\linewidth}
\centering
\textbf{\texttt{+tools}} \\[2pt]
\textit{\phantom{0}7 steps + finish, 242{,}191 input tokens, \textcolor{green!50!black}{reward 1} ($-46\%$)}
\smallskip

\footnotesize
\setlength{\tabcolsep}{3pt}
\renewcommand{\arraystretch}{1.15}
\begin{tabular}{@{}r@{\hspace{0.4em}}p{\dimexpr\linewidth-2em\relax}@{}}
\toprule
\# & Action $\to$ Result \\
\midrule
1 & \texttt{pm list packages | grep -i markor} $\to$ \texttt{net.gsantner.markor} \\
2 & \texttt{run-as markor ls shared\_prefs/; ls /sdcard/Documents/; ls /sdcard/Markor/; ls /sdcard/} (one batched probe) $\to$ \texttt{run-as} blocked; \texttt{Markor} under \texttt{/sdcard/Documents/} \\
3 & \texttt{ls -la /sdcard/Documents/Markor/} $\to$ empty \\
4 & \texttt{find-files /sdcard "*.md"} $\to$ (no output --- confirms no stray notes) \\
5 & \texttt{write-file /sdcard/.../eHwd\_helpful\_jacket.md} \quad content: \texttt{Lunch meeting with Sarah at 1 PM Cafe L'amour.} $\to$ written (wrapper base64-encodes body internally; \textbf{apostrophe never enters the shell}) \\
6 & \texttt{read-file /sdcard/.../eHwd\_helpful\_jacket.md} $\to$ \texttt{Lunch meeting ... Cafe L'amour.} (apostrophe preserved) \\
7 & \texttt{am broadcast MEDIA\_SCANNER\_SCAN\_FILE; am force-stop net.gsantner.markor} $\to$ \texttt{Broadcast completed: result=0} \\
8 & \texttt{finish "Created ... with content: ..."} $\to$ \textcolor{green!50!black}{\textbf{Reward 1}} \\
\bottomrule
\end{tabular}
\end{minipage}
\caption{\textbf{Case 2 --- Task~052 (Markor note with apostrophe), Codex.} Full \texttt{bash-only} (17-step) and \texttt{+tools} (7-step) trajectories. Both runs reach reward~$1$; the cost difference comes from \texttt{bash-only} burning steps~7--13 escaping a single apostrophe in the note body (combined with an \texttt{\&\&}-in-\texttt{adb-shell} parsing interaction). \texttt{+tools}'s \texttt{write-file} base64-encodes content internally, so neither the apostrophe nor any other content character is ever exposed to shell parsing.}
\label{tab:case_codex_052}
\end{table*}

\section{Trajectory-Level Error Analysis --- Details}
\label{app:failure_modes}

This appendix details the full taxonomy (\S\ref{app:failure_modes_taxonomy}), the construction pipeline together with its reliability and judge validation (\S\ref{app:failure_modes_construction}), and the extended per-agent results behind the trajectory-level error analysis introduced in Section~\ref{sec:failure_modes} (\S\ref{app:failure_modes_extended_results}).

\subsection{Full taxonomy}
\label{app:failure_modes_taxonomy}

This section defines the shared classification scale used by the trajectory-level error analysis in Section~\ref{sec:failure_modes} across all three comparison axes: (i)~GUI vs.\ CLI paradigms (Figure~\ref{fig:tb_failure_modes}(a)), (ii)~\texttt{bash-only} vs.\ \texttt{+tools} for GPT-5.3 Codex (Figure~\ref{fig:tb_variant_clusters}), and (iii)~model APIs and harnesses jointly (Figure~\ref{fig:tb_model_x_harness}). How the scale was instantiated on our trajectory pool --- the Docent clustering setup, the Opus~4.7 cluster$\to$leaf mapping, and the two-annotator rubric calibration --- is documented separately in Appendix~\ref{app:failure_modes_construction}.

\paragraph{Two-level paradigm-agnostic taxonomy.}
The paradigm-agnostic backbone is a two-level adaptation of MAST~\cite{cemri2026why}, single-agent subset, following Terminal-Bench's mobile adaptation~\cite{merrill2026terminalbench}. \emph{Level~1} is three top-level error classes --- \emph{Execution}, \emph{Coherence}, \emph{Verification} --- that partition the agent's cognitive loop. \emph{Level~2}, underneath each class, is a small set of \textbf{named failure modes} (the MAST \emph{leaves}); the nine MAST leaves we inherit retain their original semantics but are re-stated in Android-specific terms (Table~\ref{tab:taxonomy_leaves}). All nine leaves are defined for both paradigms with identical operational meanings; only their prevalence differs. Using the same Level-1/Level-2 scale for both paradigms is what lets us write a single classification statement about any trajectory regardless of action interface.

\paragraph{Paradigm-specific clusters underneath the taxonomy.}
Below the two-level taxonomy we further annotate each trajectory with a Docent~\cite{meng2025docent} cluster that captures the concrete mobile mechanism behind the failure. These clusters are paradigm-specific because the action interfaces themselves are paradigm-specific: a CLI \emph{wrong write surface} has no GUI analogue, and a GUI \emph{wrong-row tap} has no CLI analogue. The bottom-up clustering pass surfaced $20$ CLI clusters and $18$ GUI clusters (Tables~\ref{tab:taxonomy_cli_subclusters} and~\ref{tab:taxonomy_gui_subclusters}); every classified trajectory is assigned to exactly one cluster, and each cluster's primary MAST leaf is fixed by the mapping step in Appendix~\ref{app:failure_modes_construction}. Clusters therefore play the role of a paradigm-specific \emph{third} layer beneath the paradigm-agnostic taxonomy: they refine, but do not extend, the MAST leaves.

\paragraph{Top-level classes and leaves.}
Table~\ref{tab:taxonomy_leaves} lists the three classes, the nine MAST-derived leaves grouped under them, the operational definition of each leaf, and the primary-label prevalence we observe. Five leaves are active in at least one paradigm; four are uniformly zero in our setting (rationale at the end of this section). We retain the four inactive leaves in the published taxonomy so that future longer-horizon or multi-agent extensions can populate them without redefining the tree. The reading of the multi-label Figure~\ref{fig:tb_failure_modes}(a) vs.\ the primary-label table here is also discussed in Appendix~\ref{app:failure_modes_construction}.

\begin{table*}[t]
\centering
\small
\renewcommand{\arraystretch}{1.25}
\begin{tabular}{l >{\raggedright\arraybackslash}p{0.13\linewidth} >{\raggedright\arraybackslash}p{0.34\linewidth} c c c c}
\toprule
\textbf{Class} & \textbf{Leaf} & \textbf{Operational definition} & \textbf{CLI \%} & \textbf{\#CLI} & \textbf{GUI \%} & \textbf{\#GUI} \\
\midrule
\multirow{3}{*}{\textbf{Execution}}
& \textbf{Disobey Specification (DS)}
& Action is well-formed at the interface but does not satisfy the verifier contract --- wrong surface, wrong value, or wrong artifact for the same goal.
& \textbf{76.7} & 16 & \textbf{40.4} & 11 \\
& \textbf{Step Repetition (SR)}
& Same action repeated on the same observation with no progress; loop continues until the step budget exhausts.
& 0.0  & 0 & \textbf{35.9} & 4 \\
& Unaware of Termination
& Agent does not realize the task is done and keeps issuing actions past the success state.
& 0.0  & 0 & 0.0 & 0 \\
\midrule
\multirow{3}{*}{\textbf{Coherence}}
& \textbf{Reasoning-Action Mismatch (RAM)}
& The action does not implement the predicate the reasoning describes; the trajectory commits to a divergent action because it runs cleanly.
& \textbf{14.0} & 2 & 0.0 & 0 \\
& Context Loss
& Earlier observation or instruction is no longer reflected in the agent's state; the agent acts on stale context.
& 0.0  & 0 & 0.0 & 0 \\
& Task Derailment
& Agent veers to a related but unrequested goal in the middle of the trajectory.
& 0.0  & 0 & 0.0 & 0 \\
\midrule
\multirow{3}{*}{\textbf{Verification}}
& \textbf{Premature Termination (PT)}
& Agent calls \texttt{finish}/\texttt{submit} without a post-action read of the device state; the closing handshake is decoupled from what actually happened.
& \textbf{5.7}  & 1 & \textbf{23.7} & 3 \\
& No or Incorrect Verification
& A verification command is issued but its result is logically mis-interpreted (e.g., empty result read as success).
& 0.0  & 0 & 0.0 & 0 \\
& \textbf{Weak Verification (WV)}
& Verification reads the same surface that was just written, while the grader reads a different surface; the agent's confirmation is non-informative for the actual contract.
& \textbf{3.6} & 1 & 0.0 & 0 \\
\bottomrule
\end{tabular}
\caption{\textbf{Two-level paradigm-agnostic failure taxonomy: three top-level MAST classes and the nine MAST-derived leaves underneath.} The \textbf{CLI~\%} and \textbf{GUI~\%} columns give the primary-label share of failed trajectories assigned to each leaf (active-leaf shares sum to $100\%$ within each paradigm). The \textbf{\#CLI} and \textbf{\#GUI} columns count the paradigm-specific clusters under each leaf (Tables~\ref{tab:taxonomy_cli_subclusters} and~\ref{tab:taxonomy_gui_subclusters}). Bolded leaves are active in at least one paradigm; \textbf{DS} is the cross-paradigm dominant mode. Construction details, including how primary-label vs.\ multi-label shares relate to Figure~\ref{fig:tb_failure_modes}(a), are in Appendix~\ref{app:failure_modes_construction}.}
\label{tab:taxonomy_leaves}
\end{table*}

\paragraph{Paradigm-specific clusters --- CLI.}
Table~\ref{tab:taxonomy_cli_subclusters} enumerates the $20$ CLI clusters grouped under each active leaf (Disobey Specification, Reasoning-Action Mismatch, Premature Termination, Weak Verification). The identifiers are the canonical cluster names used throughout our released artefact, including in Figure~\ref{fig:tb_variant_clusters} (Appendix~\ref{app:failure_modes_extended_results}). The third column gives the dominant Android mechanism behind the failure; the \textbf{Share} column reports each cluster's share of the classified CLI pool.

\begin{table*}[t]
\centering
\scriptsize
\renewcommand{\arraystretch}{1.0}
\setlength{\tabcolsep}{4pt}
\begin{tabular}{c >{\raggedright\arraybackslash}p{0.23\linewidth} >{\raggedright\arraybackslash}p{0.48\linewidth} c}
\toprule
\textbf{Leaf} & \textbf{Docent cluster (canonical name)} & \textbf{Dominant Android mechanism} & \textbf{Share} \\
\midrule
\multirow{16}{*}{\textbf{DS}}
& \snake{wrote\_to\_wrong\_database\_surface}
& Persisted state to a SQLite file or content provider that the target app does not consume (e.g., wrote SMS rows to AOSP \texttt{mmssms.db} when the verifier reads Simple~SMS's own \texttt{conversations.db}).
& 20.7\% \\
& \snake{ocr\_or\_vision\_tool\_hunt\_without\_writing\_target}
& Burned the entire step budget probing for OCR / vision tools (\texttt{tesseract}, \texttt{strings}/\texttt{xxd} over image bytes, MediaStore description columns) and never executed the target \texttt{INSERT} or \texttt{write-file}.
& 8.9\% \\
& \snake{task\_required\_gui\_only\_no\_cli\_pathway}
& Task ground truth is gated by an on-screen tap or rendered DOM state with no shell write surface (Chrome canvas verifiers, Camera shutter, AudioRecorder UI). Agent computed answers offline via \texttt{awk}/\texttt{bc} or fabricated artifacts but the verifier reads the live foreground UI.
& 8.2\% \\
& \snake{fabricated\_artifact\_instead\_of\_invoking\_app\_pipeline}
& Hand-crafted the artifact (base64 JPEG, \texttt{screenrecord} MP4, GPX with invented coords, raw contacts2.db \texttt{INSERT}) instead of driving the responsible app to produce real provenance bytes/rows.
& 8.2\% \\
& \snake{permission\_or\_role\_blocked\_clipboard\_or\_sms}
& Looped on shell calls to clipboard / SMS surfaces blocked by Android~13's permission gate, often misreading a \texttt{SecurityException} parcel as ``clipboard empty'' rather than pivoting to a foreground app or the default-SMS role.
& 6.3\% \\
& \snake{time\_or\_timezone\_misinterpretation}
& Encoded ambiguous time literals as the wrong period (PM vs.\ AM, UTC vs.\ device-local), or used overlap (\texttt{start<end AND end>start}) instead of containment in time-window queries.
& 4.4\% \\
& \snake{recurrence\_or\_filter\_predicate\_omission}
& SQL or content-query predicate dropped a required column the schema exposed: \texttt{repeat\_interval} on recurring events, \texttt{dueDate} vs.\ \texttt{completed} on org.tasks, \texttt{hideUntil}/\texttt{importance}/\texttt{parent} on subtasks.
& 4.0\% \\
& \snake{byte\_exact\_file\_content\_mismatch}
& Markor merge/edit/paste produced almost-right bytes but introduced \texttt{\textbackslash n\textbackslash n} separators where a single \texttt{\textbackslash n} was required, dropped a trailing newline, or substituted file extension; \texttt{cat}-based self-checks could not catch the byte-level disagreement.
& 3.0\% \\
& \snake{apk\_static\_analysis\_loop\_without\_db\_write}
& Locked into $30$--$50$ turns of \texttt{strings}/\texttt{unzip -p}/\texttt{xxd}/\texttt{dexdump} against \texttt{base.apk} trying to recover an enum-or-ordinal, and never executed the required \texttt{INSERT}.
& 3.0\% \\
& \snake{truncated\_input\_treated\_as\_complete}
& Read a source file or dump where the harness cut output mid-row, then synthesized invented continuations rather than paginating with \texttt{sed}/\texttt{tail}/\texttt{wc -l} or re-issuing with a projection.
& 2.8\% \\
& \snake{wrong\_notebook\_root\_for\_markor}
& Inferred Markor's notebook directory from the presence of seeded \texttt{.md} files instead of reading \texttt{pref\_key\_\_notebook\_directory} from shared\_prefs, writing notes one directory level too deep or too shallow.
& 2.3\% \\
& \snake{screenrecord\_or\_screencap\_substitute\_for\_camera}
& Substituted \texttt{screenrecord} MP4 or \texttt{screencap -p} PNG (sometimes renamed \texttt{.jpg}) for the camera shutter; verifier checked \texttt{MediaStore.owner\_package\_name} and rejected the screen-capture provenance.
& 2.3\% \\
& \snake{pm\_clear\_or\_destructive\_blanket\_action}
& Used \texttt{pm clear}, \texttt{rm -rf} of an entire directory, or \texttt{DELETE FROM sqlite\_master} where the verifier expected a single-row delete; wiped sibling data that was meant to remain intact.
& 0.8\% \\
& \snake{shell\_quoting\_or\_harness\_parse\_blocked\_action}
& Command died in the wrapper due to nested-quote / apostrophe issues (\texttt{L'amour} inside an inner-shell \texttt{INSERT}), multi-JSON-per-turn parse errors, or \texttt{max\_tokens} truncation; zero commands reached the device.
& 0.6\% \\
& \snake{chrome\_firstrun\_or\_file\_scheme\_blockade}
& Burned steps fighting Chrome's \texttt{FirstRunActivity} via shared\_prefs edits and \texttt{am start ChromeTabbedActivity} hacks instead of tapping the FRE accept button through Files-app$\to$Chrome.
& 0.6\% \\
& \snake{harness\_verb\_or\_command\_prefix\_violation}
& Emitted bare Android verbs (\texttt{settings}, \texttt{cmd}, \texttt{svc}, \texttt{am}) without the \texttt{adb shell} prefix the wrapper requires; the parser rejected every command before it reached the device.
& 0.4\% \\
\midrule
\multirow{2}{*}{\textbf{RAM}}
& \snake{treated\_broadcast\_or\_intent\_dispatch\_as\_state\_change}
& Fired \texttt{am broadcast}/\texttt{am start} with an invented or AOSP-only action string and accepted \texttt{result=0} or a \texttt{dumpsys} top-activity match as proof of the state change, without inspecting persisted shared\_prefs / SQLite / UI.
& 7.6\% \\
& \snake{inverted\_or\_guessed\_enum\_mapping}
& Wrote a guessed integer for an unknown enum (\texttt{importance} where $0$ means \emph{High}; \texttt{repeat\_rule} weekday bitmask; Pro~Expense \texttt{category}; DeskClock \texttt{timer\_state\_0}) after reasoning admitted the mapping was uncertain.
& 6.3\% \\
\midrule
\textbf{PT}
& \snake{answer\_question\_through\_wrong\_provider}
& For app-specific Q\&A (Simple~Calendar~Pro, OpenTracks, org.tasks), queried the AOSP system provider, accepted \texttt{No result found.} as final, and answered \texttt{None}; never tried the app-private SQLite under \texttt{/data/data/<pkg>/databases/}.
& 5.7\% \\
\midrule
\textbf{WV}
& \snake{self\_confirming\_select\_loop}
& After any write (\texttt{sqlite3 INSERT}, \texttt{content insert}, \texttt{settings put}), verified success by re-reading the exact same surface with the inverse verb, never round-tripping through the consumer's actual read path (\texttt{content query}, app relaunch, MediaStore re-scan).
& 3.6\% \\
\bottomrule
\end{tabular}
\caption{\textbf{Paradigm-specific CLI clusters grouped under their assigned MAST leaf.} The \textbf{Share} column reports each cluster's primary-label share of the classified CLI pool; within-leaf shares sum to the leaf totals in Table~\ref{tab:taxonomy_leaves}, and cross-leaf shares sum to $100\%$. Cluster$\to$leaf assignments, the rubric and tie-breakers, and the two-annotator agreement statistics are reported in Appendix~\ref{app:failure_modes_construction}.}
\label{tab:taxonomy_cli_subclusters}
\end{table*}

\paragraph{Paradigm-specific clusters --- GUI.}
Table~\ref{tab:taxonomy_gui_subclusters} enumerates the $18$ GUI clusters grouped under each active leaf (Disobey Specification, Step Repetition, Premature Termination); none of the four Coherence or remaining Verification leaves are activated, consistent with the per-leaf zeros in Table~\ref{tab:taxonomy_leaves}. The GUI mass is more concentrated than the CLI mass: a single Step-Repetition cluster (\snake{identical\_action\_loop\_until\_step\_budget\_exhausted}) absorbs more than a quarter of all GUI failures, and three clusters above $10\%$ jointly account for over half of the pool.

\begin{table*}[t]
\centering
\scriptsize
\renewcommand{\arraystretch}{1.0}
\setlength{\tabcolsep}{4pt}
\begin{tabular}{c >{\raggedright\arraybackslash}p{0.23\linewidth} >{\raggedright\arraybackslash}p{0.48\linewidth} c}
\toprule
\textbf{Leaf} & \textbf{Docent cluster (canonical name)} & \textbf{Dominant Android mechanism} & \textbf{Share} \\
\midrule
\multirow{11}{*}{\textbf{DS}}
& \snake{skipped\_source\_image\_then\_fabricated\_destination\_entries}
& Launched the viewer app for a named source media file, switched apps within $1$--$2$ steps with no full-screen render observed, then typed invented destination values (Pro~Expense rows, Broccoli recipe lines, Markor transcriptions).
& 11.8\% \\
& \snake{missing\_two\_step\_record\_or\_count\_observation}
& Collapsed an observe-between-actions pattern (start-record/stop-record; tap-then-count) into back-to-back taps at the same coordinate with no intervening observation, so no artifact or correct aggregate is produced.
& 4.9\% \\
& \snake{stuck\_on\_search\_or\_filter\_dialog\_with\_wrong\_field\_type}
& Typed human-readable date strings or category phrases into a title-only search field (Tasks, Broccoli, OpenTracks), got empty results, then answered \texttt{0}/\texttt{""} or looped on the same wrong query.
& 4.5\% \\
& \snake{deletion\_or\_move\_targeted\_wrong\_row\_or\_first\_match\_only}
& Long-pressed a visually similar but wrong filename (\texttt{2023\_02\_13\_shy\_king\_copy.md} instead of \texttt{shy\_king\_copy.md}) or deleted only the first visible row of a multi-row category.
& 4.5\% \\
& \snake{wrong\_menu\_path\_for\_markor\_rename}
& Cycled through Markor's editor overflow menu (\textit{File settings} $\to$ \textit{Format}) instead of long-pressing the file-list row, so the filename never changes.
& 4.1\% \\
& \snake{missing\_long\_press\_gesture\_for\_selection\_or\_marker}
& Used \texttt{click} or \texttt{scroll} where the Android idiom requires \texttt{long\_press} (Markor delete, OsmAnd marker, VLC multi-select, text-selection handles) and looped on the wrong gesture indefinitely.
& 2.4\% \\
& \snake{single\_tap\_on\_html\_canvas\_instead\_of\_swipe\_stroke}
& On the \texttt{task.html} drawing subtask in Chrome, issued a single \texttt{click} at one canvas coordinate per colour instead of a \texttt{swipe}/drag, so no stroke was registered and the canvas was submitted blank.
& 2.0\% \\
& \snake{wrong\_clock\_face\_digit\_in\_time\_picker}
& On Simple~Calendar~Pro's Material 24-hour TimePicker, tapped a digit on the wrong ring leaving the default $16{:}00$ unchanged, or set only the end time without touching start.
& 2.0\% \\
& \snake{answer\_string\_emitted\_as\_unknown\_action\_or\_omitted\_entirely}
& Had the correct comma-separated titles in reasoning but emitted them under \texttt{action\_type:"unknown"}/\texttt{"status"}/\texttt{"wait"} instead of \texttt{"answer"}; verifier reports a missing answer payload.
& 1.6\% \\
& \snake{single\_clipboard\_slot\_overwritten\_during\_multi\_source\_merge}
& Chained \textit{Copy} across all source files before any \textit{Paste}, so each new Copy overwrote Android's single-slot clipboard and final Pastes delivered only the last file's content.
& 1.6\% \\
& \snake{sent\_message\_in\_wrong\_conversation\_thread}
& Read the source SMS in the sender's thread, then composed and sent the reply inside the same thread rather than backing out to start a new conversation with the named recipient.
& 0.8\% \\
\midrule
\multirow{4}{*}{\textbf{SR}}
& \snake{identical\_action\_loop\_until\_step\_budget\_exhausted}
& Fired the exact same action (click at one coordinate, scroll in one direction, \texttt{wait}, \texttt{open\_app}) for $30$--$50$ consecutive steps with byte-identical reasoning, never branching when the screen failed to advance.
& 27.3\% \\
& \snake{fixed\_coordinate\_delete\_macro\_on\_reflowing\_list}
& Recorded a $4$-tap delete sequence (row long-press $\to$ overflow $\to$ Delete $\to$ Confirm) at hard-coded $y$-coordinates on Broccoli/Markor and replayed it $5$--$9$ times without re-reading the list as rows shifted upward.
& 5.3\% \\
& \snake{horizontal\_chip\_row\_scroll\_failure\_in\_pro\_expense}
& Issued $\geq 10$ consecutive \texttt{scroll(left)} (or alternating left/right) actions against Pro~Expense's category chip strip, never tapping a visible chip or trying vertical scroll, never reaching SAVE.
& 2.0\% \\
& \snake{calendar\_chevron\_oscillation\_without\_view\_switch}
& Alternated forward and backward chevron taps in the Calendar day/month header for $30$--$44$ steps, undoing its own navigation, never switching to week/agenda view or emitting an \texttt{answer}.
& 1.2\% \\
\midrule
\multirow{3}{*}{\textbf{PT}}
& \snake{answered\_immediately\_after\_open\_app\_without\_reading\_screen}
& For retrieval tasks, emitted an \texttt{answer} action one step after \texttt{open\_app} with no scroll/click/wait in between, returning a guessed number, a fabricated comma-separated list, or \texttt{0}.
& 11.8\% \\
& \snake{declared\_complete\_before\_dialog\_confirmed}
& Filled a dialog (folder name, brightness slider, time picker, delete confirmation) and emitted \texttt{status=complete} before tapping the final OK/confirm button or screenshotting the post-tap state.
& 9.0\% \\
& \snake{filled\_only\_one\_form\_instance\_for\_multi\_item\_task}
& Task required $N$ items (recipes, transactions, expenses) but the agent saved only one entry and declared complete, or treated a multi-recipe source file as a single copy-paste into one Title field.
& 2.9\% \\
\bottomrule
\end{tabular}
\caption{\textbf{Paradigm-specific GUI clusters grouped under their assigned MAST leaf.} The \textbf{Share} column reports each cluster's primary-label share of the classified GUI pool; within-leaf shares sum to the leaf totals in Table~\ref{tab:taxonomy_leaves}, and cross-leaf shares sum to $100\%$. The GUI mass concentrates in Step Repetition and Premature Termination ($35.9 + 23.7 = 59.6\%$), whereas the CLI mass concentrates in Disobey Specification ($76.7\%$): the same paradigm-agnostic two-level tree separates the two action interfaces cleanly without redefining any leaf. Construction details are in Appendix~\ref{app:failure_modes_construction}.}
\label{tab:taxonomy_gui_subclusters}
\end{table*}

\paragraph{Departures from MAST.}
We follow Terminal-Bench in dropping MAST's inter-agent coordination subcategories (\textit{Specification Issues}, \textit{Inter-Agent Misalignment}) because all our agents are single-agent. Within the remaining tree, the four uniformly-zero leaves in Table~\ref{tab:taxonomy_leaves} have a single-agent rationale: \textit{Unaware of Termination} does not fire because both paradigms expose an explicit \texttt{finish}/\texttt{submit} action with hard semantics; \textit{Context Loss} and \textit{Task Derailment} do not fire because our tasks are short-horizon ($\leq 50$ steps) and the goal text is re-displayed every step in all three harnesses; \textit{No or Incorrect Verification} is empirically subsumed by Weak Verification (verifying on the wrong surface) in our setting, because the cases where the verification \emph{logic} is malformed turn out to coincide with cases where the verification \emph{surface} is also wrong.

\subsection{Construction}
\label{app:failure_modes_construction}
\label{app:failure_modes_pipeline}
\label{app:failure_modes_validation}

This appendix documents how the taxonomy in \S\ref{app:failure_modes_taxonomy} was instantiated on our trajectory pool and how its reliability was validated.

\paragraph{Pipeline.}
The labelling pipeline runs in three stages:
\begin{itemize}\setlength\itemsep{2pt}
\item \textbf{Per-trajectory failure summary.} Each rollout is distilled into a structured, rubric-shaped paragraph by Claude Opus~4.7, recording the agent's surface, value, and verification choices.
\item \textbf{Bottom-up clustering (Docent~\cite{meng2025docent}).} Docent agglomeratively clusters the per-trajectory summaries within each paradigm separately, yielding the paradigm-specific clusters listed in Tables~\ref{tab:taxonomy_cli_subclusters} and~\ref{tab:taxonomy_gui_subclusters}.
\item \textbf{Cluster $\to$ leaf mapping.} Claude Opus~4.7 reads each cluster's exemplars and assigns it a single MAST leaf, using a rubric that was \emph{calibrated against two human annotators} on a $20$-trajectory calibration subset before being released to the rest of the pool.
\end{itemize}

\paragraph{Docent setup.}
We use Docent~\cite{meng2025docent} (\texttt{v0.4.2}) with its default setup. The CLI and GUI failure pools are clustered separately so that paradigm-specific mechanisms surface as distinct clusters. GUI trajectories are pre-processed by rendering each rollout as an action$+$observation transcript: action JSONs are kept verbatim and the corresponding accessibility-tree screenshots are replaced by a one-sentence Claude-Opus~4.7 caption of the rendered screen, keeping the input length comparable to a CLI rollout. Docent surfaces $20$ CLI clusters and $18$ GUI clusters; singletons are discarded and the residual ``noise'' bucket (cluster $-1$) accounts for $4.1\%$ of CLI failures and $2.4\%$ of GUI failures, which we exclude from the per-cluster prevalence numbers.

\paragraph{Cluster$\to$leaf prompt (verbatim).}
The Opus~4.7 judge receives the prompt in Figure~\ref{fig:judge_prompt} for every cluster. \texttt{\{cluster\_id\}} is the canonical Docent name, \texttt{\{exemplars\}} is a random sample of eight trajectory summaries drawn from the cluster, and the two rubric files are passed verbatim in the system prompt.

\begin{figure*}[t]
\centering
\begin{minipage}{0.96\linewidth}
\footnotesize
\begin{verbatim}
You are mapping a Docent failure-mode cluster onto our two-level mobile failure taxonomy:
3 top-level classes (Execution, Coherence, Verification) and 9 named leaves underneath.
The taxonomy is single-agent MAST adapted to Android per Terminal-Bench.

INPUTS
  cluster_id     : canonical Docent cluster name
  exemplars      : 8 per-trajectory failure summaries from this cluster
  rubric         : rubric_v2.md (one-sentence operational definition of each of the
                   9 leaves; Android-specific framings)
  tie_breakers   : rubric_v2_clarifications.md (16 numbered rules that resolve the
                   most common 2-leaf ambiguities; each rule has a pattern,
                   a "maps to" leaf, and an example)

TASK
  1. Read the 8 exemplars. State the dominant Android failure mechanism in one
     sentence, in your own words (no taxonomy jargon).
  2. Walk the tree top-down. Which top-level class (Execution, Coherence, or
     Verification) best describes the failure? Justify in one sentence.
  3. Within the chosen class, pick the single best-fitting leaf.
  4. If two leaves are equally plausible, consult tie_breakers and cite the rule
     number that decides (e.g. "TB7: timezone misinterpretation -> Disobey
     Specification, wrong output protocol").
  5. Optionally record one secondary leaf, only if at least 3 of the 8 exemplars
     also support it. Never list more than one secondary.

OUTPUT  (strict JSON, no prose outside the object)
  {
    "cluster_id":     "...",
    "mechanism":      "one sentence in plain English",
    "primary_class":  "Execution | Coherence | Verification",
    "primary_leaf":   "...",
    "tie_breaker":    "TB# or null",
    "secondary_leaf": "... or null",
    "rationale":      "2-4 sentences grounded in specific exemplars"
  }
\end{verbatim}
\end{minipage}
\caption{\textbf{Verbatim Opus~4.7 prompt for the cluster$\to$leaf mapping step (Appendix~\ref{app:failure_modes_construction}).} \texttt{rubric\_v2.md} and \texttt{rubric\_v2\_clarifications.md} are passed in the system prompt. The judge runs in JSON mode with temperature $0$, reasoning effort \texttt{high}, and a maximum of $4{,}096$ output tokens.}
\label{fig:judge_prompt}
\end{figure*}

\paragraph{Rubric structure.}
The rubric (\texttt{rubric\_v2.md}) is a flat, $9$-row file: one row per MAST leaf, each row giving (i)~a one-sentence operational definition, (ii)~an Android-specific framing in the spirit of Section~\S\ref{app:failure_modes_taxonomy}, and (iii)~$2$--$3$ positive exemplars and $1$--$2$ negative exemplars drawn from the calibration subset (see below). The assignment procedure is the top-down tree walk in the prompt above: a cluster is first placed in one of three top-level classes (Execution/Coherence/Verification), then in the best-fitting leaf within that class. The companion file \texttt{rubric\_v2\_clarifications.md} contains $16$ numbered tie-breakers covering ambiguities that surfaced during calibration --- e.g.\ same-surface verification (DS vs.\ WV), timezone misinterpretation (DS, wrong value), Q\&A from the wrong provider (PT vs.\ WV), GUI long-press idiom (DS, wrong API level), single-slot clipboard during multi-source merge (DS, wrong protocol). Each tie-breaker is a single paragraph stating the pattern, the leaf it maps to, and a worked example, and is referenced by number from the judge prompt.

\paragraph{Inter-annotator agreement.}
On the $20$-trajectory calibration subset, two human annotators (one author and one external researcher with $5+$ years of Android-development experience, blinded to the cluster identities) independently assigned each trajectory a top-level class. They agreed on $18/20$ trajectories (raw agreement $P_o = 0.90$); the two disagreements were both at the Coherence/Execution boundary. Cohen's $\kappa = 0.85$ ($95\%$ CI $[0.62, 1.00]$, $1{,}000$-sample bootstrap) --- the gap from $P_o$ to $\kappa$ reflects the chance-agreement correction at three roughly balanced classes ($P_e \approx 1/3$).

\paragraph{Judge validation against human labels.}
On a disjoint, held-out validation set of $M=60$ trajectories ($20$ per top-level class, stratified by paradigm and harness), the Opus~4.7 judge agrees with the consensus human top-level-class label on $56/60$ trajectories ($P_o = 0.933$); Cohen's $\kappa = 0.90$ ($95\%$ CI $[0.80, 0.98]$, $1{,}000$-sample bootstrap), at or above the inter-human $\kappa = 0.85$ on the calibration subset, so we treat the judge as a substitute for a third human annotator.

\paragraph{Cluster$\to$leaf mapping table (released artefact).}
The full $38$-row mapping --- every Docent cluster, its assigned primary leaf, the tie-breaker invoked (if any), the secondary leaf (if any), and a flag for whether the Opus~4.7 judge agreed with both human annotators --- is too large to reproduce inline; it is released as \texttt{cluster\_to\_leaf\_mapping.csv} in the analysis bundle. Across all $38$ clusters, the judge agreed with the consensus human label on $36$ ($94.7\%$).

\subsection{Extended results}
\label{app:failure_modes_extended_results}

\begin{figure*}[t]
\centering
\includegraphics[width=\linewidth]{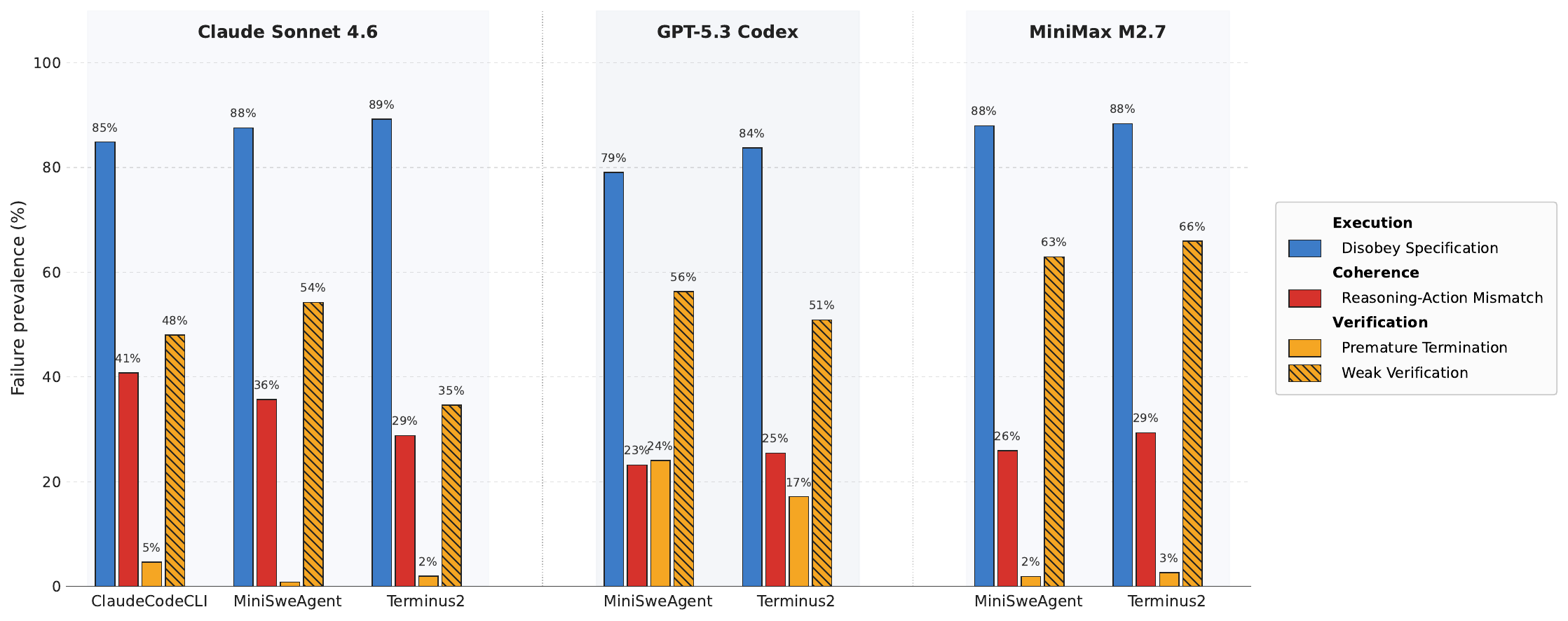}
\caption{\textbf{Failure-mode prevalence by model and harness in the CLI paradigm.} Each column corresponds to one (model $\times$ harness) cell on the fair-comparison subset (1{,}508 failed trajectories across the 103 CLI-solvable AndroidWorld tasks). Bars give the prevalence of each sub-level category. The empty Codex / MiniMax slots under ClaudeCodeCLI reflect that ClaudeCodeCLI is a Claude-only harness.}
\label{fig:tb_model_x_harness}
\end{figure*}

\begin{figure*}[t]
\centering
\includegraphics[width=\linewidth]{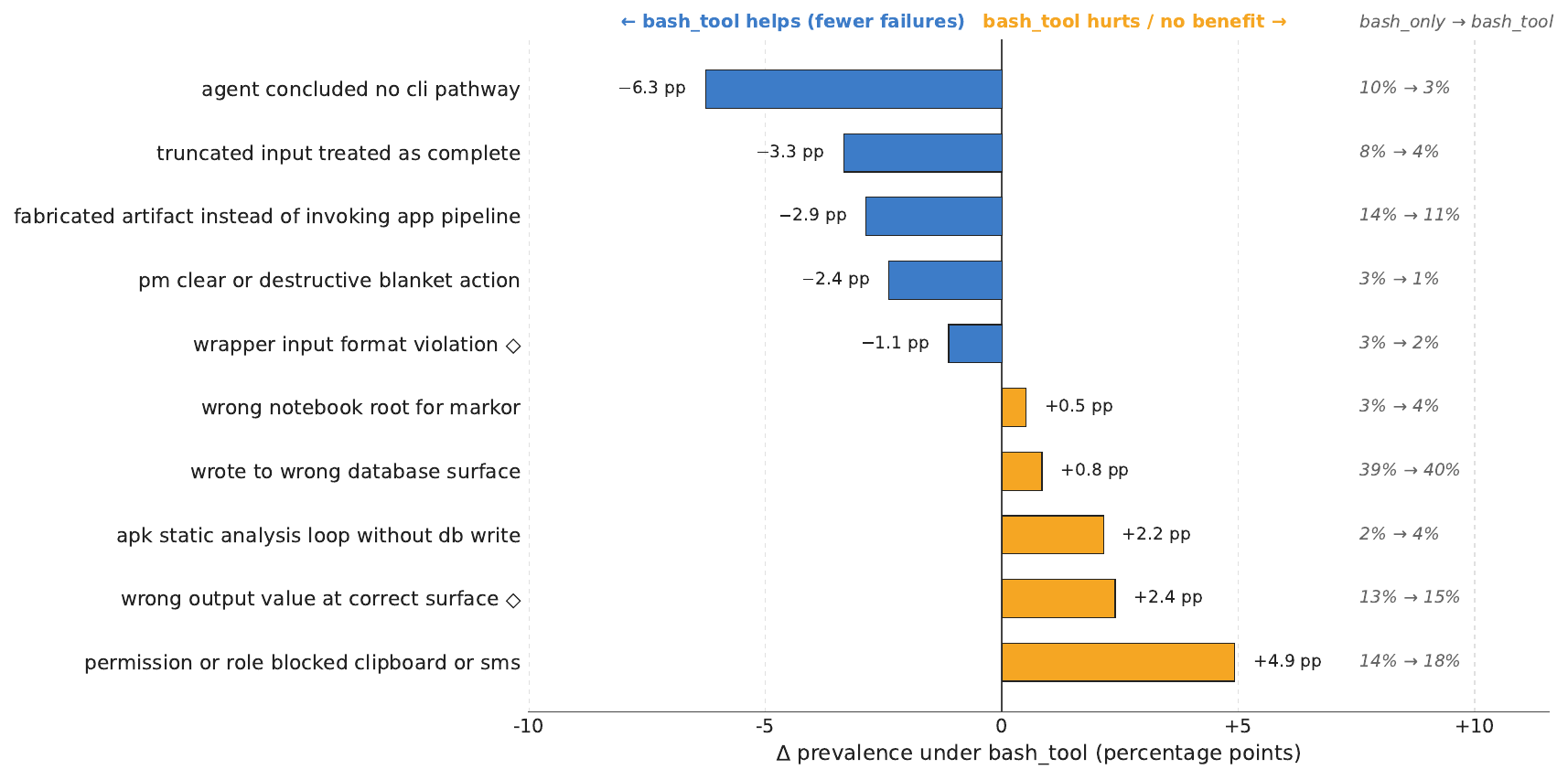}
\caption{\textbf{Disobey Specification sub-cluster shifts under the four-tool wrapper (GPT-5.3 Codex).} Each row is one fine-grained sub-cluster within Disobey Specification; bar length is the change in prevalence from \texttt{bash-only} to \texttt{+tools} on the fair-comparison subset. Blue bars (top) shrink under the wrapper; orange bars (bottom) grow. The italic column reports the absolute prevalence pair (\texttt{bash-only} $\rightarrow$ \texttt{+tools}).}
\label{fig:tb_variant_clusters}
\end{figure*}

This subsection covers the two extensions of Section~\ref{sec:failure_modes} that did not fit in the main paper: the joint model $\times$ harness breakdown that motivates Axis~(iii) of our error analysis (Figure~\ref{fig:tb_model_x_harness}), and the sub-cluster decomposition of the headline Axis~(ii) result on \emph{Disobey Specification} (Figure~\ref{fig:tb_variant_clusters}).

\paragraph{Failure-mode prevalence by model and harness (Figure~\ref{fig:tb_model_x_harness}).}
Figure~\ref{fig:tb_model_x_harness} expands the cross-paradigm view of Figure~\ref{fig:tb_failure_modes}(a) into the full model $\times$ harness matrix used by the CLI block of Table~\ref{tab:main_results}: rows of the figure are sub-level failure categories from the taxonomy of Appendix~\ref{app:failure_modes_taxonomy}, and columns are the eight CLI cells (three harnesses --- Claude Code, mini-swe-agent, Terminus-2 --- crossed with the four model APIs of Section~\ref{sec:main_results}, with the Claude Code $\times$ \{Codex, MiniMax\} cells empty because the Claude Code SDK is Claude-only). The takeaway, complementing the SR-level robustness claim of Section~\ref{sec:main_results}, is that \emph{the failure-mode distribution is shaped primarily by the model API, not the harness}: holding the model fixed, the three harness columns sit within a few percentage points of each other on every active sub-category. Holding the harness fixed and varying the model, by contrast, moves the bars substantially --- most visibly, GPT-5.3 Codex carries a much larger \emph{Premature Termination} mass than the Claude models on the same harness, which is consistent with its bash-only SR deficit in Table~\ref{tab:aw_tool_ablation} and with the Axis~(ii) mechanism described next. All bars are computed over the failed CLI trajectories on the $103$-task CLI-solvable AndroidWorld subset (Appendix~\ref{app:oracle}), averaged across three seeds.

\paragraph{Disobey Specification sub-category shifts under \texttt{+tools} for Codex (Figure~\ref{fig:tb_variant_clusters}).}
Figure~\ref{fig:tb_variant_clusters} drills into the headline Axis~(ii) result of Section~\ref{sec:failure_modes} --- the four-tool wrapper's effect on GPT-5.3 Codex --- at the sub-cluster level of the dominant \emph{Disobey Specification} mode. The figure decomposes which fine-grained sub-clusters drive that net change. Blue (top) bars are the sub-clusters that shrink under the wrapper; orange (bottom) bars are the ones that grow. The blue mass concentrates on the shell-mechanical failures the four tools are designed to neutralize --- nested-quote and heredoc errors, apostrophe-in-text escaping, unicode round-trip corruption, \texttt{adb shell} double-escaping --- consistent with the Codex case studies in Appendix~\ref{app:codex_escaping}, where \texttt{+tools} converts a bash-only failure into a success by absorbing exactly this category of quoting failure into the \texttt{sql} / \texttt{write-file} wrappers. The orange mass shifts toward higher-level specification mismatches (wrong URI, wrong column, wrong timestamp unit) that are visible only once the agent can execute the right shell at all --- in other words, removing the shell-level wall does not eliminate the trajectory's failure modes, it advances them downstream. This is the precise mechanism behind the Axis~(ii) summary in Section~\ref{sec:failure_modes} (\emph{Premature Termination} cut by half, \emph{Reasoning-Action Mismatch} and \emph{Weak Verification} both up by $\sim 10$~pp) and behind the same-direction SR / cost shift in Table~\ref{tab:aw_tool_ablation}: the tools convert a class of bash-mechanics failures into a smaller class of downstream reasoning errors, raising SR while leaving the harder semantic failures for the model to solve.

\section{LLMs Usage}
ChatGPT5 is used solely as a general-purpose writing assistant. Specifically, we apply a fixed prompt template — “Polish the writing in a concise and academic way” — to improve grammar, clarity, and style of text written by the authors. The LLM did not contribute to research ideation, methodology design, experimental setup, analysis, or result interpretation.

\section{Potential Risks and Impact Statement}
\label{app:impact}

We flag two privacy risks specific to deploying CLI mobile agents
outside the benchmark setting. First, on a real device the CLI
surface reads user-private storage directly through
\texttt{adb shell} and \texttt{content} queries, without the
permission prompts, SAF intents, or provider-authority checks that
mediate ordinary app access; any deployed CLI agent is therefore
a privileged on-device process that sees the user's data in clear
form by construction. Second, the agents we evaluate are driven by
frontier model APIs hosted in the cloud, which means that whatever
the on-device CLI reads --- including the user-private data above
and any task context that quotes it --- is sent off-device to the
model provider on every step; a careless deployment leaks user
information to a third party.

\end{document}